\documentclass[11pt,tightenlines,eqsecnum,floats,aps,amsmath,amssymb,nofootinbib,prd,shownopacs,floatfix]{revtex4}
\usepackage[english]{babel}
\usepackage[utf8x]{inputenc}
\usepackage{amsmath,amsfonts,amssymb,extarrows,mathrsfs}
\usepackage{xcolor}
\usepackage{physics}
\usepackage{graphicx}
\usepackage{float}
\usepackage{empheq}
\usepackage{hyperref}
\usepackage{bbm}
\usepackage{pdfpages}
\usepackage{makecell}
\usepackage{tensor}
\usepackage{dsfont}
\usepackage{tikz}
\usetikzlibrary{trees}

\newcommand{\be}{\begin{equation}}
\newcommand{\ee}{\end{equation}}
\newcommand{\ba}{\begin{eqnarray}}
\newcommand{\ea}{\end{eqnarray}}

\newcommand{\defeq}{\mathrel{\mathop:}=}
\newcommand{\eqdef}{\mathrel{\mathop=}:}
\newcommand{\ch}{\,\textrm{ch}}
\newcommand{\sh}{\,\textrm{sh}}
\renewcommand{\d}{\mathrm{d}}
\newcommand{\q}{\mathbf{q}}
\newcommand{\p}{\mathbf{p}}
\newcommand{\vecq}{\mathbf{q}}
\newcommand{\vecp}{\mathbf{p}}
\newcommand{\vecQ}{\mathbf{Q}}
\newcommand{\vecP}{\mathbf{P}}
\newcommand{\aDsq}{\hat{A}^\dagger \hat{A}^\dagger}
\newcommand{\Hankelf}{H^{(1)}_\nu}
\newcommand{\Hankels}{H^{(2)}_\nu}
\newcommand{\BesselJ}{J_\nu}
\newcommand{\BesselY}{Y_\nu}
\newcommand\ddfrac[2]{\frac{\displaystyle #1}{\displaystyle #2}}

\DeclareMathAlphabet{\mathpzc}{OT1}{pzc}{m}{it}
\newcommand*\bigcdot{\mathpalette\bigcdot@{.5}}
\newcommand{\Norm}[1]{\left\lVert#1\right\rVert}
\begin{document}

\title{{Dynamical Properties of the Mukhanov-Sasaki Hamiltonian in the context of adiabatic vacua and the Lewis-Riesenfeld invariant}}

\author{Max Joseph Fahn}
\thanks{max.j.fahn@gravity.fau.de}
\author{Kristina Giesel}
\thanks{kristina.giesel@gravity.fau.de}
\author{Michael Kobler}
\thanks{michael.kobler@gravity.fau.de}
\affiliation{Institute for Quantum Gravity, Department of Physics,  \\ FAU Erlangen -- N\"urnberg,
Staudtstr. 7, 91058 Erlangen, Germany}

\begin{abstract}
We use the method of the Lewis-Riesenfeld invariant to analyze the dynamical properties of the Mukhanov-Sasaki Hamiltonian and, following this approach, investigate whether we can obtain possible candidates for initial states in the context of inflation considering a quasi-de Sitter spacetime. Our main interest lies in the question to which extent these already well-established methods at the classical and quantum level for finitely many degrees of freedom can be generalized to field theory. As our results show, a straightforward generalization does in general not lead to a unitary operator on Fock space that implements the corresponding time-dependent canonical transformation associated with the Lewis-Riesenfeld invariant. The action of this operator can be rewritten as a time-dependent Bogoliubov transformation and we show that its generalization to Fock space has to be chosen appropriately 
in order that the Shale-Stinespring condition is not violated, where we also compare our results to already existing ones in the literature. Furthermore, our analysis relates the Ermakov differential equation that plays the role of an auxiliary equation, whose solution is necessary to construct the Lewis-Riesenfeld invariant, as well as the corresponding time-dependent canonical transformation to the defining differential equation for adiabatic vacua. Therefore, a given solution of the Ermakov equation directly yields a full solution to the differential equation for adiabatic vacua involving no truncation at some adiabatic order. As a consequence, we can interpret our result obtained here as a kind of non-squeezed Bunch-Davies mode, where the term non-squeezed refers to a possible residual squeezing that can be involved in the unitary operator for certain choices of the Bogoliubov map.
\end{abstract}

\maketitle

\section{Introduction}
In the framework of linear cosmological perturbation theory the Mukhanov-Sasaki equation plays a central role. It encodes the dynamics of the Mukhanov-Sasaki variable, which is a linearized and gauge invariant quantity that is build from a specific combination of matter and gravitational perturbations such that the resulting expression is gauge invariant up to linear order. A way to derive this equation is to consider the Einstein-Hilbert action together with a scalar field minimally coupled to gravity and expand this action up to second order in the perturbations around an FLRW background. One decomposes the perturbations into scalar, vector and tensor perturbations since these decouple at linear order. In the scalar sector, we are left with one physical degree of freedom that can for instance be expressed in terms of the Mukhanov-Sasaki variable denoted by $v(\eta,\mathbf{x})$. Given this, we can express the scalar part of the perturbed action entirely in terms of the Mukhanov-Sasaki variable and the corresponding equation of motion takes the following form \cite{Mukhanov:1990me}:
\begin{equation*}
v''(\eta, \mathbf{x}) - \bigg( \Delta 
+ \frac{z''(\eta)}{z(\eta)} \bigg) v(\eta, \mathbf{x}) = 0, \quad \quad 
z(\eta) = \frac{a}{\mathcal{H}} \frac{\d \bar{\phi}}{d \eta}, \quad \quad
\eta \defeq \int^t \frac{\d \tau}{a(\tau)},
\end{equation*}
where $\Delta$ is the spatial Laplacian, $\eta$ denotes conformal time, $a$ the scale factor, $\bar{\phi}
(\eta)$ the isotropic background scalar field and $\mathcal{H}:=\frac{a'}{a}$ the Hubble parameter with respect to conformal time. Contrary to the background quantities, the linear perturbations carry a position dependence breaking the spatial symmetries of the FLRW background spacetime. Throughout this article we will work with the Fourier transform of this differential equation. For each Fourier mode $v_{\bf k}(\eta)$, this leads to a differential equation given by:
\begin{equation}
\label{eq:Mukhanov-Sasaki}
v''_{\mathbf{k}}(\eta) +  \bigg( \Norm{\mathbf{k}}^2
- \frac{z''(\eta)}{z(\eta)} \bigg) v_{\mathbf{k}}(\eta) = 0,
\end{equation}
where quantities with $\mathbf{k}$-label corresponds to the associated Fourier transforms.  The quantity in the brackets of the Fourier transformed equation is called the Mukhanov-Sasaki frequency $\omega_\mathbf{k}(\eta)$ and reflects the backreaction of the matter degrees of freedom with the background spacetime. Further commonly used gauge invariant quantities in the context of linear cosmological perturbation theory are the Bardeen potential $\Phi_B$ as well as the comoving curvature perturbation $\mathcal{R}$. The latter is related to the Mukhanov-Sasaki variable $v$ by $v= z \mathcal{R}$. Whether one considers a specific gauge invariant quantity is often influenced by the choice of a particular gauge in which these variables simplify and have an obvious physical interpretation. For the Bardeen potential this is the longitudinal gauge, whereas the Mukhanov-Sasaki variable naturally arises in the spatially flat gauge, where it is directly related to the perturbations of the inflaton scalar field. More details about the construction of these gauge invariant variables as well as the derivation of their dynamics from the perturbed Einstein equations in the Lagrangian framework can for instance be found in \cite{Mukhanov}. A similar derivation in the canonical approach is for example presented in \cite{Langlois:1994ec,Giesel:2017roz,Giesel:2018opa,Giesel:2018tcw}. Here we will not work in a particular gauge but take the form of the Mukhanov-Sasaki equation in (\ref{eq:Mukhanov-Sasaki}) as our starting point. As far as a comparison with experimental data is concerned, the relevant quantity is the power spectrum that is defined as the (dimensionless) Fourier transform of the real space two-point correlation function, that is in the case of the quantized Mukhanov-Sasaki variable $\langle 0|\, \hat{v}(\eta,\mathbf{x}), \hat{v}(\eta,\mathbf{y})\,|0\rangle$. 

Obviously the power spectrum can only be determined if some initial state has been chosen with respect to which the correlation functions are defined. The most common choice for the initial state is the Bunch-Davies vacuum that can be uniquely selected by the conditions that it is de Sitter invariant and satisfies the Hadamard condition. The latter requires that the corresponding two-point function has a specific behavior in the ultraviolet, that is for short distances. If we drop the Hadamard condition, we obtain the family of so-called $\alpha$-vacua that include the Bunch-Davies vacuum. Other choices for the initial conditions than the ones for the Bunch-Davies vacuum have been considered and their possible fingerprints on the power spectrum have been investigated, see for instance \cite{Danielsson:2002mb,ArmendarizPicon:2003gd,Handley:2016ods} and references therein. The Bunch-Davies vacuum is selected by requiring that in the limit of $\eta \to -\infty$ the mode functions take the form of the usual Minkowski mode functions. Another method to choose an initial state is the so-called Hamiltonian diagonalization method, where one minimizes the expectation value $\langle 0_{\eta_0}\,|\, \hat{H}(\eta_0)\,|0_{\eta_0}\rangle$ of the Mukhanov-Sasaki Hamiltonian at one moment in time, say $\eta_0$.  Hamiltonian diagonalization refers to the fact that at $\eta_0$ the coefficients of the off-diagonal terms involving second powers of annihilation and creation operators, respectivley, vanish for all modes. That is, at $\eta_0$ the Mukhanov-Sasaki Hamiltonian is given by the field theoretical generalization of the standard harmonic oscillator. Considering this, a natural question to ask is whether such a Hamiltonian diagonalization can be obtained not only instantaneously but for each moment in time and particularly how this aspect is related to the choice of initial states. In order to work into that direction we take into account that the Mukhanov-Sasaki equation represents a time-dependent harmonic oscillator in each Fourier mode, whereas the specific form of the time dependence reflects the properties of the expanding background spacetime. What we are aiming at is a transformation that maps the time-dependent harmonic oscillator to the time-independent harmonic oscillator for each mode and all times. Defining such a transformation will only work if we consider time-depedendent canonical transformations, that are adapted specifically to the two systems of the time-dependent and time-independent harmonic oscillator, respectively. This is conveniently done in the extended phase space framework outlined below.

There has been considerable interest in the study of the time-dependent harmonic oscillator, both in a purely classical and quantum mechanical context. A distinct role in all of these considerations is played by the Lewis-Riesenfeld invariant, which is a constant of motion with respect to the evolution governed by a time-dependent harmonic oscillator. At the classical level, this invariant has been considered in the context of a canonical transformation in the extended phase space \cite{Struckmeier,Chung} that involves time and its momentum as canonical phase space variables among the usual position and momentum variables. Such an extended phase space provides a convenient platform to implement time-dependent canonical transformations. The obtained canonical transformation allows to map the system of the time-dependent harmonic oscillator onto the system of a harmonic oscillator with constant frequency and thus completely removes the time dependence of the Hamiltonian, which drastically simplifies the task of finding solutions of the equations of motion after applying the transformation. As shown in \cite{Hartley:1982ygx}, the invariant can also be defined in the context of quantum mechanics. In this case the eigenstates of the invariant can be used to construct solutions of the Schr\"odinger equation involving the original time-dependent Hamiltonian. Further application are to construct coherent states of the time-dependent harmonic oscillator by means of the eigenstates of the Lewis-Riesenfeld invariant as for instance discussed in \cite{Hartley:1982ygx,Villasenor}.

If we aim at relating the framework of the Lewis-Riesenfeld invariant to the notion of initial states associated with the Mukhanov-Sasaki equation, we need to generalize this approach to the field theory context. There exists already some work in this direction, see for example in \cite{Villasenor,invariant_vacuum} and references therein, although with a slightly different focus than we want to consider here, because both of them do not apply this techniques directly to the Mukhanov-Sasaki equation in the framework of the extended phase space, meaning that they consider different time-dependent frequencies in general, and particularly the generalization to field theory has not been analyzed in very much detail in \cite{invariant_vacuum}. The strategy we want to follow in our work is that first we consider the Lewis-Riesenfeld invariant and the corresponding canonical transformation at the classical level for finitely many degrees of freedom in the extended phase space, building on former work of \cite{Struckmeier,Chung}, who however did not consider the quantization of the canonical transformation. In order to be able to implement the corresponding unitary map at the quantum level, we also construct the corresponding generator of the canonical transformation. For the reason that in the extended phase space the physical system of the time-dependent harmonic oscillator is described as a constrained system, we construct Dirac observables and use the technique of reduced phase space quantization to implement this unitary map on the physical Hilbert space in a quantum mechanical setting, where it can also be formulated in terms of a time-dependent Bogoliubov transformation. Given this setup, we could take the vacuum of the time-independent harmonic oscillator, apply the constructed unitary map to it and obtain a in this sense natural candidate for a vacuum state for the time-dependent harmonic oscillator, that has then been determined directly by means of the unitary map.

The question we want to address in this article is whether we can carry this idea over from finitely many degrees of freedom to field theory and use the Lewis-Riesenfeld invariant approach to obtain possible candidates for initial states. In particular, we are interested in the physical properties of such initial states and their relation to the Bunch-Davies vacuum and other adiabatic vacua. As we will show, the most straightforward generalization to field theory is not possible because the so constructed map involves an infrared divergence, hence the Shale-Stinespring condition is violated. As we will discuss, a suitable modification of the map in the infrared range can be obtained to cure the infrared divergenes. Furthermore, as we will show, if this map is not chosen carefully for all but the infrared modes it can also involve ultraviolet divergences striclty permitting a unitary implementation on Fock space. Interestingly, in the context of the Mukhanov-Sasaki Hamiltonian, different choices at this level can be related to different choices for the initial conditions of the associated mode functions. Moreover it becomes clear that we can recover the defining differential equation for adiabatic vacua from the Ermakov equation, where the latter is an auxiliary differential equation whose solution is needed to explicitly construct the Lewis-Riesenfeld invariant and the corresponding canonical transformation. This allows us to interpret the initial conditions and the result for the Fourier modes we obtain using the method of the Lewis-Riesenfeld invariant in the context of adiabatic vacua.
\\

This article is structured as follows: In section \ref{sec:Extended_PS} we introduce the framework of the extended phase space and rederive the canonical transformation that maps the system of the time-dependent harmonic oscillator to the time-independent one generalizing the approach in \cite{Chung}. The time-rescaling that is involved in this canonical transformation naturally occurs in the extended phase space and the physical interpretation of the Lewis-Riesenfeld invariant can be easily understood. In order to deal with the constrained system in the extended phase space later on, we want to choose reduced phase space quantization and thus derive the reduced phase space in terms of Dirac observables. Their dynamics is generated by the Dirac observable associated with the time-dependent Hamiltonian. As the next step in section \ref{sec:Quant}, we consider the quantization of the system and show that the canonical transformation can be implemented as a unitary map on the one-particle physical Hilbert space, where our results agree with already existing results in the literature for finitely many degrees of freedom. 
In order to simplify the actual application of the unitary operator we perform a generalized Baker-Campbell-Hausdorff decomposition by means of which we then rewrite the unitary transformation as a time-dependent Bogoliubov map. 

Afterwards we consider the generalization of our results obtained so far to field theory, discussing the two most common cases in the literature, where one maps from a time-dependent harmonic oscillator to a harmonic oscillator with either frequency $\omega_\mathbf{k}=k$ or $\omega_\mathbf{k}=1$. As far as the implementation on Fock space is considered, the first choice can be implemented unitarily, whereas the second cannot due to an ultraviolet divergence. This ultraviolet divergence is caused by a residual squeezing operation by which the two maps differ. 
To avoid issues that occur for the infrared modes, we discuss a possible modification of the map using the Arnold transformation discussed in \cite{Arnold_trafo:Guerrero}. Section \ref{sec:applications} presents practical applications of this formalism by considering the case of a quasi-de Sitter spacetime and the corresponding Mukhanov-Sasaki equation in a slow-roll approximation. We construct the Lewis-Riesenfeld invariant and, in the context of a quantum mechanical toy model, compute the lowest and next to lowest eigenvalue eigenstates associated to it and analyze their properties. Finally we summarize and conclude in section \ref{sec:Conclusions}.
\newpage
\section{Extended phase space formulation and time-dependent canonical transformations}
\label{sec:Extended_PS}
A convenient framework for implementing time-dependent canonical transformations is the extended phase space in which also time and its conjugate momentum are treated as phase space variables and thus transformations of them can be naturally formulated. Motivated by the Mukhanov-Sasaki equation, we first investigate a single mode of the equation in a classical context. This corresponds to a harmonic oscillator with time-dependent frequency. We will consider the single mode Mukhanov-Sasaki Hamiltonian as a  mechanical toy model and later generalize the results obtained in this case to the field theory context. Our goal is to remove this explicit time dependence by a time-dependent canonical transformation. This transformation will be defined on the extended phase space as a symplectic map that also includes the time variable and its associated conjugate momentum as phase space degrees of freedom.

\subsection{Time-dependent Hamiltonians on extended phase space}
 As a first step we reformulate the dynamics encoded in the single mode Mukhanov-Sasaki Hamiltonian on the extended phase space, where it becomes a constrained system.  Let us consider a system with finitely many degrees of freedom where we denote  all configuration variables as $\q=(q^1,\cdots,  q^n)$ and the configuration space by $\Sigma$. A time-dependent Lagrangian is then defined as a function $L:T\Sigma\times\mathbb{R}\to\mathbb{R}$. Because we want to include time among the elementary configuration variables, closely following the work in \cite{Struckmeier} and \cite{Chung}, we extend the configuration manifold $\Sigma$ to $M:=\Sigma\times\mathbb{R}$ and rewrite the action as 
    \begin{align}
S[L] &= \int_\mathbb{R} \d s \, L \bigg( \tilde{\q}(s), t(s), \Big( \frac{\d t}{\d s} \Big)^{-1} \frac{\d \tilde{\q}}{\d s} \bigg) \frac{\d t(s)}{\d s} \\[0.5em]
&\defeq \int_\mathbb{R} \d s \, \mathfrak{L} \bigg( \tilde{\q}(s), t(s), \Big( \frac{\d t}{\d s} \Big)^{-1} \frac{\d \tilde{\q}}{\d s}, \frac{\d t(s)}{\d s} \bigg) 
\nonumber \eqdef  S[\mathfrak{L}],
\label{eq:ext_action}
\end{align}
where we will refer to $\mathfrak{L}$ as the \textit{extended} Lagrange function now understood as a function on the extended tangent bundle $TM$ that is even-dimensional and associated to the extended configuration manifold, including the former system evolution parameter commonly referred to as time. A non-degenerate symplectic structure on the corresponding cotangent bundle  $T^*M$, whose elementary variables are $(\tilde{\q},t,\tilde{\p},p_t)$ can be defined as usual. This allows to establish a one-to-one correspondence between smooth phase space functions and Hamiltonian vector fields. In complete analogy to the conventional case, one can formulate the Euler-Lagrange equations in terms of the extended variables by means of the variational principle, which results in equivalent equations of motion as derived from the original action $S[L]$. The equations of motion for the time variable are just given by $\frac{\d t}{\d s}=\lambda(s)$  where $\lambda(s)$ is an arbitrary real parameter reflecting the rescaling symmetry of the action, this reflects the arbitrary parametrization of time and has no physical significance. If we perform a Legendre transform, we realize that $p_t=-H(\tilde{\q},\tilde{p},t)$ becomes a primary constraint since it cannot be solved for the velocities $\frac{dt}{ds}$ with\footnote{In general we could also take into account a time dependent mass in the Hamiltonian, however in the case of the Mukhanov-Sasaki equation it is sufficient to set the mass parameter $m$ equal to $m=1$.} 
\begin{equation*}
H(\tilde{\q},\tilde{\p},t)=\frac{{\tilde{\p}}^2}{2}+\frac{1}{2}\omega^2(t)\tilde{\q}^2,
\end{equation*}
where the Hamiltonian is a function $H:T^*M\to\mathbb{R}$ that is independent of $p_t$. We denote this constraint by $C:=p_t+H(\tilde{\q},\tilde{\p},t)$. Therefore, we apply the Legendre transform for singular systems and obtain the following Hamiltonian on the extended phase space $T^*M$:
\begin{align*}
\mathfrak{H} &= \tilde{p}_a \frac{\d \tilde{q}^{a}}{\d s} 
+ p_t \frac{\d t}{\d s} - \mathfrak{L}\Big|_{\dot{q}^a(\tilde{\q},\tilde{\p},t,\lambda), \frac{\d t}{\d s}=\lambda}
= \Big( H(\tilde{\q}, \tilde{\p}, t) + p_t \Big) \lambda(s)
= \lambda(s) C \approx 0,
\end{align*}
where we used $\approx$ to denote weak equivalence and used the definition of $\lambda(s)$ from above. Due to reparametrization invariance of the extended action, there is no true Hamiltonian but a Hamiltonian \textit{constraint} $C$. From now on we will neglect the tilde on the top of the variables $\q,\p$ to keep our notation more compact. For the time-dependent harmonic oscillator the so-called Lewis-Riesenfeld invariant $I_{LR}$ has played a pivotal role, particularly in the construction of solutions for the corresponding equations of motion. $I_{LR}$  is a phase space function being quadratic in the elementary variables $(\q,\p)$ and its time dependence is encoded in a function $\xi:\rm{I}\subseteq\mathbb{R}\to\mathbb{R}$. Explicitly, it is given by:
\begin{equation}
I_{LR} \big(\vecq, \vecp, t \big) \defeq \frac{1}{2} \bigg( \big( \xi(t) \vecp - \dot{\xi(t)} \vecq \big)^2 
+ \frac{\omega_0^2 \, \vecq^2}{\xi^2(t)} \bigg). \label{eq:LR_invariant}
\end{equation}
Since $I_{LR}$ is an invariant, it has to commute with the constraint $C$ on the extended phase space\footnote{The symplectic form associated with the Poisson bracket $\{.,.\}_{\rm ext}$ on the extended phase space has the form $\Omega=d\tilde{q}^a\wedge d\tilde{p}_a+dt\wedge dp_t$.}:
\begin{equation}
\{ I_{LR}, C \}_{ext} 
= \{ I_{LR}, H(t) \} + \frac{\partial I_{LR}}{\partial t} 
= 0. 
\label{eq:PBILRCons}
\end{equation}
This carries over to a condition on the function $\xi$ that has to satisfy the following non-linear, ordinary, second-order differential equation
\begin{equation}
\bigg( \frac{\d^2}{\d t^2} + \omega(t)^2 \bigg) \xi(t) - \omega_0^2 \, \xi(t)^{-3} = 0,
\label{eq:Ermakov}
\end{equation}
known as the Ermakov equation. It has been shown that $I_{LR}$ is an invariant both at the classical level \cite{Struckmeier} and at the quantum level \cite{Hartley:1982ygx}, \cite{Schroedinger_1D}. In the following, the explicit form of $I_{LR}$ will be our guiding line for finding an extended canonical transformation that removes the time dependence from the Hamiltonian of the harmonic oscillator with a time-dependent frequency.

\subsection{Extended canonical transformations and Hamiltonian flows}
In the framework of the extended phase space formalism we can now regard time as a configuration degree of freedom and consequently apply a canonical transformation to implement a time-rescaling. It is worth noting that the Mukhanov-Sasaki equation in \textit{conformal} time (commonly denoted $\eta$) takes the form of a time-dependent harmonic oscillator, however we shall refer to the time variable as $t$ in the context of the classical and one-particle quantum theory, respectively. We aim at finding a symplectic map $\Phi$ such that the explicitly time-dependent Mukhanov-Sasaki Hamiltonian is mapped into an autonomous one, that is, one with time-independent frequency that we denote by $\omega_0$. During this procedure, the Hamiltonian constraint together with the Poisson structure on the extended phase space remain invariant by construction.

\begin{align}
\Phi: T^*M \to T^*M, \quad
\mathfrak{H} \; \mapsto \; \Phi^* \mathfrak{H} = \mathfrak{H}.
 \label{eq:def_ext_canonical}
\end{align}

Correspondingly, the symplectic form $\Omega$ on $T^*M$ is invariant under the Hamiltonian flow of the associated Hamiltonian vector field of $\Phi$ that infinitesimally generates this transformation. In order to apply this procedure to the case of the single-mode Mukhanov-Sasaki Hamiltonian, we need to impose conditions on the explicit form of $\Phi$. We would like to preserve the functional dependence of the Hamiltonian constraint on the one hand and keep the quadratic order in both momentum and configuration variables on the other hand. For this purpose, we make the following ansatz, closely related to the work presented in \cite{Chung}:

\begin{align}
\Phi: \; \left(
\begin{array}{c}	q^{a}\\	p_a\\	t\\ 	p_t\\ \end{array}
\right) \quad \mapsto \quad
\left(
\begin{array}{c}	Q^{a}(\vecq,t)\\	F(\vecq,t)p_a + G_a(\vecq,t)\\	T(\vecq,t) \\ 	P_T(\vecq, t, \vecp, p_t)\\ \end{array}
\right) \quad \textrm{s.t.} \quad \Phi^* \Omega = \Omega
\label{eq:trafo_ansatz}
\end{align}

Note the ansatz  $\vecp \propto \vecP + \mathbf{G}(\vecq,t)$ ensures that the transformed Hamiltonian is again quadratic in the new momentum, whereas the prefactor allows for a time-rescaling of the momentum variable. Additionally, the only variable that carries a dependence on $p_t$ is the new momentum conjugate to $T$ denoted by $P_T$, which is a choice that preserves the form of the Hamiltonian constraint being linear in the conjugate momentum of the time variable. We employ the ansatz in (\ref{eq:trafo_ansatz}) for the symplectic map $\Phi$ and from subsequent comparison of coefficients of the two-form basis elements we obtain a set of five coupled differential equations that determine the form of $\Phi$ to be a canonical transformation. This system of differential equations corresponds to a generalization to $n+1$ configuration degrees of freedom of the set of equations presented in \cite{Chung}, where only the case for $n=1$ was presented. It explicitly reads:
\begin{align}
\frac{\partial Q^{a}}{\partial t} \bigg( p_a \frac{\partial F}{\partial q^b} 
+ \frac{\partial G_a}{\partial q^b} \bigg) 
+ \frac{\partial P_T}{\partial q^b} \frac{\partial T}{\partial t}
&= \frac{\partial Q^{a}}{\partial q^b} \bigg( p_a \frac{\partial F}{\partial t} 
+ \frac{\partial G_a}{\partial t} \bigg) 
+ \frac{\partial P_T}{\partial t}\frac{\partial T}{\partial q^b},
\nonumber \\[0.6em]
F \frac{\partial Q^{a}}{\partial t} 
+ \frac{\partial P_T}{\partial p_a} \frac{\partial T}{\partial t} = 0, &\qquad
\frac{\partial P_T}{\partial p_a} \frac{\partial T}{\partial q^b} 
+ F \frac{\partial Q^{a}}{\partial q^b} = \tensor{\delta}{^a _b}, \label{eq:canonical_system} \\[0.6em]
\frac{\partial P_T}{\partial p_t} \frac{\partial T}{\partial q^{a}} = 0, &\qquad
\frac{\partial P_T}{\partial p_t} \frac{\partial T}{\partial t} = 1. \nonumber
\end{align}
We can get a first hint how a solution could look like when we consider the Lewis-Riesenfeld invariant $I_{LR}$ from equation (\ref{eq:LR_invariant}) in section \ref{sec:Extended_PS} above.  Hence, there is a natural starting point for finding the favored canonical transformation we are aiming at, by fixing the transformations of $\vecq$ and $\vecp$ according to a factorization of $I_{LR}$. This leads to
\begin{align}
Q^{a}(\vecq,t) \defeq \frac{q^{a}}{\xi(t)} \quad &\Longleftrightarrow \quad 
q^{a}(\vecQ,T) = \xi(t(T)) Q^{a}, \label{eq:qQ_trafo}\\[0.3em]
P_a(\vecq, \vecp,t) \defeq \xi(t) p_a - \dot{\xi} (t) q_a \quad &\Longleftrightarrow \quad
p_a(\vecQ, \vecP, T) = \frac{P_a}{\xi(t(T))} + \dot{\xi}(t(T)) Q_a,
\label{eq:pP_trafo}
\end{align}
where $\dot{\xi} = \partial_t \xi$ is the derivative with respect to the dynamical time variable. In order to proceed, we need to find a suitable transformation for the time variable $T(t)$ that is consistent with the system of equations (\ref{eq:canonical_system}) previously found. A convenient possibility is to use Euler's time scaling transformation for the three-body problem, recently introduced by Struckmeier \cite{Struckmeier} in the context of the time-dependent harmonic oscillator. However, the approach in \cite{Struckmeier} differs from the one outlined in this work in the sense that we derive the explicit form of the transformation instead of making use of the corresponding generating function. The relevant transformation of $t$ is given by:
\begin{align}
T(t) \defeq \int_{t_0}^t \frac{\d \tau}{\xi^2(\tau)} \quad \Longleftrightarrow
\quad \frac{\partial T}{\partial t} = \frac{1}{\xi^2(t)},
\label{eq:timescaling}
\end{align}
where $\xi(t) \in C^2(\mathbb{R})$ is an up-to-now arbitrary function with the only restriction that the above integral needs to be well-defined. Given the explicit form of (\ref{eq:timescaling}), we can require mutual consistency of the transformations in (\ref{eq:canonical_system}) in order to fix the form of the transformed canonical momentum $P_T$. Solving the first equation in (\ref{eq:canonical_system}) for $\partial_b P_T$ and subsequently integrating the obtained expression yields the following result:
\begin{equation}
P_T(\vecq, \vecp, t, p_t) = \xi^2(t) p_t + \xi(t) \dot{\xi}(t) \vecq \cdot \vecp
- \frac{1}{2} \Big( \xi (t) \ddot{\xi}(t) + \dot{\xi}^2(t) \Big) \vecq^2,
\end{equation}
with the term $\xi^2(t) p_t$ arising from an arbitrary additive constant with respect to $\vecq$ and the requirement of inverse scaling behavior between $t$ and $p_t$ according to (\ref{eq:canonical_system}). Now that we have fixed the transformation to the new canonical coordinates, we can use the invariance of the Hamiltonian constraint $\mathfrak{H}$ under the change of canonical coordinates to derive an autonomous Hamiltonian from the original, time-dependent one:
\begin{equation}
\Phi^* \mathfrak{H} = \Big( \Phi^*H + P_T \Big) \frac{\d T}{\d s}
= \Big( \Phi^* H + P_T \Big) \frac{\partial T}{\partial t} \frac{\d t}{\d s}
= \Big( H(\vecq, \vecp, t) + p_t \Big) \frac{\d t}{\d s} = \mathfrak{H}.
\label{eq:constraint_invariance}
\end{equation}
In fact, using the one before the last equality sign in (\ref{eq:constraint_invariance}) we find an expression for $\Phi^* H$:
\begin{equation}
H_0 \defeq \Phi^* H
= \xi^2(t) \Big( H(\vecq, \vecp, t) + p_t \Big)
\Big|_{\big(\Phi \big)(\vecq, \vecp, t)}  - P_T, \label{eq:constraint}
\end{equation}
with $\vecq, \vecp$ and $t$ considered as functions of the new variables $\vecQ, \vecP$ and $T$ via the extended canonical transformation $\Phi(\vecq, \vecp, t)$ defined in equation (\ref{eq:trafo_ansatz}). Analogous to the treatment displayed in \cite{Struckmeier}, we would also like to point out the crucial property that not the bare constraint $C$ but the product with the Lagrange multiplier $\mathfrak{H} = \lambda C(\q, \p, t, p_t)$ is invariant under this transformation by construction. As a consequence, the canonical momentum $P_T$ drops out in $H_0$. If we evaluate all expressions using the inverse of $\Phi$ to express $\q, \p$ in terms of $\vecQ, \vecP$, we finally obtain:
\begin{align}
H_0(\vecQ, \vecP, T) &= \bigg[ \frac{\xi^2(t)}{2} \Big( \vecp^2 + \omega(t)^2 \vecq^2 \Big) 
- \xi(t) \dot{\xi}(t) \vecq \cdot \vecp + \frac{1}{2} \Big( \xi (t) \ddot{\xi}(t) 
+ \dot{\xi}^2(t) \Big) \vecq^2 \bigg] \Bigg
|_{\big(\Phi^{-1}\big)\big(\vecQ, \vecP, T \big)} \nonumber \\[0.5em]
&= \frac{\xi^2}{2} \bigg( \frac{\vecP^2}{\xi^2} 
+ 2 \frac{\dot{\xi}}{\xi} \vecQ \cdot \vecP + \dot{\xi}^2 \vecQ^2 + \omega(t(T))^2 \xi^2 \vecQ^2 \bigg) - \xi \dot{\xi} \vecQ \cdot \vecP - \frac{1}{2} \bigg( \xi^2 \dot{\xi}^2
- \xi^3 \ddot{\xi} \bigg) \vecQ^2 \nonumber \\[1.2em]
&= \frac{1}{2} \bigg( \vecP^2 + \xi^3 \Big( \ddot{\xi} 
+ \omega \big(t(T) \big)^2 \xi \Big) \vecQ^2 \Big)
= \frac{1}{2} \Big( \vecP^2 + \omega_0^2 \vecQ^2 \bigg), \label{eq:autonomous_H}
\end{align}
where we designed the symplectic map $\Phi$ in such a way that the requirement that the term in the brackets multiplying $\vecQ^2$ in equation (\ref{eq:autonomous_H}) equals $\omega_0^2 \in \mathbb{R}$ is respected. This leads to the condition that $\xi(t)$ needs to satisfy the Ermakov differential equation, which we already encountered during the discussion of the Lewis-Riesenfeld invariant $I_{LR}$ in section \ref{sec:Extended_PS} in (\ref{eq:Ermakov}). The so constructed map $\Phi$ describes a time-dependent canonical transformation that maps a harmonic oscillator with time-dependent frequency $\omega(t)$ onto a time-independent harmonic oscillator with constant frequency $\omega_0$. The explicit form of the map of course depends on the time dependence of $\omega(t)$  but can be determined from the Ermakov equation once $\omega(t)$ is given. While in principle we could fix the frequency $\omega_0$ to one, as it has been done for the form of the Ermakov equation for instance in \cite{Hartley:1982ygx,Villasenor}, we would like our transformation $\Phi$ to correspond to the identity for an already time-independent harmonic oscillator Hamiltonian. This can only be achieved if not all time-dependent frequencies are mapped to unity, as even a constant $\omega_0$ would then be transformed non trivially, resulting in a residual transformation analogous to a squeezing operation in quantum theory. 
\subsection{The reduced phase space associated with $T^*M$ and the infinitesimal generator of $\Phi$}
\label{sec:ReducedPS}
In this section we want to derive the infinitesimal generator corresponding to the finite canonical transformation $\Phi$ on the extended phase space $T^*M$ that we presented in the last section. This will be relevant later on when we discuss the implementation of $\Phi$ in the quantum theory. As we have discussed, the system under consideration can be understood as a constrained system in the context of the extended phase space. Consequently, we have two options for handling the constraint, either we solve it in the quantum theory via Dirac quantization or we reduce with respect to this constraint already classically and quantize the reduced phase space only. In the first place, both approaches are equally justified from the physical perspective, so this is a choice one makes for each given model. In our case this goes along with the selection whether we want to implement the canonical transformation $\Phi$ on the extended or reduced phase space, respectively. Firstly, as the transformation from $t$ to $T(t)$ in (\ref{eq:timescaling}) involves a time-rescaling in form of an integral, if we are not able to obtain the antiderivative of the integrand in closed form, it will be problematic to formulate this kind of canonical transformation in the quantum theory based on the extended phase space where $t$ becomes an operator. Secondly, following Dirac quantization, we need to construct a physical inner product for physical states and this is non-trivial if the  constraint is of the form $C=p_t+H({\q},{\p},t)$ with $H$ being explicitly time-dependent, a similar situation that occurs in loop quantum cosmology if we consider the inflaton as reference matter. The final physical sector of the theory should be related in both approaches and in the best case yield the same physical predictions. This might not be the case in general but yields some restrictions on possible choices in the quantization procedure to match the models based on Dirac and reduced quantization respectively. In the following we choose the reduced phase space approach for which the initial phase space $T^*\Sigma$ can be naturally identified with the reduced phase space of our system. In order to show this we 
construct Dirac observables for our constrained system by means of the formalism presented in \cite{Dittrich:Observables}, \cite{Thiemann:ReducedPS} and references therein, that is based on the relational formalism originally introduced in \cite{RovelliPartial,RovelliObservable}. In the extended phase space, we consider the configuration variable $t$ as the reference field (clock) for time and introduce the following  gauge fixing condition $G_\tau := t-\tau \approx 0$. $G_\tau$ together with the first class constraint $C$ build a second class pair since $\{G_\tau,C\}=1$. The Dirac observables for all degrees of freedom except the clock degrees of freedom $(t,p_t)$ are given by
\begin{equation}
\mathcal{O}^C_{q^a, t}(\tau) = \sum_{n=0}^\infty \frac{G^n_\tau}{n!} \big\{ H(\q, \p, t), q^a \big\}_{(n)}, \quad
\mathcal{O}^C_{p_a, t}(\tau) = \sum_{n=0}^\infty \frac{G^n_\tau}{n!} \big\{ H(\q, \p, t), p_a \big\}_{(n)},
\end{equation}
where $\{A, B \}_{(n)}$ denotes the iterated Poisson bracket defined via $\{A, B \}_{(n)} \defeq \{ A, \{ A, B \}_{(n-1)} \}$ and $\{A, B \}_{(0)} \defeq B$ and we have used that $q^a$ and $p_a$ both commute with the conjugate momentum $p_t$. The observable map can also be applied to the clock degrees of freedom, leading to
\begin{equation}
\mathcal{O}^C_{t, t}(\tau) = \sum_{n=0}^\infty \frac{G^n_\tau}{n!} \big\{ C(\q, \p, t,p_t), t \big\}_{(n)}=\tau, \quad
\mathcal{O}^C_{p_t, t}(\tau) = \sum_{n=0}^\infty \frac{G^n_\tau}{n!} \big\{ C(\q, \p,t,p_t), p_t \big\}_{(n)}.
\end{equation}
We realize that the clock $t$ is mapped to the parameter $\tau$ as expected, whereas contrary to the deparametrized models presented in \cite{Giesel:2007wi,Giesel:2007wk,Giesel:2007wn,Domagala:2010bm,Giesel:2012rb,Husain:2011tk,Han:2015jsa,Giesel:2016gxq}, the physical Hamiltonian retains its time dependence, hence $p_t$ is not yet a Dirac observable by itself. Using the properties of the observable map we have that $p_t=-\mathcal{O}^C_{H(\q,\p,t),t}=-H(\mathcal{O}^C_{\q, t},\mathcal{O}^C_{\p, t},\tau)$ and hence $p_t$ can be expressed as a function of $\mathcal{O}^C_{\q, t},\mathcal{O}^C_{\p, t}$ only, where we introduced the abbreviation $\mathcal{O}^C_{\q, t}:=(\mathcal{O}^C_{q^1, t},\cdots,\mathcal{O}^C_{q^n, t})$ and likewise for the momenta. This shows that $(\mathcal{O}^C_{\q, t},\mathcal{O}^C_{\p, t})$ are the elementary variables of the reduced phase space and the degrees of freedom encoded in $(t,p_t)$ have been reduced, which leaves us with $2n$ \textit{true} degrees of freedom in the physical sector of the phase space. As a consequence, we can identify the reduced phase space with $T^*\Sigma$ and the Hamiltonian can be understood as a function from $T^*\Sigma\times\mathbb{R}$ to the real numbers. In order to analyze the Poisson algebra of the observables we have to construct the corresponding Dirac bracket, denoted by $\{.,.\}^*$,  associated to the second class system $(G_\tau,C)$. However, for the reason that all variables $(\q,\p)$ commute with the gauge fixing condition, their Dirac bracket reduces to the usual Poisson bracket. Given this and considering the result in \cite{Thiemann:ReducedPS}, the algebra of our Dirac observables reads:
\begin{align*}
\big\{ \mathcal{O}^C_{q^a, t}(\tau), \mathcal{O}^C_{p_b, t}(\tau) \big\}
= \mathcal{O}^C_{ \{q^a, p_b\}^*,  t}(\tau) = \delta^a_b.
\end{align*}
Thus, the kinematical Poisson algebra of $(\q,\p)$ and the algebra of their corresponding Dirac observables are isomorphic, which is a big advantage for finding representations of the observable algebra in the context of the quantum theory in section \ref{sec:Quant}. The observable map applied to a generic phase space function $f$ returns the values of $f$ at those values where the clock takes the value $\tau$. Therefore, the natural evolution parameter for these Dirac observables is $\tau$. If the constraint is linear in the clock momenta as in our case where $C=p_t+H(\q,\p,t)$, then as shown in \cite{Thiemann:ReducedPS} and \cite{Giesel:2012rb} the so-called physical Hamiltonian generating the $\tau$-evolution is given by the Dirac observable corresponding to $H(\q,\p,t)$. Thus, in our case the evolution on the reduced phase space is given by the following Hamilton's equations:
\begin{equation}
\frac{\d}{\d \tau} \mathcal{O}^C_{q^a, t} = \big\{ \mathcal{O}^C_{q^a, t}, H(\mathcal{O}^C_{\q, t}, \mathcal{O}^C_{\p, t}, \tau) \big\}, \quad
\frac{\d}{\d \tau} \mathcal{O}^C_{p_a, t} = \big\{ \mathcal{O}^C_{p_a, t}, H(\mathcal{O}^C_{\q, t}, \mathcal{O}^C_{\p, t}, \tau) \big\}.
\end{equation}
Lastly, by an abuse of notation we replace $\tau$ by $t$ as well as   $\mathcal{O}^C_{q^a, t}$ by $q^a$ and $\mathcal{O}^C_{p_a, t}$ by $p_a$ in order to be closer to the notation used in previous works in the literature and emphasize that the generator of $\Phi$ acts as a one-parameter family of transformations on configuration and momentum degrees of freedom in $T^*\Sigma$. When we have a look at the form of $\Phi$, we immediately recognize that the generator $\mathcal{G} \in C^\infty (T^*\Sigma \times \mathbb{R})$ needs to be a polynomial of second order in the original configuration and momentum variables, where $T^*\Sigma \times \mathbb{R}$ corresponds to the presymplectic space for explicitly time-dependent systems as for instance used in \cite{Struckmeier}. This ensures that the action of the associated Hamiltonian vector field $X_\mathcal{G}$ with $X_\mathcal{G}(f):=\{\mathcal{G}, f \}$ onto the elementary phase space variables $\q$ and $\p$ results in a linear combination of those quantities. The explicit form of $\Phi$ suggests an ansatz in order to find $\mathcal{G}$, which naturally depends on $\xi, \dot{\xi}$, incorporating the parametric dependence on $t$:

\begin{equation}
\mathcal{G}(\xi, \dot{\xi}, \vecq, \vecp) 
\defeq f(\xi, \dot{\xi}) \vecq \cdot \vecp + \tfrac{1}{2} g(\xi, \dot{\xi}) \vecq^2,
\end{equation}
where the factor in front of $g(\xi, \dot{\xi})$ was introduced for later convenience. Application of the exponentiated Hamiltonian vector field $X_\mathcal{G}$ onto $\vecq$ and $\vecp$ leads to the following results:
\begin{align}
\exp \{ X_\mathcal{G} \} q^{a} &\defeq \sum_{n=0}^\infty \frac{1}{n!} \{ \mathcal{G}, q^{a} \}_{(n)} 
= \sum_{n=0}^\infty \frac{(-1)^n}{n!} 
\big( f (\xi, \dot{\xi}) \big)^{n} q^{a}
= e^{-f(\xi, \dot{\xi})} q^{a},
\label{eq:qcanonical_flow} \\[0.5em]
\exp \{ X_\mathcal{G} \} p_a &\defeq \sum_{n=0}^\infty 
\frac{1}{n!} \{ \mathcal{G}, p_a \}_{(n)} 
= e^{f(\xi, \dot{\xi})} p_a + \frac{1}{2} \Big( e^{f(\xi, \dot{\xi})}
- e^{-f(\xi, \dot{\xi})} \Big) 
\frac{g(\xi, \dot{\xi})}{f(\xi, \dot{\xi})} q_a,
\label{eq:pcanonical_flow}
\end{align}
with the iterated Poisson bracket defined as above. A direct comparison of the results in (\ref{eq:qcanonical_flow}) and (\ref{eq:pcanonical_flow}) to the solutions of the system of equations in (\ref{eq:canonical_system}) yields the dependencies of $f(\xi, \dot{\xi})$ and $g(\xi, \dot{\xi})$ on $\xi$ and $\dot{\xi}$, respectively:
\begin{equation}
f(\xi, \dot{\xi}) = \ln(\xi), \quad
g(\xi, \dot{\xi}) = \frac{2 \ln(\xi) \xi \dot{\xi} }{1 - \xi^2}.
\end{equation}
Finally, we are able to explicitly write down the generator of the extended canonical transformation $\Phi$ restricted to the constraint hypersurface  $T^*\Sigma \times \mathbb{R}$, that is the physical sector.  We call this restriction of $\Phi$, which is a time-dependent canonical transformation on the reduced phase space, $\Gamma_\xi$ from now on. In a convenient notation, it has the following form:

\begin{equation}
\label{eq:ClassGen}
\mathcal{G}(\xi, \dot{\xi}, \vecq, \vecp) 
= \frac{1}{2} \ln(\xi) \bigg( \vecq \cdot \vecp 
+ \vecp \cdot \vecq + h(\xi, \dot{\xi}) \vecq^2 \bigg), \quad
h(\xi, \dot{\xi}) \defeq \frac{2 \xi \dot{\xi} }{1 - \xi^2}.
\end{equation}

In fact, this classical generator precisely corresponds to the exponential operator found in \cite{Schroedinger_1D} for a quantized version of the time-dependent harmonic oscillator. It is worth noting that, regardless of the choice of coordinates, $\mathcal{G}$ takes the same form in either $\vecq, \vecp$ or $\vecQ, \vecP$, that is it holds that $\mathcal{G}\big(\vecq (\vecQ, \vecP), \vecp(\vecQ, \vecP) \big) = \mathcal{G} \big(\vecQ, \vecP \big)$. Not surprisingly, we can switch between the autonomous Hamiltonian and the Lewis-Riesenfeld invariant in this framework, using the action of $\Gamma_\xi$ on (analytical) phase space functions, leading to:
\begin{equation}
H_0 \big( \Gamma_\xi (\vecq), \Gamma_\xi (\vecp) \big)
=H_0 \big( e^{ X_\mathcal{G} } \vecq, e^{ X_\mathcal{G} } \vecp \big)
= \frac{1}{2} \bigg( \Big( \xi (t) \vecp  - \dot{\xi} (t) \vecq \Big)^2 
+ \frac{\omega_0^2 \, \vecq^2}{\xi (t)^2} \bigg) \eqdef I_{LR},
\label{eq:lewis_invariant}
\end{equation}
Of course this was how $\Gamma_\xi$ or rather $\Phi$ was constructed in the first place. However, relation (\ref{eq:lewis_invariant}) will be of importance in the quantum theory, where it is part of the time evolution operator (i.e. the Dyson series) associated to the time-dependent Hamiltonian. Furthermore, this will allow us to make contact to previous work and strictly derive the phase factor that was introduced by hand in \cite{Hartley:1982ygx} in order to construct eigenfunctions of the time-dependent Schr\"odinger equation. Referring to the relational formalism outlined in for example \cite{Dittrich:Observables} and \cite{Thiemann:ReducedPS}, we reconsider the fact that the Lewis-Riesenfeld invariant strongly commutes with the constraint $C$ as shown in (\ref{eq:LR_invariant}). Hence, in this language $I_{LR}$ is a strong Dirac observable with respect to the constraint $C(\vecq, \vecp, t, p_t)$ if and only if $\xi(t)$ satisfies the Ermakov equation (\ref{eq:Ermakov}), connecting to the results presented in \cite{Hartley:1982ygx} in the context of quantization. As a concluding remark, let us introduce $e_+ \defeq \tfrac{1}{2} \vecp^2$, $e_- \defeq - \tfrac{1}{2} \vecq^2$ and $h \defeq q^{a} p_a$, which amount to the generators of the classical canonical transformation $\Gamma_\xi$ we derived in the preceding section. Then these three generators form a basis of the $\mathfrak{sl}(2, \mathbb{R})$ algebra, which is evident due to the structure constants of their Poisson brackets. Hence, the exponential of these generators (or a subset thereof) constitutes a group element of $\textrm{SL}(2, \mathbb{R})$ and consequently the classical canonical transformation $\Gamma_\xi$ is a real representation of $\textrm{SL}(2, \mathbb{R})$ on the space of phase space polynomials or everywhere-analytic phase space functions, respectively.
\\

Let us briefly summarize what we have established in the previous section. Starting from an explicitly time-dependent Hamiltonian and its associated Lewis-Riesenfeld invariant, we systematically constructed a time-dependent canonical transformation on an extended phase space, which removes the time dependence of the original Hamiltonian. Let us stress at this point that $H(t)$, $I_{LR}(t)$ and $H_0$ are in fact the \textit{same object} in \textit{different coordinates} on the extended phase space. Consequently, we were able to construct the associated infinitesimal generator of this symplectic map and established the notion of a reduced phase space with the prospect of a corresponding unitary transformation in the one-particle quantum theory. The construction of the latter will be the content of the next section.
\newpage
\section{Quantization: One-particle Hilbert space}
\label{sec:Quant}
In this section we will present the quantization of the time-dependent canonical transformation derived in the last section on the one-particle Hilbert space. This allows to transform each mode of the single-mode Mukhanov-Sasaki Hamiltonian into a harmonic oscillator with constant frequency. In section \ref{sec:FockSpace} we will discuss in which sense the results obtained in this section can be generalized to field theories. The unitary implementation of the symplectic transformation we considered can be used  for constructing an analytic solution to the time-dependent Schr\"odinger equation in the form of a unitary time evolution operator.

\subsection{Canonical Quantization of the time-dependent canonical transformation}
\label{sec:CanQuantCanTrafo}

From the classical theory, the relevant algebra  is $\mathcal{P} = \big( C^\infty(T^*\mathbb{R}^d), \{.,.\}, \cdot \big)$ equipped with the Poisson bracket and pointwise multiplication, which is the algebra of elementary variables of a classical point-particle in d-dimensional Euclidean space. This algebra can be further extended by an involution operation leading to the Poisson  *-algebra that will be our starting point for the canonical quantization. In the following we can restrict our discussion to the case d=1 which is sufficient for the quantization of the single mode Mukhanov-Sasaki system. As a first step we define a quantization map $\mathcal{Q}$ that maps elements of $\mathcal{P}$ into an abstract operator algebra $\mathcal{Q}(\mathcal{P})$. Given any two smooth phase space functions $f,g\in\mathcal{P}$ we have
\begin{equation}
\mathcal{Q}: \mathcal{P} \to \mathcal{Q}(\mathcal{P}), \qquad
\{ f, g \}  \, \mapsto \, \mathcal{Q} \big( \{ f, g \} \big) 
= -i \big[ \mathcal{Q}(f), \mathcal{Q}(g) \big] \, \in \mathcal{Q}(\mathcal{P}),
\label{eq:quant_prescript}
\end{equation}
where we have set $\hbar=1$. Requiring that $\mathcal{Q}$ is function-preserving, that is $\mathcal{Q} \big( F(q,p) \big) = F \big( \mathcal{Q}(q), \mathcal{Q}(p) \big)$ for any real function $F$ as usually required for any quantization map, we can now directly write down the quantum version of the generator for the one-parameter family (i.e. time-dependent) of canonical transformations $\Gamma_\xi$ on $T^*\Sigma$ and its exponential:
\begin{equation}
\mathcal{Q}(\mathcal{G}) \defeq \hat{\mathcal{G}}
= \frac{1}{2} \ln(\xi) \bigg( 
\hat{q} \hat{p} + \hat{p} \hat{q}
+ h(\xi, \dot{\xi}) \hat{q}^2 \bigg),
\label{eq:canonical_trafo}
\end{equation}
where $\hat{q},\hat{p}$ denote elements of the abstract operator algebra $\mathcal{Q}(\mathcal{P})$. For later convenience  we quantize the inverse of $\Gamma_\xi$ and hence the inverse map, that is due to the minus sign in the quantization prescription and to be closer to existing results in the literature, since the mapping to the autonomous Hamiltonian is classically achieved by the inverse of $\Phi$:
\begin{align*}
\mathcal{Q} \big( \Gamma^{-1}_\xi \big)
= \exp \Big\{ i \big[ \mathcal{Q}(\mathcal{G}),. \big] \Big\}
= \exp \Big\{ i \big[ \hat{\mathcal{G}} ,. \big] \Big\} 
= \exp \Big\{ i \textrm{ ad}_{\hat{\mathcal{G}}} \Big\} 
\eqdef \textrm{Ad}_{\hat{\Gamma}_\xi},
\end{align*}
with '$\textrm{ad}$' and '$\textrm{Ad}$' denoting the adjoint representation of the Lie algebra and the corresponding Lie group, respectively. Now since we want to define the action of $\hat{\Gamma}_\xi$ on some Hilbert space we need a representation that maps the abstract operators into the set of linear operators on a Hilbert space respecting the commutator relations of the abstract algebra, that is $\pi:\mathcal{Q}(\mathcal{P})\to {\cal L}(\cal H)$ such that  $\pi(\mathcal{Q}(\mathbbm{1}_\mathcal{P}))=\mathbbm{1}_{\cal{H}}$, $\pi({\mathcal{Q}(\{q,p\})})=-i [\pi({\mathcal{Q}
(q)}),\pi(\mathcal{Q}(p))]$ as well as $\pi({\mathcal{Q}(\{q,q\})})=-i [\pi({\mathcal{Q}(q)}),\pi(\mathcal{Q}(q))]$ and $\pi({\mathcal{Q}(\{p,p\})})=-i [\pi({\mathcal{Q}(p)}),\pi(\mathcal{Q}(p))]$. If not otherwise stated we will work with the standard Schr\"odinger position representation given by $(\pi,{\cal H}=L_2(\mathbb{R},dx))$ with
\begin{eqnarray*}
\pi_q(\mathcal{Q}(q))&=&\pi_q(\hat{q}):{\cal S}(\mathbb{R})\to {\cal S}(\mathbb{R}),\quad (\pi_q(\hat{q})\Psi)(q)=q\Psi(q),\\
\pi_q(\mathcal{Q}(p))&=&\pi_q(\hat{p}):{\cal S}(\mathbb{R})\to {\cal S}(\mathbb{R}),\quad (\pi_q(\hat{p})\Psi)(q)=-i\frac{d\Psi}{dq}(q).
\end{eqnarray*}
Here $\cal{S}(\mathbb{R})$ denotes the space of Schwartz functions on $\mathbb{R}$. Given the representation we can define the action of $\hat{\Gamma}_\xi$ on both operators and elements $\Psi$ in $\mathcal{S}(\mathbb{R})$ lying dense in $L_2(\mathbb{R}, dq)$ according to the prescription:
\begin{equation}
\label{eq:GammaXi}
\pi_q(\hat{O}) \mapsto \textrm{Ad}_{\hat{\Gamma}_\xi} (\pi_q(\hat{O}))
\defeq \hat{\Gamma}_\xi \,\pi_q(\hat{O}) \, \hat{\Gamma}^\dagger_\xi, \qquad
\Psi \mapsto \hat{\Gamma}_\xi \Psi
\defeq \sum_{n=0}^\infty \frac{\big( i \pi_q(\hat{\mathcal{G}}) \big)^n}{n!} \, \Psi,
\end{equation}
where we used the abbreviation $\hat{\Gamma}_\xi:=\pi_q(\hat{\Gamma}_\xi)$ to keep our notation compact. Let us briefly check that the  the adjoint action of $\hat{\Gamma}_\xi$ on $\pi_q(\hat{q})$ and $\pi_q(\hat{p})$ is consistent. We have:
\begin{equation}
\textrm{Ad}_{\hat{\Gamma}_\xi} (\pi_q(\hat{q})) 
= \sum_{n=0}^\infty \frac{(-i^2)^n}{n!} \Big( \ln(\xi) \Big)^n \pi_q(\hat{q})
= \xi \pi_q(\hat{q}), 
\label{eq:Gamma_Q}
\end{equation}
where the iterated commutator $[\pi_q(\hat{A}),\pi_q(\hat{B})]_{(n)}$ is defined similarly to the iterated Poisson bracket with an identity at the zeroth order. For $\pi_q(\hat{p})$ we get as expected:
\begin{equation}
\textrm{Ad}_{\hat{\Gamma}_\xi} (\pi_q(\hat{p}))
= \sum_{n=0}^\infty \frac{i^{2n}}{n!} \Big( \ln(\xi) \Big)^n \pi_q(\hat{p})
+ \sum_{n=0}^\infty \frac{(i^2)^{2n+1}}{(2n+1)!} \Big( \ln(\xi) \Big)^{2n+1}
h(\xi, \dot{\xi}) \pi_q(\hat{q})
=\frac{\pi_q(\hat{p})}{\xi} + \dot{\xi} \pi_q(\hat{q}),
\end{equation}
which precisely corresponds to the inverse of the transformation of $q$ and $p$ generated by the classical Hamiltonian vector field $X_{\cal G}$. As discussed in section \ref{sec:ReducedPS}, the dynamics of the classical theory is generated by the physical Hamiltonian $H(\q,\p,t)$. Thus, we can directly consider the corresponding Schr\"odinger equation in the one dimensional case that is given by
\begin{equation*}
i\frac{\partial}{\partial t}\Psi(q,t)=\frac{1}{2} \Big( \pi_q (\hat{p})^2 
+ \omega^2(t) \pi_q (\hat{q})^2 \Big)\Psi(q,t)   
\end{equation*}
and unitarily equivalent to the corresponding Heisenberg equations for $\pi_q(q)$ and $\pi_q(p)$. If we apply the transformation induced by $\Gamma_\xi$ on the Hamiltonian and $\Psi$, which is the natural choice since classically, the replacement of $\q, \p$ in terms of $\vecQ, \vecP$ (the inverse extended map $\Phi$) achieved our aim of mapping $H(t)$ to $H_0$, we end up with:
\begin{align}
&\qquad \; \hat{\Gamma}_\xi \bigg( \frac{1}{2} \Big( \pi_q (\hat{p})^2 
+ \omega^2(t) \pi_q (\hat{q})^2 \Big) 
- i \frac{\partial}{\partial t}\bigg) 
\hat{\Gamma}^\dagger_\xi \hat{\Gamma}_\xi \Psi(q, t) = 0 \nonumber \\[0.5em]
\Longleftrightarrow& \qquad
\bigg[ \frac{1}{2} \hat{\Gamma}_\xi \bigg( \pi_q (\hat{p})^2 
+ \omega^2(t) \pi_q (\hat{q})^2 \bigg) \hat{\Gamma}^\dagger_\xi - i \hat{\Gamma}_\xi \frac{\partial \hat{\Gamma}^\dagger_\xi}{\partial t} - i \frac{\partial}{\partial t} \bigg] \hat{\Gamma}_\xi\Psi(q, t) = 0 \nonumber \\[0.5em]
\Longleftrightarrow& \qquad
\bigg[ \frac{1}{2} \bigg( \frac{\pi_q(\hat{p})^2}{\xi^2} 
+ \xi \Big( \omega^2(t) \xi + \ddot{\xi} \Big) \pi_q(\hat{q})^2 \bigg)  - i \frac{\partial}{\partial t} \bigg] \hat{\Gamma}_\xi\Psi(q, t) = 0 \nonumber \\[0.5em]
\Longleftrightarrow& \qquad
\bigg[ \frac{1}{2 \xi^2} \bigg( \pi_q(\hat{p})^2 + \omega_0^2 \pi_q(\hat{q})^2 \bigg)
- i \frac{\partial}{\partial t} \bigg] \hat{\Gamma}_\xi\Psi(q, t) = 0
\\[0.5em]
\Longleftrightarrow& \qquad
\bigg[ \frac{1}{\xi^2} \hat{H}_0
- i \frac{\partial}{\partial t} \bigg] \hat{\Gamma}_\xi\Psi(q, t) = 0,
\label{eq:transformed_Schroedinger}
\end{align}
with $\hat{H}_0:= \tfrac{1}{2} \big(\pi_q(\hat{p})^2 + \omega_0^2 \pi_q(\hat{q})^2 \big)$.
In the second step, we used the $t$-derivative of the one-parameter family of transformations $\hat{\Gamma}^\dagger_\xi$, which has already been derived in \cite{Schroedinger_1D}, \cite{Villasenor}. We can rediscover their result by using the explicit form of the generator $\pi_q(\hat{\mathcal{G}})$ using (\ref{eq:canonical_trafo}) and a Baker-Campbell-Hausdorff decomposition of $\hat{\Gamma}_\xi$ in the position representation. Later a similar but slightly generalized procedure for the occupation number representation will be discussed in section \ref{sec:BCH}. We realize that $\hat{H}_0$ in equation (\ref{eq:transformed_Schroedinger}) does \textit{not} carry any explicit time dependence, hence we can construct a solution of the Schr\"odinger equation in (\ref{eq:transformed_Schroedinger}) by integration. Further note that the inverse square of the time scaling function $\xi(t)$ precisely corresponds to the Lagrange multiplier $\lambda(s) = \d t/\d s$ that is involved in the extended classical Hamiltonian (\ref{eq:constraint}). Given this result we can now give an explicit solution of the Schr\"odinger equation as was already shown in \cite{Schroedinger_1D}:
\begin{align}
\Psi(q,t) = \hat{\Gamma}^\dagger_\xi \, \exp \bigg\{ -i \pi_q\big( \hat{H}_0 \big) \int_{t_0}^t \frac{\d \tau}{\xi^2(\tau)} \bigg\} \hat{\Gamma}_{\xi,0} \Psi(q, t_0), \quad \Psi(q,t_0) \in \mathcal{S}(\mathbb{R})_{t_0},
\label{eq:solution_constraint}
\end{align}
with $\mathcal{S}(\mathbb{R})_{t_0}$ denoting a one-parameter family of Schwarz spaces, each corresponding to a different initial time $t_0$. In a cosmological context, this behavior is a very natural one, as the \textit{instantaneous vacuum} on cosmological backgrounds shows an analogous behavior. 
Using that $\hat{I}_{LR}(t) = \hat{\Gamma}^\dagger_\xi \, \hat{H}_0 \hat{\Gamma}_\xi$, the time evolution in equation (\ref{eq:solution_constraint}) can also be rewritten as:
\begin{equation}
\hat{U}(t_0,t) = \hat{\Gamma}^\dagger_\xi \, \exp \bigg\{ -i \pi_q\big( \hat{H}_0 \big) \int_{t_0}^t 
\frac{\d \tau}{\xi^2(\tau)} \bigg\} \hat{\Gamma}_{\xi,0}
= \exp \bigg\{ -i \pi_q\big( \hat{I}_{LR} \big) \int_{t_0}^t \frac{\d \tau}{\xi^2(\tau)} \bigg\} \hat{\Gamma}^\dagger_\xi \hat{\Gamma}_{\xi,0} 
\label{eq:time_evolution_op}
\end{equation}
At this point let us further discuss the result in the quantum theory: Firstly, the integrand in the exponential corresponds exactly to our time-rescaling transformation in (\ref{eq:timescaling}) that we naturally obtained in the extended phase space approach of the classical theory. Secondly, if we compare the result here to the one in \cite{Hartley:1982ygx}, they use the eigenstates of the Lewis-Riesenfeld invariant multiplied by a phase factor to construct the solutions of the time-dependent Schr\"odinger equation. Now,  if $\hat{\Gamma}^\dagger_\xi \hat{\Gamma}_{\xi,0} \Psi(q, t_0)$ corresponds to an eigenstate of the Lewis-Riesenfeld invariant, then this reproduces precisely the phase factor that was introduced in \cite{Hartley:1982ygx} in a rather ad hoc manner. In fact, it can be easily shown that $\hat{\Gamma}^\dagger_\xi \Psi_0 (q, t_0)$ for the time-independent vacuum $\Psi_0$ corresponds to the time-dependent vacuum state of the Lewis-Riesenfeld invariant as we will see later.  The expression in equation (\ref{eq:time_evolution_op}) corresponds to the unique time evolution operator, that is the Dyson series associated to the time-dependent Hamiltonian $\hat{H}(t)$, since it satisfies identical initial conditions. Moreover, $\hat{U}(t_0,t)$ is closely related to the unitary operator found in \cite{Villasenor} (see the equation above (3.15) in that reference). In our framework, it is very natural to find the time-independent Hamiltonian in the central exponential operator on the left-hand-side of (\ref{eq:time_evolution_op}). The reason for this is  twofold: Firstly, the extended canonical transformation $\Phi$ maps the Hamiltonian $\hat{H}(t)$ into the time-independent one $\hat{H}_0$ by transforming the Schr\"odinger equation via $\hat{\Gamma}_\xi$. Secondly, the time-rescaling that is used in the extended phase space appears as a Lagrange multiplier in the extended Hamiltonian constraint and consequently as the integrand in the time-evolution operator. Lastly, let us mention that compared to \cite{Villasenor} we use a slightly different Ermakov equation here because the prefactor of $\xi^{-3}(t)$ in the Ermakov equation in (\ref{eq:Ermakov}) corresponds to the squared frequency $\omega_0^2$ of the time-independent oscillator. In the prospects of a field theoretical treatment of this transformation, it is rather unnatural to map every time-dependent mode $\omega_\mathbf{k}(t)$ onto the Minkowski case $\omega^{(0)}_{\mathbf{k}}=1$ for all $\mathbf{k}$ as done in \cite{Villasenor}. As we will discuss later on, our choice of mapping $\omega_\mathbf{k}(t)$ onto $\omega^{(0)}_\mathbf{k}=k$ is of advantage when we analyze the implementation of the unitary map on the bosonic Fock space in section \ref{sec:FockSpace}. 

\subsection{Baker-Campbell-Hausdorff decomposition} 
\label{sec:BCH}
Explicit calculations involving the evolution operator derived in the last section turn out to be rather tedious, even for simple initial conditions. This is due to the structure of the exponential in $\hat{\Gamma}^\dagger_\xi$ and the associated generator, respectively. As we will show in this section, we can perform a generalized Baker-Campbell-Hausdorff (BCH) decomposition of the operator $\hat{\Gamma}_\xi$ that brings it into a form that is more suitable for actual practical computations. For this purpose it is of advantage to change the representation and henceforth work in the occupation number basis, that is with the usual ladder operators defined as (where we omit the explicit mentioning of the representation from now on):
\begin{equation}
\pi_q(\hat{q}) = \frac{1}{\sqrt{2 \omega_0}} (\hat{A}^\dagger + \hat{A}), \quad
\pi_q(\hat{p}) = i \sqrt{\frac{\omega_0}{2}} (\hat{A}^\dagger - \hat{A}), \quad
[\hat{A}, \hat{A}^\dagger] = \mathds{1}_\mathcal{H},
\label{eq:changeofrep}
\end{equation}
where we set as before $\hbar=1$ and $m=1$. Inserting these identities into the generator $\pi_q(\hat{\mathcal{G}})$ from (\ref{eq:canonical_trafo}) and $\hat{\Gamma}_\xi$, we obtain (again without the explicit representation):
\begin{align}
\hat{\Gamma}_\xi
&= \exp \bigg \{ \frac{i}{2} \ln(\xi) \bigg( i \big( \aDsq - \hat{A}\hat{A} \big)
+ \frac{h(\xi)}{2 \omega_0} \big( \aDsq 
+ \hat{A}^\dagger \hat{A} + \hat{A} \hat{A}^\dagger 
+ \hat{A}\hat{A} \big) \bigg) \bigg \} \nonumber
\\[0.5em]
&= \exp \bigg \{ \frac{1}{2} \ln(\xi) \bigg( \Big(1 
+ \frac{i h(\xi)}{2 \omega_0} \Big) \hat{A}\hat{A} 
- \Big(1- \frac{i h(\xi)}{2 \omega_0} \Big) \aDsq
+ \frac{i h(\xi)}{\omega_0} \big( \hat{A}^\dagger \hat{A} 
+ \frac{\mathds{1}}{2} \big) \bigg) \bigg \} \label{eq:Gamma_generator} \\[0.5em]
&= \exp \bigg \{ \overline{\alpha}(\xi) \frac{\hat{A}\hat{A}}{2} 
- \alpha(\xi) \frac{\hat{A}^\dagger \hat{A}^\dagger}{2}
+ i \lambda(\xi) \big( \hat{A}^\dagger \hat{A} 
+ \frac{\mathds{1}}{2} \big) \bigg \} \nonumber \\[0.5em]
&\eqdef \exp \Big \{ \overline{\alpha}(\xi) \hat{\sigma}_-
- \alpha(\xi) \hat{\sigma}_+
+ i \lambda(\xi) \hat{\sigma}_3 \Big \}, \nonumber
\end{align}

where we made the following redefinitions for later notational convenience: 

\begin{align}
\hat{\sigma}_+ \defeq \frac{1}{2} \hat{A}^\dagger \hat{A}^\dagger, \quad
\hat{\sigma}_- \defeq \frac{1}{2} \hat{A} \hat{A}, \quad 
\hat{\sigma}_3 \defeq \hat{A}^\dagger \hat{A} + \frac{\mathds{1}}{2}, \quad
[\hat{\sigma}_3, \hat{\sigma}_{\pm}] = \pm 2 \hat{\sigma}_{\pm}, \quad 
[\hat{\sigma}_-, \hat{\sigma}_+] = \hat{\sigma}_3.
\label{eq:sigma_generators}
\end{align}

The coefficients are in fact explicitly time-dependent functions $\alpha(\xi)$, $\lambda(\xi)$, where the time dependency is carried by the solution $\xi$ of the Ermakov equation (\ref{eq:Ermakov}) as we have seen in the discussion of the classical setup. They are defined as:

\begin{align}
\alpha &= \ln(\xi)\bigg(1 - \frac{i h(\xi)}{2 \omega_0} \bigg), \quad
\lambda = \frac{h(\xi)}{2 \omega_0} \ln(\xi), \quad |\alpha|^2 > \lambda^2 \; \forall \; \xi: \mathbb{R} \supseteq \mathbb{I} \to \mathbb{R}. \label{def:alpha_lambda}
\end{align}

After this replacement the resulting expression for $\hat{\Gamma}_\xi$ takes the form of a generalized, time-dependent squeezing operation. The commutation relations in equation (\ref{eq:sigma_generators}) are those of $\mathfrak{sl}(2, \mathbb{R})$, which was already evident in the classical sector of the theory. It is straightforward to see that the standard Baker-Campbell-Hausdorff decomposition does not work, since the iterated commutator structure leads to infinitely many non-vanishing contributions in the well-known formula. However, a BCH decomposition of SL$(2, \mathbb{R})$ elements has been performed using analytic techniques as shown in \cite{BCH_Decomp}. This was done by introducing a parametric rescaling of $\hat{\Gamma}_\xi$ and allowing a corresponding dependence of the coefficient functions in the decomposition on this parameter. In our case, a rescaling of the original $\hat{\Gamma}_\xi$ leads to:
\begin{align}
\hat{\Gamma}_\xi (\mu) \defeq \exp \Big \{ \mu \hat{\mathcal{G}} \Big \} =
 \exp\Big \{\mu\Big( \overline{\alpha}(\xi) \hat{\sigma}_-
- \alpha(\xi) \hat{\sigma}_+
+ i \lambda(\xi) \hat{\sigma}_3\Big) \Big \},
\end{align}
with an arbitrary rescaling by some parameter $\mu \in \mathbb{R}$. Let us denote the decomposed version of $\hat{\Gamma}_\xi(\mu)$ by $\tilde{\hat{\Gamma}}_\xi(\mu)$, with a semicolon representing a parametric dependence:
\begin{align}
\tilde{\hat{\Gamma}}_\xi(\mu)\eqdef
\exp \Big \{ \beta_+(\xi; \mu) \hat{\sigma}_+ \Big \}
\exp \Big \{ \gamma(\xi; \mu) \hat{\sigma}_3 \Big \}
\exp \Big \{ \beta_-(\xi; \mu) \hat{\sigma}_- \Big \}
\end{align}
Then we aim at determining the coefficient functions $\beta_+(\xi; \mu),\gamma(\xi; \mu)$ and $ \beta_-(\xi; \mu)$ such that we have $\hat{\Gamma}_\xi(\mu)=\tilde{\hat{\Gamma}}_\xi(\mu)$. This rescaling allows us to differentiate $\hat{\Gamma}_\xi(\mu)$ and $\tilde{\hat{\Gamma}}_\xi(\mu)$ with respect to $\mu$. Considering this we start with a consistency requirement for $\hat{\Gamma}_\xi(\mu)$ and $\tilde{\hat{\Gamma}}_\xi(\mu)$ given by:
\begin{align}
\bigg( \frac{\partial}{\partial \mu} \hat{\Gamma}_\xi(\mu) \bigg)
\bigg(\hat{\Gamma}_\xi(\mu) \bigg)^\dagger
= \bigg( \frac{\partial}{\partial \mu} \tilde{\hat{\Gamma}}_\xi(\mu) \bigg)
\bigg( \tilde{\hat{\Gamma}}_\xi(\mu) \bigg)^\dagger. 
\label{eq:BCH_Idea}
\end{align}
In the next step we will omit the arguments of the coefficient functions for the sake of a more compact notation. Explicitly evaluating the differentials and using the unitarity of $\hat{\Gamma}_\xi$, we end up with three contributions. A closer look reveals that these contributions contain the adjoint action of $\hat{\Gamma}_\xi$ onto the three generators of the algebra in (\ref{eq:sigma_generators}), which can be easily computed due to the simple structure of their commutators. The linear independence of the generators then leads to a coupled system of differential equations for the coefficient functions:
\begin{align}
\overline{\alpha} &= \exp\{ - 2 \gamma \} \frac{\partial \beta_-}{\partial \mu} 
\label{eq:BCHalpha_bar} \\[0.7em]
i \lambda &= \frac{\partial \gamma}{\partial \mu} - \beta_+ 
\exp\{ - 2 \gamma \} \frac{\partial \beta_-}{\partial \mu} \label{eq:BCHlambda}\\[0.7em]
\alpha &= 2\beta_+ \frac{\partial \gamma}{\partial \mu}
- \frac{\partial \beta_+}{\partial \mu} - \beta_+^2 \exp\{ - 2 \gamma \}
\frac{\partial \beta_-}{\partial \mu}. 
\label{eq:BCHalpha}
\end{align}
Performing a number of substitutions, this system of differential equations can be cast into the form of a complex Riccati-type ordinary differential equation, for more details we refer the reader to the explicit computations done in \cite{BCH_Decomp}. An appropriate ansatz for this equation yields a solution, subsequent resubstitution then leads to the desired BCH coefficient functions of the normal-ordered decomposition of $\hat{\Gamma}_\xi$. A similar procedure can be performed for the normal and anti-normal ordering of both $\hat{\Gamma}_\xi$ and $\hat{\Gamma}^\dagger_\xi$, respectively, while we have chosen that $\hat{\sigma}_3$ remains in the middle for computational convenience. Although depending on the given initial state $\Psi(q, t_0)$ at our disposal, the most useful forms of the coefficients (or operator orderings, respectively) regarding computational convenience are given by:
\begin{align*}
\begin{split}
\delta_+(\mu) &= + \frac{\alpha \sh(\Delta \mu)}{\Delta \ch(\Delta \mu) 
+ i \lambda \sh(\Delta \mu)} \\[0.7em]
\delta_-(\mu) &= - \frac{\overline{\alpha} \sh(\Delta \mu)}{\Delta \ch(\Delta \mu) 
+ i \lambda \sh(\Delta \mu)} \\[0.7em]
\nu(\mu) &= \underbrace{- \ln \bigg( \ch(\Delta \mu) + \frac{i \lambda}{\Delta} \sh(\Delta \mu) \bigg),}_{\textrm{normal ordering of $\hat{\Gamma}^\dagger_\xi$}}
\end{split} \qquad
\begin{split}
\tau_+(\mu) &= + \frac{\alpha \sh(\Delta \mu)}{\Delta \ch(\Delta \mu) 
- i \lambda \sh(\Delta \mu)} \\[0.7em]
\tau_-(\mu) &= - \frac{\overline{\alpha} \sh(\Delta \mu)}{\Delta \ch(\Delta \mu) 
- i \lambda \sh(\Delta \mu)} \\[0.7em]
\rho(\mu) &= \underbrace{\ln \bigg( \ch(\Delta \mu) - \frac{i \lambda}{\Delta} \sh(\Delta \mu) \bigg).}_{\textrm{anti-normal ordering of $\hat{\Gamma}^\dagger_\xi$}}
\end{split}
\end{align*}

\begin{align*}
\begin{split}
\beta_+(\mu) &= - \frac{\alpha \sh(\Delta \mu)}{\Delta \ch(\Delta \mu) 
- i \lambda \sh(\Delta \mu)} \\[0.7em]
\beta_-(\mu) &= + \frac{\overline{\alpha} \sh(\Delta \mu)}{\Delta \ch(\Delta \mu) 
- i \lambda \sh(\Delta \mu)} \\[0.7em]
\gamma(\mu) &= \underbrace{-\ln \bigg( \ch(\Delta \mu) 
- \frac{i \lambda}{\Delta} \sh(\Delta \mu) \bigg),}_{\textrm{normal ordering of $\hat{\Gamma}_\xi$}}
\end{split} \qquad
\begin{split}
\varepsilon_+(\mu) &= - \frac{\alpha \sh(\Delta \mu)}{\Delta \ch(\Delta \mu) 
+ i \lambda \sh(\Delta \mu)} \\[0.7em]
\varepsilon_-(\mu) &= + \frac{\overline{\alpha} \sh(\Delta \mu)}{\Delta \ch(\Delta \mu) 
+ i \lambda \sh(\Delta \mu)} \\[0.7em]
\iota(\mu) &= \underbrace{\ln \bigg( \ch(\Delta \mu) + \frac{i \lambda}{\Delta} \sh(\Delta \mu) \bigg).}_{\textrm{anti-normal ordering of $\hat{\Gamma}_\xi$}}
\end{split}
\end{align*}
with $\Delta^2 \defeq |\alpha|^2 - \lambda^2$ and $\Delta^2 > 0$ for all real solutions $\xi(t)$ of the Ermakov equation (\ref{eq:Ermakov}). By fixing  the parameter $\mu = 1$, we recover the unitary transformation we initially started with. Let us note that for an initially time-independent Hamiltonian, the decomposed transformation reproduces the identity operator, as expected. This is due to the fact that in this case $\xi(t)=1$, which in turn leads to a vanishing generator. Furthermore one can explicitly check that the adjoints of the decompositions of $\hat{\Gamma}_\xi$ and $\hat{\Gamma}^\dagger_\xi$ are the decompositions of the adjoints, which illustrates mutual consistency and conservation of unitarity among the obtained results. To briefly summarize this chapter, we have used analytical techniques to perform a decomposition of the exponentiated generator in (\ref{eq:Gamma_generator}) into three individual contributions. Due to the fact that we are working with unitary representations of the algebra of non-compact Lie group with mutually non-commuting elements, this result is nontrivial and enables the realization of computations in a compact form. Examples of applications of the Baker-Campbell-Hausdorff decomposition of $\hat{\Gamma}_\xi$ can be found in sections \ref{sec:Bogoliubov_one_part} and \ref{sec:applications}, respectively.

\subsection{Time-dependent Bogoliubov maps} \label{sec:Bogoliubov_one_part}
In this section we will show that the transformation induced by the operator $\hat{\Gamma}_\xi$ can be understood as a time-dependent Bogoliubov transformation when applied to the ladder operators. Given the action of $\hat{\Gamma}_\xi$ on the elementary position and momentum operators, it can naturally be extended to the ladder operators as well. The same applies also to the adjoint action, which is however tedious to evaluate in the original form of the generator. Due to the possibility of decomposing the operator $\hat{\Gamma}_\xi$ and its adjoint, we can take advantage of the result in the last section and compute the action on $\hat{A}$ and $\hat{A}^\dagger$ with a normal and anti-normal ordered decomposition, respectively. Using the commutator structure of the generators in (\ref{eq:sigma_generators}), we obtain:

\begin{align}
\textrm{Ad}_{\hat{\Gamma}_\xi} (\hat{A})
&= e^{-\gamma(\xi)} \big( \hat{A} - \beta_+(\xi) \hat{A}^\dagger \big)
\label{eq:Bogoliubov_A}, \\[0.5em]
\textrm{Ad}_{\hat{\Gamma}_\xi} (\hat{A}^\dagger)
&= e^{\iota(\xi)} \big( \hat{A}^\dagger 
+ \varepsilon_-(\xi) \hat{A} \big) \label{eq:Bogoliubov_AD}, \\[0.5em]
\textrm{Ad}_{\hat{\Gamma}_\xi^\dagger} (\hat{A})
&= e^{-\nu(\xi)} \big( \hat{A} - \delta_+(\xi) \hat{A}^\dagger \big)
\label{eq:inv_Bogoliubov_A}, \\[0.5em]
\textrm{Ad}_{\hat{\Gamma}_\xi^\dagger} (\hat{A}^\dagger)
&= e^{\rho(\xi)} \big( \hat{A}^\dagger + \tau_- (\xi) \hat{A} \big),
\label{eq:inv_Bogoliubov_AD}
\end{align}

with the corresponding coefficient functions derived in the preceding section. These functions carry an explicit time dependence via $\xi(t)$, which is a solution of the Ermakov equation (\ref{eq:Ermakov}) with the time-dependent frequency of the initial Hamiltonian. In fact, the transformations of the ladder operators in (\ref{eq:inv_Bogoliubov_AD}) look already close to that of a time-dependent Bogoliubov transformation. Whether this is indeed the case depends on the coefficient functions involved and will be analyzed in the following. For this purpose, let us rewrite the action of $\hat{\Gamma}_\xi$ on these operators as a $2 \times 2$ matrix representation, considering the equations (\ref{eq:Bogoliubov_A}) - (\ref{eq:inv_Bogoliubov_AD}):

\begin{align}
\begin{pmatrix}
g(\xi, \dot{\xi}) & \overline{h}(\xi, \dot{\xi}) \\[0.5em]
h(\xi, \dot{\xi}) & \overline{g}(\xi, \dot{\xi})
\end{pmatrix}
\begin{pmatrix}
\hat{A} \\[0.5em] \hat{A}^\dagger
\end{pmatrix} =
\begin{pmatrix}
g(\xi, \dot{\xi}) \hat{A} + \overline{h}(\xi, \dot{\xi}) \hat{A}^\dagger \\[0.5em]
h(\xi, \dot{\xi}) \hat{A} +\overline{g}(\xi, \dot{\xi}) \hat{A}^\dagger
\end{pmatrix} \defeq
\begin{pmatrix}
\hat{B} \\[0.5em] \hat{B}^\dagger
\end{pmatrix}, \label{eq:Bogoliubov_matrix}
\end{align}

with the additional requirement that if $[\hat{A}, \hat{A}^\dagger] = \mathds{1}_\mathcal{H}$, then similarly $[\hat{B}, \hat{B}^\dagger] = \mathds{1}_\mathcal{H}$ needs to hold, usually required for a Bogoliubov transformation.  In order to achieve this, we need to impose the condition that the determinant of the matrix on the left-hand side of (\ref{eq:Bogoliubov_matrix}) involving  $g(\xi, \dot{\xi})$ and $h(\xi, \dot{\xi})$ is equal to one, which amounts to:
\begin{align*}
[\hat{B}, \hat{B}^\dagger] = \mathds{1}_\mathcal{H} 
\quad \Longleftrightarrow \quad 
\big|g(\xi, \dot{\xi})\big|^2 - \big|h(\xi, \dot{\xi})\big|^2 = 1.
\end{align*}
Applying this to the transformation in (\ref{eq:Bogoliubov_A}), we get:
\begin{align*}
\begin{pmatrix}
g(\xi, \dot{\xi}) & \overline{h}(\xi, \dot{\xi}) \\[0.5em]
h(\xi, \dot{\xi}) & \overline{g}(\xi, \dot{\xi})
\end{pmatrix} = 
\begin{pmatrix}
e^{-\gamma(\xi)}  & - e^{ -\gamma(\xi)} \beta_+(\xi) \\[0.5em]
- e^{ -\overline{\gamma}(\xi)} \overline{\beta}_+(\xi) & e^{ -\overline{\gamma}(\xi)}
\end{pmatrix}
\quad \Longrightarrow \quad
e^{-(\gamma + \overline{\gamma})} \Big( 1 - \big|\beta_+ \big|^2 \Big) \stackrel{!}{=} 1.
\end{align*}
Given the explicit functional form of the BCH coefficients, it can be easily shown that $\hat{\Gamma}_\xi$ indeed describes a time-dependent Bogoliubov transformation and the expression above equals one. For all remaining cases this can be also shown using the same method. We are now in a situation where we can formulate the time evolution of $\hat{A}, \hat{A}^\dagger$ in the Heisenberg picture, using the time-evolution operator $\hat{U}(t_0,t)$. It consists of the aforementioned Bogoliubov map together with an exponential operator involving the Lewis-Riesenfeld invariant or the autonomous Hamiltonian, respectively. The additional exponent also carries the information of the time-rescaling encoded in the function $\xi(t)$ and hence is sensitive to the underlying spacetime geometry. In the following, we introduce the following notation for the coefficient functions $\beta_+(\xi(t)) = \beta_+(\xi)$ and $\beta_+(\xi(t_0)) \defeq \beta_{+}(\xi_0)$ involved in the decomposition of $\hat{\Gamma}_\xi$ and $\hat{\Gamma}_{\xi,0}$,  respectively. Carefully applying $\hat{U}(t_0, t)$ and collecting everything together, we obtain:
\begin{align}
\hat{A}_H(t_0, t) &= \exp \Big\{ - \Big(\gamma(\xi) 
+ i \omega_0 \int_{t_0}^t \frac{\d \tau}{\xi^2(\tau)} + \nu(\xi_0) \Big) \Big\}
\Big( \hat{A} - \delta_{+}(\xi_0) \hat{A}^\dagger \Big) \nonumber \\
&- \exp \Big\{ - \Big(\gamma(\xi) - i \omega_0 \int_{t_0}^t \frac{\d \tau}{\xi^2(\tau)} 
+ \bar{\nu}(\xi_0) \Big) \Big\} \beta_+(\xi) \Big( \hat{A}^\dagger - \bar{\delta}_+(\xi_0) \hat{A} \Big), \label{eq:Heisenberg_A} \\[1.0em]
\hat{A}^\dagger_H(t_0, t) &= \exp \Big\{ - \Big(\bar{\gamma}(\xi) 
- i \omega_0 \int_{t_0}^t \frac{\d \tau}{\xi^2(\tau)} + \bar{\nu}(\xi_0) \Big) \Big\}
\Big( \hat{A}^\dagger - \bar{\delta}_{+}(\xi_0) \hat{A} \Big) \nonumber \\
&- \exp \Big\{ - \Big(\bar{\gamma}(\xi) + i \omega_0 \int_{t_0}^t \frac{\d \tau}{\xi^2(\tau)} 
+ \nu(\xi_0) \Big) \Big\} \bar{\beta}_+(\xi) \Big( \hat{A} 
- \delta_+(\xi_0) \hat{A}^\dagger \Big). \label{eq:Heisenberg_AD}
\end{align}
Although the expressions (\ref{eq:Heisenberg_A}) and (\ref{eq:Heisenberg_AD}) look rather complicated at first, as expected they reduce to $\hat{A}, \hat{A}^\dagger$ in the limit $t \to t_0$. This can be seen by replacing $t$ by $t_0$ in the expression above and using the definitions of the Baker-Campbell-Hausdorff coefficients from section \ref{sec:BCH}. One finally observes that all contributions apart from $\hat{A}$ or $\hat{A}^\dagger$, respectively, cancel upon inserting the definition of $\Delta$ and using the fact that $\cosh^2(\Delta)-\sinh^2(\Delta)=1$. In principle, these results allow to compute expectation values for various initial conditions and investigate the behavior of these operators for the single-mode Mukhanov-Sasaki equation. We will discuss some application of this framework in section (\ref{sec:applications}), where we consider the derived unitary map for the single-mode Mukhanov-Sasaki equation in the context of quasi-de Sitter spacetimes. Prior to that, in the next section we will discuss whether the results obtained so far can be carried over to field theory, that is whether the obtained unitary map can be extended to the bosonic Fock space.

\section{Implementing the time-dependent canonical transformation  as a unitary map on the bosonic Fock space}
\label{sec:FockSpace}
For the reason that we were able to construct a unitary map for the toy model of the single-model Mukhanov-Sasaki Hamiltonian, the next obvious step is to aim at a unitary implementation of the time evolution operator $\hat{U}(t_0, t)$ on the full Fock space $\mathcal{F}$. Since every mode of the Mukhanov-Sasaki equation is a time-dependent harmonic oscillator, we need to treat every mode separately and with a different frequency, depending on the absolute value of $\mathbf{k}$. Hence it is natural to equip the solution of the Ermakov equation, which also differs from mode to mode for precisely this reason, with a corresponding mode label, which in turn carries over to the time-dependent Bogoliubov transformation $\hat{\Gamma}_\xi$. In the conventional formalism, the Mukhanov-Sasaki Hamiltonian and the mode expansion of the Mukhanov-Sasaki variable and its conjugate momentum are of the form:
\begin{align}
    \hat{H}(\eta) &= \frac{1}{2} \int \d ^3x \Big( \hat{\pi}_v^2(\eta, \mathbf{x}) + \big( \partial_a \hat{v}(\eta, \mathbf{x}) \big) \big( \partial^a \hat{v}(\eta, \mathbf{x}) \big) - \frac{z''(\eta)}{z(\eta)} \hat{v}^2(\eta, \mathbf{x}) \Big), 
    \label{eq:MS_Ham_def} \\[0.5em]
    \hat{v}(\eta, \mathbf{x}) &= \int \frac{\d ^3k}{(2 \pi)^{3}} \Big( v_\mathbf{k}(\eta) \hat{a}_\mathbf{k} \exp \{ i \mathbf{k} \cdot \mathbf{x} \} + \overline{v}_\mathbf{k}(\eta) \hat{a}^\dagger_\mathbf{k} \exp \{ - i \mathbf{k} \cdot \mathbf{x} \} \Big), \label{eq:MS_variable_mode}\\[0.5em]
    \hat{\pi}_v(\eta, \mathbf{x}) &= \int \frac{\d ^3k}{(2 \pi)^{3}} \Big( \partial_0 v_\mathbf{k}(\eta) \hat{a}_\mathbf{k} \exp \{ i \mathbf{k} \cdot \mathbf{x} \} + \partial_0 \overline{v}_\mathbf{k}(\eta) \hat{a}^\dagger_\mathbf{k} \exp \{ - i \mathbf{k} \cdot \mathbf{x} \} \Big), \label{eq:MS_momentum_mode}
\end{align}
with $\partial_0$ denoting a derivative with respect to conformal time $\eta$, $a(\eta)$ is the scale factor, $\mathcal{H}$ is the conformal Hubble function and $\bar{\phi}$ stands for the homogeneous and isotropic part of the inflaton scalar field. 
Given the canonical commutator $[\hat{v}(\eta, \mathbf{x}), \hat{\pi}_v(\eta, \mathbf{y})] = i \delta^{(3)}(\mathbf{x}, \mathbf{y}) \mathbbm{1}_{\cal{H}}$ 
together with the mode expansion of $\hat{v}(\eta,\mathbf{x})$ and $\hat{\pi}_v(\eta,\mathbf{x})$ as well as the following choice for the Wronskian
\begin{equation}
    W(v_\mathbf{k}, \overline{v}_\mathbf{k}) \defeq v_\mathbf{k} \overline{v}'_\mathbf{k} - v_\mathbf{k}' \overline{v}_\mathbf{k} = i,
\end{equation}
the corresponding annihilation and creation operators satisfy the commutator algebra
\begin{equation*}
[\hat{a}_\mathbf{k}, \hat{a}^\dagger_\mathbf{m}]=(2 \pi)^3 \delta^{(3)}(\mathbf{k}, \mathbf{m}) \mathbbm{1}_{\cal{H}},
\end{equation*}
where all remaining commutators vanish. Compared to the one-particle case we obtain an additional factor of $(2\pi)^3$ here, which in principle needs to be considered when deriving the corresponding Bogoliubov coefficients in (\ref{eq:Bogoliubov_A}) - (\ref{eq:inv_Bogoliubov_AD}) for the field theory case. In order to avoid to include appropriate powers of $2\pi$ in the derivation of the Bogoliubov coefficients, as an intermediate step we rescale the creation and annihilation operators such that they satisfy a commutator algebra that involves just the $\delta$-function. This yields:
\begin{align}
    \hat{A}_\mathbf{k} \defeq (2 \pi)^{- \frac{3}{2}} \hat{a}_\mathbf{k}, \quad
    \hat{A}^\dagger_\mathbf{k} \defeq (2 \pi)^{- \frac{3}{2}} \hat{a}^\dagger_\mathbf{k}, \quad
    [\hat{A}_\mathbf{k}, \hat{A}^\dagger_\mathbf{m}] = \delta^{(3)}(\mathbf{k}, \mathbf{m}) \mathbbm{1}_{\cal{H}}.
\end{align}
Note that we consider a quantization of the inflaton perturbation in the context of quantum field theory on a curved background, where the background quantities are considered as external quantities and we thus neglect any backreaction effects. 

Now we can let the Bogoliubov transformation act on the rescaled operators $\hat{A}_\mathbf{k}, \hat{A}^\dagger_\mathbf{k}$ and all results obtained  in the previous section \ref{sec:Bogoliubov_one_part} can be easily carried over to the field theoretic case, where the Bogoliubov transformation maps $\hat{A}_\mathbf{k}, \hat{A}^\dagger_\mathbf{k}$ to a new set of creation and annihilation operators  $\hat{B}_\mathbf{k}, \hat{B}^\dagger_\mathbf{k}$ that fulfill the same rescaled commutation relation. Due to the linearity of the Bogoliubov transformation, the rescaling affects both sides of the equation and thus can be easily removed and the standard algebra we started with is restored. In the field theory the generator in the exponential of $\hat{\Gamma}^{\mathbf{k}}_\xi$ is smeared with the Baker-Campbell-Hausdorff coefficient functions that act as the smearing functions. The action on the operators $\hat{A}_\mathbf{k}, \hat{A}^\dagger_\mathbf{k}$ is then diagonal, because at each order of the iterated commutator, the to be found Dirac distributions can be absorbed into the integral involved due to the smearing. Hence, the generalization of Bogoliubov coefficients we obtained in the one-particle case in (\ref{eq:Bogoliubov_matrix}) to the field theory case just consists of equipping them with a mode label. The questions that still needs to answered is whether the so defined extension of $\hat{\Gamma}^{\mathbf{k}}_\xi$ to Fock space describes a unitary map on the latter. Fortunately, there exists a criterion whether a given Bogoliubov transformation can be unitarily implemented on Fock space, called the  \textit{Shale-Stinespring} condition. A review on the Shale-Stinespring condition with a sketched proof can be for example found in \cite{Solovej_Diagonalization}. The theorem essentially states that the anti-linear part of the Bogoliubov transformation under consideration needs to be a Hilbert-Schmidt operator. In our case, this condition carries over to the product of the off-diagonal coefficients in equation (\ref{eq:Bogoliubov_matrix}) being bounded when integrated over all of $\mathbb{R}^3$:
\begin{align}
\int_{\mathbb{R}^3} \d^3k \, h_{\mathbf{k}}(\xi, \xi^\prime) 
\overline{h}_{\mathbf{k}}(\xi, \xi^\prime) < \infty, \quad
\overline{h}_{\mathbf{k}}(\xi, \xi^\prime) = - \exp \big\{ - \gamma(\xi_\mathbf{k}) \big\} 
\beta_+(\xi_\mathbf{k}), \label{eq:Shale_condition}
\end{align}
where $\gamma(\xi_\mathbf{k})$ and $\beta_+(\xi_\mathbf{k})$ are the Baker-Campbell-Hausdorff coefficients from the decomposition in section \ref{sec:BCH}, $\xi_\mathbf{k}(t)$ is the mode-dependent solution of the Ermakov equation and $\xi^\prime_\mathbf{k}$ is the derivative with respect to conformal time. At this point the advantage of rescaling the operator algebra becomes evident, since we can copy our results from previous computations of the Bogoliubov coefficients. A discussion on the initial conditions regarding the solutions $\xi_\mathbf{k}(t)$ can be found in section \ref{sec:applications} below, which will provide the basis for the investigations of the finiteness of the integral over the anti-linear part of $\hat{\Gamma}_\xi^\mathbf{k}$. Explicitly inserting the coefficients while still keeping $\xi_\mathbf{k}(\eta)$ in the arguments and considering the rescaled operators such that they satisfy the same commutator algebra as in section \ref{sec:Bogoliubov_one_part} leads to:
\begin{align*}
| h_{\mathbf{k}}(\xi, \xi^\prime) |^2 = \bigg( \ch^2(\Delta_\mathbf{k}) 
+ \frac{\lambda_\mathbf{k}^2}{\Delta_\mathbf{k}^2} \sh^2(\Delta_\mathbf{k}) \bigg) 
 \frac{|\alpha_\mathbf{k}|^2 \sh^2(\Delta_\mathbf{k})}{\Delta_\mathbf{k}^2 
 \ch^2(\Delta_\mathbf{k}) + \lambda_\mathbf{k}^2 \sh^2(\Delta_\mathbf{k})}
 = \frac{|\alpha_\mathbf{k}|^2 \sh^2(\Delta_\mathbf{k})}{|\alpha_\mathbf{k}|^2 
 - \lambda_\mathbf{k}^2}.
\end{align*}
Given the former definition of  $\alpha,\lambda$ and $\Delta$ we consider the extension of these quantities  to the multi-mode case given by $\Delta_\mathbf{k} = \sqrt{|\alpha_\mathbf{k}|^2 - \lambda_\mathbf{k}^2}$. Inserting the explicit form of $\alpha$ and $\lambda$ from equation (\ref{def:alpha_lambda}) we end up with:
\begin{align*}
\Delta^2_\mathbf{k} = |\alpha_\mathbf{k}|^2 - \lambda^2_\mathbf{k} 
= \bigg| \ln(\xi_\mathbf{k})\bigg(1 - \frac{i h(\xi_\mathbf{k})}{2 \big(\omega^{(0)}_\mathbf{k}\big)^2} \bigg) \bigg|^2
- \bigg( \frac{h(\xi_\mathbf{k})}{2 \big(\omega^{(0)}_\mathbf{k}\big)^2} \ln(\xi_\mathbf{k}) \bigg)^2
= \ln^2(\xi_\mathbf{k}).
\end{align*}
Note that $\ln^2(\xi_\mathbf{k}) > 0$ for all modes $\mathbf{k} \in \mathbb{R}^3$ with $\Norm{\mathbf{k}} \neq 0$ and all conformal times $\eta \in \mathbb{R}_- \setminus \{ 0 \}$, from which we can conclude that $\Delta_\mathbf{k} = \ln(\xi_\mathbf{k})$ since we already know that $\Delta_\mathbf{k} > 0$ holds. Explicitly substituting the definitions of $\alpha_k, \lambda_k$ and $\Delta_k$ into $v_k(\xi,\dot{\xi})$, we arrive at the following integral for the de Sitter case with $\xi(\eta)$ as derived in the succeeding section \ref{sec:applications}:
\begin{align*}
\int_{\mathbb{R}^3} \d^3k \, |h_{\mathbf{k}}(\xi, \xi^\prime) |^2
&= \int_{\mathbb{R}^3} \d^3k \, \bigg( 1 + \frac{1}{\big(\omega^{(0)}_\mathbf{k}\big)^2} \bigg[ \frac{\xi_\mathbf{k}(\eta) \xi^\prime_\mathbf{k}(\eta)}{1 - \xi^2_\mathbf{k}(\eta)} \bigg]^2 \bigg)
\bigg( \frac{1}{2} \frac{\xi^2_\mathbf{k}(\eta) - 1}{\xi_\mathbf{k}(\eta)} \bigg)^2 \\[0.6em]
&= \frac{1}{4} \int_{\mathbb{R}^3} \d^3k \,  \bigg[ \bigg(\frac{\xi^2_\mathbf{k}(\eta) - 1}{\xi_\mathbf{k}(\eta)} \bigg)^2 + \bigg( \frac{\xi^\prime_\mathbf{k}(\eta)}{\omega^{(0)}_\mathbf{k}} \bigg)^2 \bigg] \\[0.6em]
&= \frac{1}{4} \int_{\mathbb{R}^3} \d^3k \, \bigg[ \bigg( \frac{(k \eta)^2}{1+(k \eta)^2} \frac{1}{(k \eta)^4} \bigg) + \frac{1}{k^2} \bigg( \frac{(k \eta)^2}{1+(k \eta)^2} \frac{1}{k^4 \eta^6} \bigg) \bigg].
\end{align*}
This expression allows us to consider a simple power-counting procedure of the individual contributions. For large $k$, the first term behaves as $k^{-4}$ whereas the second term decays as $k^{-6}$, so there is no divergence in the ultraviolet. For small $k$ we observe the first term to be proportional to $k^{-2}$ and the second contribution to $k^{-4}$, which leads to an infrared divergence of the latter, which in turn shows that the integral above is not finite. Consequently, the Shale-Stinespring condition is not satisfied in our case and the Bogoliubov transformation $\hat{\Gamma}_\xi$ cannot be unitarily implemented on Fock space by simply extending the toy model of the single-mode case to the multi-mode case due to 'infinite particle production' between mutually different vacuum states. Interestingly, there is no issue with the ultraviolet here but just in the infrared sector, showing that next to the large $k$ behavior one also needs to check whether there are occurring singularities in the infrared, as they can equally add a diverging contributions to the number operator expectation value with respect to different vacua. It is not obvious to us that this aspect has been considered in the recent work of \cite{invariant_vacuum}, where a similar Bogoliubov transformation is used on Fock space. As can be seen form our analysis, the behavior of the BCH coefficients is different for small $k$ than it is for large $k$, hence it is not obvious that even if the Bogoliubov coefficients are finite for large $k$ this is a sufficient check in order to conclude that the Bogoliubov transformation under consideration can be unitarily implemented on Fock space.

As discussed before at the end of section \ref{sec:CanQuantCanTrafo}, the most common transformation in the literature in the context of the Lewis-Riesenfeld invariant is the one where the time-dependent Hamiltonian is mapped to the Hamiltonian of a harmonic oscillator with frequency $\omega_k^{(0)}=1$. For this reason we also analyze what happens to the Shale-Stinespring condition if we do not require the time-independent frequency to be just $\omega^{(0)}_{\mathbf{k}} = k$ but unity instead. This changes the solution $\xi_\mathbf{k}(\eta)$ by an additional factor of $k^{-\frac{1}{2}}$ if we impose similar initial conditions, that is $\xi^{(sq)}_\mathbf{k}(\eta)=k^{-\frac{1}{2}}\xi_\mathbf{k}(\eta)$ and leads to the residual squeezing transformation in $\hat{\Gamma}^\mathbf{k}_\xi$ in the limit of past conformal infinity already mentioned in section \ref{sec:CanQuantCanTrafo}. Considering this modification in $\xi_{\bf k}^{sq}$ compared to $\xi_{\bf k}$, we can also analyze whether the Shale-Stinespring conditions is satisfied here. We have:
\begin{align*}
    \int_{\mathbb{R}^3} \d^3k \, |h_{\mathbf{k}}(\xi^{(sq)}, \xi^{(sq)\prime}_{\bf k}) |^2
&= \frac{1}{4} \int_{\mathbb{R}^3} \d^3k \,  \bigg[ \bigg(\frac{\big( \xi^{(sq)}_\mathbf{k}(\eta) \big)^2 - 1}{\xi^{(sq)}_\mathbf{k}(\eta)} \bigg)^2 + \Big( \xi^{(sq)\prime}_\mathbf{k}(\eta) \Big)^2 \bigg] \\[0.6em]
&= \frac{1}{4} \int_{\mathbb{R}^3} \d^3k \, \bigg[ \big( \xi^{(sq)}_\mathbf{k} \big)^{-2} \bigg( \frac{1}{k^3 \eta^2} - \frac{k-1}{k} \bigg)^2 + \frac{1}{k} \bigg( \frac{(k \eta)^2}{1+(k \eta)^2} \frac{1}{k^4 \eta^6} \bigg)\bigg] \\[0.6em]
&= \frac{1}{4} \int_{\mathbb{R}^3} \d^3k\left[ \, k \bigg( \frac{(k \eta)^2}{1 + (k \eta)^2} \bigg) \bigg( \frac{1}{k^3 \eta^2} - \frac{k-1}{k} \bigg)^2 + \frac{1}{k} \bigg( \frac{(k \eta)^2}{1+(k \eta)^2} \frac{1}{k^4 \eta^6} \bigg)\right].
\end{align*}
We apply a similar power counting to the two terms involved in the last line separately. For small $k$ the second summand in that line decays as $k^{-3}$, whereas it is proportional to $k^{-5}$ in the limit of large $k$, which yields a finite contribution in the ultraviolet and an infrared divergence. Similarly, in the small $k$ region at lowest order, the first summand behaves as $k^{-3}$ and thus is divergent in the infrared. Furthermore, it increases linearly in $k$ in the large $k$ limit causing a divergence in the  ultraviolet. As a consequence, also the transformation associated with $\xi^{(sq)}_{\bf k}$ is not unitarily implementable on Fock space, just as it was the case with $\xi_{\bf k}$. However, there is a subtle distinction between the two cases. For $\xi_{\bf k}$ we found that the infrared modes lead to a divergence, whereas this time both small and large values of $\Norm{\bf k}$ are problematic. Let us understand a bit more in detail why it is expected that the infrared modes can be problematic in the case of the map corresponding to $\xi_{\bf k}$. This map transforms the time-dependent harmonic oscillator Hamiltonian with frequency $\omega^2_{\bf k}(\eta)=k^2-\frac{z''(\eta)}{z(\eta)}$ into the Hamiltonian of the harmonic oscillator with constant frequency $\omega_{\bf k}^{(0)}=k$. Hence, for $\mathbf{k}=0$ the latter corresponds to the Hamiltonian of a free particle because here the frequency just vanishes. This aspect has not been carefully taken into account in the map constructed so far. Therefore, in the next section we will discuss how the map constructed up to now could be modified for the low $\norm{\mathbf{k}}$ modes such that the infrared singularity can be avoided. For the map corresponding to $\xi_{\bf k}$, this attempt is possible because the problematic behavior of these modes constitutes only a compact domain in the space of modes, in contrast to the additional ultraviolet divergence involved in the map associated with $\xi^{(sq)}_{\bf k}$. For this purpose, we will introduce the so-called Arnold transformation that has been used already in \cite{Arnold_trafo:Guerrero} at the quantum level, which is designed to perform a mapping to the free particle Hamiltonian.

\subsection{Proposal of a modified map for the infrared modes: The Arnold transformation}
\label{sec:ZeroMode}
In the previous section we have observed that the map $\hat{\Gamma}_\xi$ that maps the time-dependent harmonic oscillator system onto the system of a harmonic oscillator with constant frequency $\omega_{\bf k}^{(0)}=k$ with $k \defeq \Norm{\bf k}$ is not a unitary operator on Fock space due to an infrared divergence that occurs in the off-diagonal trace of the Bogoliubov coefficients for the infrared modes. Given the fact that no ultraviolet singularities arise, the strategy we will follow in this section is to consider a modification of the map induced by $\hat{\Gamma}_\xi$ for a finite spherical neighbourhood $1 \gg \norm{{\bf k}_\epsilon} > 0$ of the infrared modes in such a way that no infrared singularities occur. As mentioned above, the natural target Hamiltonian we should map to in the case of the zero mode is the Hamiltonian of a free particle. At the classical level this so-called Arnold transformation \cite{Arnold:1978} was introduced in order to transform a generic second-order differential equation, that physically describes a driven harmonic oscillator with time-dependent friction coefficient and time-dependent frequency, into the differential equation corresponding to the motion of a free particle. Its implementation as a unitary map at the quantum level has been investigated for instance in \cite{Arnold_trafo:Guerrero}. From our approach we can  make an immediate connection to this formalism by going back to equation (\ref{eq:autonomous_H}) that has played an important role in deriving our classical transformation. Now if we aimed at mapping the original time-dependent Hamiltonian $H(\eta)$ onto the free particle Hamiltonian, $\xi(\eta)$ would need to satisfy the harmonic equation of motion with the frequency $\omega_{\mathbf{k}}(\eta)$ for each mode instead of the Ermakov equation, as it was presented in our case before. As already discussed in \cite{Arnold_trafo:Guerrero}, one can recover the Ermakov equation when considering three physical systems, a time-dependent and a time-independent harmonic oscillator together with the free particle. Then one constructs the two Arnold transformations that relate the time-dependent and the time-independent harmonic oscillator to the free particle. From combining one of these Arnold transformation with the inverse of the second one, one obtains a map that relates the systems of the time-dependent harmonic oscillator with the time-independent one via a time-rescaling. For more details regarding this aspect we refer the reader to the presentation in \cite{Arnold_trafo:Guerrero}. In order to be able to discuss  the approach from \cite{Arnold_trafo:Guerrero} and ours in parallel, we will denote the time-rescaling function associated with the Arnold transformation by $\Theta_{{\bf k}_\epsilon}(\eta)$. The non-zero $\norm{{\bf k}_\epsilon}$ modes lead to the well-known solutions of the Mukhanov-Sasaki equation for finite $\norm{\bf k}$, whereas for the $\mathbf{k}=0$ mode we need to find appropriate solutions. By construction, $\Theta_{{\bf k}_\epsilon}(\eta)$ satisfies the Ermakov equation with vanishing $\omega_{\bf k}^{(0)}$, that is the time-dependent harmonic oscillator equation of the associated mode, given by:
\begin{align*}
    \Theta''_{{\bf k}_\epsilon}(\eta) + \omega^2_{\mathbf{k}_\epsilon}(\eta) \Theta_{{\bf k}_\epsilon}(\eta) = 0,
\end{align*}
where we are interested in the case where $\omega_{\mathbf{k}_\epsilon}(\eta)$ is determined by the Mukhanov-Sasaki equation. If we compare the time rescaling in equation (\ref{eq:timescaling}) with the one given in \cite{Arnold_trafo:Guerrero}, we obtain an exact agreement if we take into account that the Wronskians of two solutions of the time-dependent and time-independent harmonic oscillator, respectively, are constant and can be chosen to be identical. Since in our work the physical system under consideration is described by a time-dependent harmonic oscillator (that is, the Mukhanov-Sasaki equation), let us first consider this equation for arbitrary modes $\mathbf{k}$ and prior to any gauge-fixing:
\begin{align}
    v''_{\mathbf{k}}(\eta) +  \bigg( \Norm{\mathbf{k}}^2
- \frac{z''(\eta)}{z(\eta)} \bigg) v_{\mathbf{k}}(\eta) = 0,
\label{eq:Mukhanov_Sasaki}
\end{align}
For the particular case of a quasi-de Sitter spacetime, the explicit form of this equation can be given in terms of the so-called slow-roll parameters. The Friedmann equations together with the Klein-Gordon equation describing the dynamics of the background scalar field $\phi$ on a slow-rolling quasi-de Sitter background can be used to rewrite the Mukhanov-Sasaki equation in a convenient way. For this purpose we define a set of three slow-roll parameters $\varepsilon, \tau$ and $\kappa$, that describe the fractional change of $\dot{H}$ per Hubble time, the fractional chance of $\varepsilon$ per Hubble time as well as the fractional change of $\tau$ per Hubble time, respectively:
\begin{align*}
    \varepsilon = - \frac{\dot{H}}{H^2}, \quad
    \tau = \frac{\dot{\varepsilon}}{\varepsilon H}, \quad
    \kappa = \frac{\dot{\tau}}{\tau H},
\end{align*}
where a dot denotes a derivative with respect to cosmological time in these expressions. Inserting the Klein-Gordon equation and using  the Friedmann equations along with an assumed subdominance of the second-order derivatives of $\phi$, we can rewrite $z(\eta)$ and its derivatives in terms of $\varepsilon$, $\tau$ and $\kappa$. By truncating the resulting expressions after the first order in the slow-roll parameters,  the time-dependent part of the frequency $\omega_k(\eta)$ becomes:
\begin{align*}
    z = \frac{a^2 \varepsilon}{4 \pi}, \quad 
    \frac{z'}{z} = \mathcal{H} \Big(1+ \frac{\tau}{2} \Big), \quad
    \frac{z''}{z} = \frac{1}{\eta^2} \big(1+\varepsilon \big)^2
    \Big( 2-\varepsilon + \frac{3 \tau}{2} \Big) \approx
    \frac{4\nu^2 - 1}{4\eta^2}, \quad \nu = \frac{3}{2} + \varepsilon + \frac{\tau}{2}.
\end{align*}
Thus, the Mukhanov-Sasaki equation up to first order in the slow-roll parameters for a quasi-de Sitter background reads:
\begin{align}
    v''_{\mathbf{k}}(\eta) +  \bigg( \Norm{\mathbf{k}}^2
- \frac{4\nu^2 - 1}{4\eta^2} \bigg) v_{\mathbf{k}}(\eta) = 0.
\label{eq:Mukhanov_slow_roll}
\end{align}
Given the equation above we can read off the time-dependent frequency that we considered for the time-dependent harmonic oscillator in our single-mode toy model approach. This is also precisely the equation that $\Theta(\eta)$ needs to satisfy for a given but finite $\Norm{{\bf k}_\epsilon}$. Let us emphasize that the solutions to equation (\ref{eq:Mukhanov_slow_roll}) need to be computed separately for vanishing and non-vanishing $\norm{\bf k}$, respectively. The real-valued solutions for $\norm{\bf k} > 0$ are given by the Bessel functions of first and second kind, for details the reader is referred to section \ref{sec:applications}. If we consider the limiting case of $\mathbf{k}=0$ in the context of the quantum Arnold transformation, we obtain the following linear differential equation with time-dependent coefficients for the rescaling function $\Theta_{{\bf k}_\epsilon}(\eta)$, omitting the label for the zero mode:
\begin{equation*}
  \Theta''(\eta) - \frac{4\nu^2 - 1}{4\eta^2} \Theta(\eta) = 0 \qquad
  \Longleftrightarrow \qquad
  \eta^2 \Theta''(\eta) - \Big(\nu^2 - \frac{1}{4} \Big) \Theta(\eta) = 0 \qquad {\rm for}
  \qquad \norm{{\bf k}} = 0.
\end{equation*}
This differential equation with time-dependent coefficients can be transformed into an equation with constant coefficients, which then again can be solved by means of the substitution $y = \ln(|\eta|)$ and an exponential ansatz of this new variable incorporating the dependence on the effective slow-roll parameter $\nu$. The general solution of this differential equation is given by:
\begin{equation}
    \Theta(\eta) = c_1 |\eta|^{r} + c_2 |\eta|^{s}, \quad \textrm{with} \quad r,s=\frac{1}{2} \Big(1 \pm \sqrt{4\nu^2} \Big) \quad \textrm{for} \quad \nu^2 > 0
    \label{eq:Arnold_solution}
\end{equation}
From this solution we readily obtain two linearly independent solutions $\Theta_1, \Theta_2$ that can be used to construct the Arnold transformation for the $\mathbf{k}=0$ mode. Note that the differential equation above can be also solved for $\nu=0$ or $\nu^2<0$, respectively. However, according to the parameter space of the slow-roll parameters in \cite{Planck_inflation:2018}, this range is not physically reasonable and hence we only use the result for strictly positive, real-valued slow-roll parameters. Due to the range of conformal time ($\eta \in \mathbb{R}_-\setminus \{0\}$), the relevant solution here is the growing branch proportional to $|\eta|^s$ with $s<0$, since the decreasing branch diverges in the limit of past conformal infinity. This then coincides with the choice of the final time-rescaling transformation suggested in \cite{Arnold_trafo:Guerrero}. We reconsider the form of the transformation $\hat{\Gamma}_\xi$ and insert the corresponding solution for $\Theta_{{\bf k}_\epsilon}(\eta)$ to obtain an analogous transformation $\hat{\Gamma}_{\Theta_{{\bf k}_\epsilon}}$ by means of which we can transform the Schr\"odinger equation similar to equation (\ref{eq:transformed_Schroedinger}) and according to:
\begin{align*}
    i \frac{\partial}{\partial t} \Psi(q,t) = \frac{1}{2} \Big( \pi_q (\hat{p})^2 + \omega_{{\bf k}_\epsilon}^2(t) \pi_q (\hat{q})^2 \Big)\Psi(q,t) \quad \Longleftrightarrow \quad
    \bigg( \frac{1}{2 \Theta_{{\bf k}_\epsilon}^2} \pi_q (\hat{p})^2 - i \frac{\partial}{\partial t} \bigg) \hat{\Gamma}_{\Theta_{{\bf k}_\epsilon}} \Psi(q,t) = 0.
\end{align*}
This means that $\hat{\Gamma}_{\Theta_{{\bf k}_\epsilon}}$ maps the time-dependent Hamiltonian for the ${\bf k}_\epsilon$ mode into the time-independent Hamiltonian of the free particle modulo a rescaling of the momentum operator. It is important to emphasize that this unitary map can only be performed at the level of the full Schr\"odinger equation, as otherwise the spectrum of the two related operators would have to be equivalent, which is clearly not the case for the time-dependent harmonic oscillator and the free particle. It is the time-derivative in the Schr\"odinger equation that is crucial for removing the term proportional to $\pi_q(\hat{q})^2$ altogether. Furthermore we would like to stress that the time rescaling function $\Theta_{{\bf k}_\epsilon}$ is different for $\norm{{\bf k}_\epsilon} > 0$ and $\norm{{\bf k}_\epsilon}=0$, respectively. In the first case, depending on the imposed initial conditions, it is given by the Bessel functions of first and second kind $J_\nu(-{\bf k}_\epsilon \eta)$ and $Y_\nu(-{\bf k}_\epsilon \eta)$. In the latter case, $\Theta_{{\bf k}_\epsilon}$ corresponds to the above power-law solution. Unfortunately, there are some drawbacks of the implementation of the quantum Arnold transformation with the help of $\hat{\Gamma}_{\Theta_{{\bf k}_\epsilon}}$. Firstly, the limit of past conformal infinity is not well-defined in terms of the generator $\hat{\mathcal{G}}$ as depicted in (\ref{eq:canonical_trafo}), for neither of the two cases. This especially means that we do not get an asymptotic identity map for an already free particle (i.e. the $\norm{{\bf k}_\epsilon}=0$ case in the limit of past conformal infinity) as we do with the initially time-independent harmonic oscillator in the case of the original transformation $\hat{\Gamma}_{\xi}$. Secondly, the attempt to relate the free particle with a Lewis-Riesenfeld type invariant does not work as smoothly as in the case of the previous map. If we construct a similar invariant in this case here for the initially time-dependent oscillator Hamiltonian, it can be trivially factorized and has the following form:
\begin{align}
    I_{LR}=\frac{1}{2} \hat{\Gamma}^\dagger_{\Theta_{{\bf k}_\epsilon}} \pi_q (\hat{p})^2 \hat{\Gamma}_{\Theta_{{\bf k}_\epsilon}} = \frac{1}{2} \big(\Theta_{{\bf k}_\epsilon} \pi_q (\hat{p}) - \Theta_{{\bf k}_\epsilon}^\prime \pi_q (\hat{q}) \big)^2 = \hat{a}_{{\bf k}_\epsilon}^\dagger \hat{a}_{{\bf k}_\epsilon}, \quad \hat{a}_{{\bf k}_\epsilon}=\frac{i}{\sqrt{2}}\big( \Theta_{{\bf k}_\epsilon} \pi_q (\hat{p}) - \Theta'_{{\bf k}_\epsilon} \pi_q (\hat{q}) \big).
    \label{eq:Arnold_invariant}
\end{align}
This quantity has for example been already obtained in \cite{Schroedinger_1D} as a quantum invariant based on orthogonal functions in a similar context. It is immediate that the above factorization is in this sense pathological, as one can immediately see that the occurring operators $\hat{a}_{{\bf k}_\epsilon},\hat{a}_{{\bf k}_\epsilon}^\dagger$ can not be interpreted as ladder operators due to $[\hat{a}_{{\bf k}_\epsilon}, \hat{a}_{{\bf k}_\epsilon}^\dagger] = 0$, since they only differ by a global sign. This can be also seen by looking at the original invariant (\ref{eq:lewis_invariant}) which has an additional term proportional to $\hat{q}^2 \xi^{-2}$ that is absent in the case of the Arnold transformation for $\mathbf{k}_\epsilon$ by construction, simply because of the lack of a term proportional to $\pi_q (\hat{q})$ in the transformed free particle Hamiltonian. Nevertheless, the transformation $\hat{\Gamma}_{\Theta_{{\bf k}_\epsilon}}$ is unitary for all finite times $\eta$ and all considered modes. However, due to the non-preservation of the commutator structure between $\hat{a}_{{\bf k}_\epsilon}$ and $\hat{a}_{{\bf k}_\epsilon}^\dagger$, it is not a Bogoliubov transformation, hence it does not qualify as an infrared continuation of the map $\hat{\Gamma}_\xi$ used throughout this work.
\\

In summary, it was not possible to find a transformation similar to $\hat{\Gamma}_\xi$ for the infrared modes. Regarding predictions in inflationary comsology, we are naturally interested in the large $k$ modes, which are properly implementable in the context of our symplectic transformation. Hence, as an alternative to the proposed maps for the infrared, we suggest the identity map as a proper choice, that is:

\begin{equation}
    \hat{\Gamma}_\xi \defeq
       	\begin{cases} 
            \exp \Big( -i \Big[ \int_{V_\epsilon} \d^3k \; 
            \hat{\mathcal{G}} (\xi_{\bf k}, \dot{\xi}_{\bf k}), \cdot \Big ] \Big) \quad \, {\rm for} \quad \norm{\bf k} > \norm{{\bf k}_\epsilon}, \\
           	\mathds{1}_\mathcal{H} \qquad \qquad \qquad \qquad
           	\qquad \qquad \quad \; {\rm for} \quad
           	\norm{\bf k} \leq \norm{{\bf k}_\epsilon},
        \end{cases} \label{eq:full_bogoliubov}
\end{equation}

where $V_\epsilon \defeq \{ {\bf k} \in \mathbb{R}^3: \norm{\bf k} > \norm{{\bf k}_\epsilon} \}$ is the smearing domain and $\hat{\mathcal{G}}_{\bf k}$ denotes the mode-dependent generator of the Bogoliubov transformation depicted in equation (\ref{eq:canonical_trafo}) and especially in (\ref{eq:Gamma_generator}) in terms of annihilation and creation operators, respectively. This is possible and well-defined since the occurring coefficients in the generator are smooth for $\norm{\bf k} > \norm{{\bf k}_\epsilon}$ and moreover lie in $L^1(V_\epsilon, \d^3k)$ as can be checked by explicit integration. The reasons for choosing the identity map are twofold. Firstly, this trivially constitutes a Bogoliubov transformation with the off-diagonal coefficients in (\ref{eq:Bogoliubov_matrix}) vanishing, rendering the Shale-Stinespring integral finite, thus allowing for unitary implementability of $\hat{\Gamma}^{\bf k}_\xi$ on Fock space. Secondly, the functions multiplying the off-diagonal elements in the Mukhanov-Sasaki Hamiltonian remain unchanged compared to the standard case, which means that they can be neglected for sufficiently early times. This is due to the fact that the effective friction term in the equation of motion for these functions is subdominant in this regime. For details the reader is referred to the discussion in the succeeding section.

\section{Relation of the Lewis-Riesenfeld invariant approach to the Bunch-Davies vacuum and adiabatic vacua}
\label{sec:RelAdVacua}
In the context of the results  of the previous sections, it is a natural question  whether there exists a relation of the mode functions obtained in the framework of the formalism in this work and the ones obtained in the standard approach in cosmology. As we will show by taking time-rescaling transformation into account, we can relate the solutions $\xi_\mathbf{k}$ of the Ermakov equation to the mode functions associated with the Bunch-Davies vacuum and other adiabatic vacua. For this purpose we consider the following form of the Mukhanov-Sasaki mode function
\begin{equation}
 v_{\mathbf{k}}(\eta) = N_{\mathbf{k}} \xi_{\mathbf{k}}(\eta) \exp \Big\{ -i \omega^{(0)}_{\mathbf{k}} \int^\eta \frac{\d \tau}{\xi_{\mathbf{k}}^2(\tau)} \Big\},
    \label{eq:polar_modes}   
\end{equation}
corresponding to a polar representation of the complex mode $v_k$ into a real function $\xi_{\bf k}$ and a complex phase that was in a similar form already mentioned in \cite{invariant_vacuum}.  $N_\mathbf{k}$ is time-independent for each mode, $\xi_{\mathbf{k}}(\eta)$ remains arbitrary at this point and $\omega^{(0)}_{\mathbf{k}}$ can take the values $k$ or $1$ depending on the choice of map that is considered. We want to show that the Mukhanov-Sasaki equation
\begin{equation}
    v^{\prime\prime}_{\mathbf{k}}(\eta)+\omega_{\mathbf{k}}^2(\eta) v_{\mathbf{k}}(\eta) = 0\end{equation}
expressed in terms of the polar representation exactly coincides with the Ermakov equation. Starting from this polar representation of the mode functions we compute the second derivative and reinsert it into the Mukhanov-Sasaki equation to obtain:
\begin{equation}
    v^{\prime\prime}_{\mathbf{k}}(\eta) = N_{\mathbf{k}} \exp \Big\{ -i \omega^{(0)}_{\mathbf{k}} \int^\eta \frac{\d \tau}{\xi_{\mathbf{k}}^2(\tau)} \Big\} \bigg( \xi_{\mathbf{k}}'' - i \omega^{(0)}_{\mathbf{k}} \frac{\xi_{\mathbf{k}}'}{\xi_{\mathbf{k}}^2} + i \omega^{(0)}_{\mathbf{k}} \frac{\xi_{\mathbf{k}}'}{\xi_{\mathbf{k}}^2} - \frac{(\omega^{(0)}_{\mathbf{k}})^2}{\xi_\mathbf{k}^3} \bigg).
    \label{eq:polar_modes_derivative}
\end{equation}
We realize that summands involving $\xi_{\bf k}^\prime$ cancel each other and the Mukhanov-Sasaki equation can be rewritten as:
\begin{equation}
    \quad \xi^{\prime\prime}_{\mathbf{k}} + \omega_\mathbf{k}^2(\eta) \xi_\mathbf{k} - \frac{(\omega^{(0)}_{\mathbf{k}})^2}{\xi_\mathbf{k}^3} = 0\quad 
    \Longleftrightarrow\quad 
    v^{\prime\prime}_{\mathbf{k}}(\eta)+\omega_{\mathbf{k}}^2(\eta) v_{\mathbf{k}}(\eta) = 0.
\end{equation}
That is, we recover the Ermakov equation for the radial part of the polar representation in equation (\ref{eq:polar_modes}). The polar representation of the mode functions can also be obtained if we consider how the Fourier modes transform under the time-dependent canonical transformation that relates the time-dependent and time-independent harmonic oscillator. The mode functions in the system of the harmonic oscillator written as a function of conformal time are given by $u_{\bf k}(\eta)=N_{\bf k} \exp \big( -i\omega^{(0)}_{\bf k}\int\limits^\eta \frac{d{\tau}}{\xi^2_{\bf k}(\tau)} \big)$. Here $u_{\bf k}(T)$ satisfies the standard harmonic oscillator differential equation with respect to the time variable $T$. Using that for each mode we have $T^\prime_{\bf k}=\xi^{-2}_{\bf k}$, one can easily derive the corresponding differential equation that $u_{\bf k}(\eta)$ fulfills with respect to conformal time $\eta$. Now the time-dependent canonical transformation rescales the spatial coordinate by $\xi^{-1}_{\bf k}$. Considering this as well as the fact that the mode $u_{\bf k}(\eta)$ depends on $k$ only, the corresponding Fourier mode after the transformation is given by $v_{\bf k}(\eta)=\xi_{\bf k} u_{\bf k}(\eta)$, yielding again the polar representation of the Fourier mode shown in (\ref{eq:polar_modes}), where we used how the Fourier transform changes under a scaling of the coordinates. 
\\

A second way to obtain this result is via the explicit form of the Bogoliubov transformation associated with the time-dependent canonical transformation. We denote the time-dependent annihilation and creation operators of the harmonic oscillator system by $\hat{b}_{\bf k}(T)=u_{\bf k}(T)\hat{b}_{\bf k}$ and $\hat{b}^\dagger_{\bf k}(T)=\overline{u}_{\bf k}(T)\hat{b}^\dagger_{\bf k}$ respectively, where the time-dependent annihilation and creation operators satisfy the Heisenberg equation associated with the Hamiltonian of the harmonic oscillator. Once more considering the relation between $T$ and $\eta$ for each mode, we can also understand $\hat{b}_{\bf k}(\eta)$ and $\hat{b}^\dagger_{\bf k}(\eta)$ as operator-valued functions of conformal time $\eta$. The mode expansion in the system of the time-dependent harmonic oscillator can be written in terms of time-dependent annihilation and creation operators $\hat{a}_{\bf k}(\eta)=v_{\bf k}(\eta)\hat{a}_{\bf k}$ and  $\hat{a}^\dagger_{\bf k}(\eta)=\overline{v}_{\bf k}(\eta)\hat{a}^\dagger_{\bf k}$ which both satisfy the Heisenberg equation associated to the Mukhanov-Sasaki Hamiltonian. As shown in section \ref{sec:FockSpace}, the time-dependent canonical transformation corresponds to a time-dependent Bogoliubov map at the quantum level. In the notation of the last section, this relates the two sets of annihilation and creation operators as follows\footnote{Note that the roles of $\hat{a}_{\bf k}, \hat{a}_{\bf k}^\dagger$ and $\hat{b}_{\bf k}, \hat{b}_{\bf k}^\dagger$ are interchanged in comparison to the one-particle case considered in (\ref{eq:Bogoliubov_matrix}) for notational convenience, whereas the coefficients are named analogously. Here the first set of operators belongs to the Mukhanov-Sasaki Hamiltonian, whereas the second set is associated to the time-independent harmonic oscillator. In contrast, in (\ref{eq:Bogoliubov_matrix}) the operators $\hat{B},\hat{B}^\dagger$ belong to  the time-dependent system, whereas $\hat{A},\hat{A}^\dagger$ are associated with the time-independent harmonic oscillator.}:
\begin{equation*}
\hat{a}_{\bf k}(\eta)=g_{\bf k}(\xi,\xi')\hat{b}_{\bf k}(\eta)+\overline{h}_{\bf k}(\xi,\xi')\hat{b}^\dagger_{\bf k}(\eta),\quad
\hat{a}^\dagger_{\bf k}(\eta)=\overline{g}_{\bf k}(\xi,\xi')\hat{b}^\dagger_{\bf k}(\eta)+h_{\bf k}(\xi,\xi')\hat{b}_{\bf k}(\eta).
\end{equation*}
The explicit form of these coefficients is given by:
\begin{equation}
g_{\bf k}(\xi,\xi')=\frac{1}{2}\left(\xi_{\bf k}+\frac{1}{\xi_{\bf k}}\right)+\frac{i}{2\omega^{(0)}_{\bf k}}\xi^\prime,\quad\quad 
h_{\bf k}(\xi,\xi')=\frac{1}{2}\left(\xi_{\bf k}-\frac{1}{\xi_{\bf k}}\right)-\frac{i}{2\omega^{(0)}_{\bf k}}\xi^\prime.
\end{equation}
Given this time-dependent Bogoliubov map, the Fourier modes in the two systems are related via
\begin{equation}
v_{\bf k}(\eta)=\left(g_{\bf k}(\xi,\xi^\prime)+ h_{\bf k}(\xi,\xi^\prime)\right)u_{\bf k}
=\xi_{\bf k}u_{\bf k}(\eta) = N_{\bf k}\xi_{\bf k} \exp \Big\{ -i \omega^{(0)}_{\mathbf{k}} \int^\eta \frac{\d \tau}{\xi_{\mathbf{k}}^2(\tau)} \Big\}.
\end{equation}
Hence, we again recover the polar representation of the Fourier mode. At this point we did not yet clarify the purpose of the $N_\mathbf{k}$, which is intricately connected with the commutator algebra of annihilation and creation operators as we will see. Recall the well-known (off-diagonal) form of the Mukhanov-Sasaki Hamiltonian if we insert the mode expansions into the Hamiltonian density:
\begin{equation}
\label{eq:MSHam}
    H = \int \frac{\d ^3k}{(2 \pi)^3} \bigg[ F_\mathbf{k}(\eta) \hat{a}_\mathbf{k} \hat{a}_\mathbf{-k}
    + \overline{F}_\mathbf{k}(\eta) \hat{a}^\dagger_\mathbf{k} \hat{a}^\dagger_\mathbf{-k} 
    + E_\mathbf{k}(\eta) \Big( 2 \hat{a}^\dagger_\mathbf{k} \hat{a}_\mathbf{k} + (2 \pi)^3 \delta^{(3)}(0) \Big) \bigg], 
\end{equation}
where we used the isotropy of the mode functions due to the high degree of symmetry of the spacetime, the invariance of the measure under reflection and the following definitions:
\begin{equation}
    F_\mathbf{k}(\eta) \defeq \big(v^{\prime}_\mathbf{k}\big)^2 + \omega^2_\mathbf{k}(\eta) v^2_\mathbf{k}, \quad
    E_\mathbf{k}(\eta) \defeq v^{\prime}_\mathbf{k} \overline{v}^{\prime}_\mathbf{k} + \omega^2_\mathbf{k}(\eta) v_\mathbf{k} \bar{v}_\mathbf{k}.
\end{equation}

Regarding the normalization of the mode functions $v_\mathbf{k}$, we can transfer this condition to the polar representation given in equation (\ref{eq:polar_modes}) by just inserting the definition into the Wronskian. This removes the dependence on $\xi_\mathbf{k}$ completely and we can explicitly give a relation between $N_\mathbf{k}$ and the Wronskian of the original mode functions:
\begin{equation}
    W(v_\mathbf{k}, \overline{v}_\mathbf{k}) = 2i \omega^{(0)}_\mathbf{k} N^2_\mathbf{k}.
\end{equation}
This is not surprising upon closer inspection. Recall that the $\omega^{(0)}_\mathbf{k}$ in the Ermakov equation corresponds to the time-independent frequency in the transformed Schr\"odinger equation. We conveniently chose to map the Mukhanov-Sasaki frequency into just the $\mathbf{k}$-dependent part, completely removing the time dependence. This has the effect that $\hat{\Gamma}_\xi$ becomes the identity transformation for the case of an initially time-independent oscillator, whereas we would obtain a residual squeezing if we mapped every mode to unity. This freedom of choice is reflected in the explicit form  of the normalization constant $N_\mathbf{k}$, which depends on the choice of the oscillator frequency in the target system in order to preserve the normalization of the mode functions and hence the standard commutator algebra of annihilation and creation operators. Given the Mukhanov-Sasaki Hamiltonian in the form of annihilation and creation operators in (\ref{eq:MSHam}), we can discuss the assumptions for the initial condition regarding the Fourier modes associated with the Bunch-Davies vacuum and the ones obtained in our work and compare them, consequently. First, we rewrite the Fourier mode associated to the Bunch-Davies vacuum given by
\begin{equation}
\label{eq:BD_mode_function}
v^{\rm BD}_{\bf k}(\eta)=\frac{1}{\sqrt{2 k}}\left(1-\frac{i}{k\eta}\right)e^{-ik\eta}
\end{equation}
in the polar representation as shown in (\ref{eq:polar_modes}) for our general solution. This yields:
\begin{equation}
v^{\rm BD}_{\bf k}(\eta)=i|v_{\bf k}^{\rm BD}|\exp \Big\{ -i \omega^{(0)}_{\mathbf{k}} \int^\eta \frac{\d \tau}{\xi_{\mathbf{k}}^2(\tau)} \Big\} 
=\frac{i}{\sqrt{2 k}}\sqrt{1+\frac{1}{(k\eta)^2}} \exp \Big\{ -i \omega^{(0)}_{\mathbf{k}} \int^\eta \frac{\d \tau}{\xi_{\mathbf{k}}^2(\tau)} \Big\},
\end{equation}
which corresponds exactly to the $\xi_{\bf k}^{(sq)}$ that we obtained from the Ermakov equation by requiring appropriate initial conditions for $\xi_{\bf k}$ which carry over to initial conditions on the Fourier mode and the additional factor $i$ comes from the phase of (\ref{eq:BD_mode_function} compared to the one arising from the integral.

As far as the Hamiltonian diagonalization (HD) of the Mukhanov-Sasaki Hamiltonian is concerned, one diagonalizes this Hamiltonian instantaneously at some time $\eta_0$ which requires the coefficients $F_{\bf k}$ and $\overline{F}_{\bf k}$ to vanish at $\eta_0$. In addition it can be shown that the state satisfying this requirement also minimizes the energy at that time $\eta_0$, so that requiring both does not yield to further conditions on the state. If the requirement $F_{\bf k}=0$ and the normalization of the Wronskian $W(v_{\bf k},\overline{v}_{\bf k})$ to $W(v_{\bf k},\overline{v}_{\bf k})=i$ is satisfied, we choose the following initial conditions for the model:
\begin{equation}
{\rm HD\, (I)}\quad |v_{\bf k}|(\eta_0)=\frac{1}{\sqrt{2 \omega_{\bf k}(\eta_0)}} \quad {\rm and}  \quad
{\rm HD \, (II)}\quad v^\prime_{\bf k}(\eta_0)=-i\omega_{\bf k}(\eta_0)v_{\bf k}(\eta_0).
\end{equation}
If we consider the specefic choice $\eta_0\to-\infty$ in this context we exactly end up with the initial conditions usually chosen to obtain the Bunch-Davies vacuum:
\begin{equation}
{\rm BD\, (I)}\quad |v_{\bf k}|(-\infty)=\frac{1}{\sqrt{2 k}} \quad {\rm and}  \quad
{\rm BD \, (II)}\quad v^\prime_{\bf k}(-\infty)=-ikv_{\bf k}(-\infty),
\end{equation}
where we used that $\omega_k=k$ at $\eta_0\to-\infty$, meaning that the modes become the standard Minkowski modes in this limit. Looking closer into the condition $F_{\bf k}=0$ we can rewrite this non-linear differential equations as:
\begin{equation}
\label{eq:Falltimes}
F_{\bf k}=0\quad\Longleftrightarrow\quad v^\prime_{\bf k}(\eta)\left(v^{\prime\prime}_{\bf k}-\left(\frac{\omega_{\bf k}^\prime(\eta)}{\omega_{\bf k}(\eta)}\right)v^\prime_{\bf k}(\eta)+\omega^2_{\bf k}(\eta)v_{\bf k}(\eta)\right)=0.
\end{equation}
We realize that $F_{\bf k}=0$ at all times $\eta$ requires that $v_k$ satisfies a differential equations that looks like the Mukhanov-Sasaki equation but with an additional friction term included. For a constant frequency $\omega_{\bf k}$ the friction term vanishes, which for the case of de Sitter where $\omega^2_{\bf k}(\eta)=k^2-\frac{2}{\eta^2}$ is given in the limit of large $k$. For de Sitter the friction coefficient reads $\frac{\omega^\prime_{\bf k}}{\omega_{\bf k}}=\frac{2}{\eta}\frac{1}{(k\eta)^2-2}$ and thus, depending on the values of $k$ and $\eta$ it will not always be negligible, which is the reason why in the case of Bunch-Davies one can only achieve an instantaneous Hamiltonian diagonalization. This is due to the fact that $v_k$ satisfies the Mukhanov-Sasaki equation and at the same time needs to fulfill $F_{\bf k}=0$, generally being in conflict already for the simple case of a de Sitter universe. Note that in our work the Hamiltonian diagonalization of the Mukhanov-Sasaki Hamiltonian can be obtained for each instant in time and is not obtained by setting $F_{\bf k}(\eta)$ equal to zero but by a time-dependent unitary transformation that involves also a time-rescaling. Now since we fixed our initial condition in the limit $\eta\to -\infty$ and, as we will show below, the solution we obtained satisfies the differential equation for adiabatic vacua without any approximation, it is very natural that our initial conditions at $\eta_0=-\infty$ are given by the following:
\begin{equation}
{\rm LR\,\, (I)}\quad |v_{\bf k}(\eta_0)|=|N_{\bf k} \xi_{\bf k}(\eta_0)|=\frac{1}{\sqrt{2 k}} \quad{\rm and}\quad
{\rm LR\,\, (II)}\quad v_{\bf k}^\prime(\eta_0)=-i\omega^{(0)}_{\bf k}v_{\bf k}(\eta_0),
\end{equation}
where we used again the same normalization of the Wronskian for the condition LR (II) and that $\lim\limits_{\eta_0\to-\infty}\xi_{\bf k}(\eta_0)=1$. Thus the initial conditions obtained here coincide with the initial conditions one chooses for adiabatic vacua to any order as well as the ones chosen for the Bunch-Davies vacuum where we fix them in the large $k$ limit and for $\eta_0\to -\infty$. However, in our work the latter was necessary in order that the unitary operator that implements the Bogoliubov transformation (see \ref{eq:Gamma_generator}) becomes the identity operator for an already time-independent harmonic oscillator and is hence considerably natural. Now let us discuss how the results obtained in our work are related to the notion of adiabatic vacua. In the framework of adiabatic vacua one uses the following ansatz for the mode functions:
\begin{equation}
\label{eq:AdiabVacua}
v_{\bf k}(\eta)=\frac{1}{\sqrt{W_{\bf k}}}\exp{-i\int\limits^\eta \d \tilde{\eta}W_{\bf k}(\tilde{\eta})},    
\end{equation}
where $W_{\bf k}(\eta)$ is defined through the following differential equation
\begin{equation}
\label{eq:DefWk}
W_{\bf k}^2(\eta)=\omega^2_{\bf k}(\eta)-\frac{1}{2}\left(\frac{W_{\bf k}^{\prime\prime}(\eta)}{W_{\bf k}(\eta)}-\frac{3}{2}\left(\frac{W_{\bf k}^\prime(\eta)}{W_{\bf k}(\eta)}\right)^2\right),
\end{equation}
where $\omega_{\bf k}(\eta)$ is the time-dependent frequency, so in our case the one involved in the Mukhanov-Sasaki Hamiltonian. If we compare the ansatz in (\ref{eq:AdiabVacua}) with the form of the solution for the Mukhanov-Sasaki equation in (\ref{eq:polar_modes}), we realize that we can map the two expression for $v_{\bf k}$ into each other by the substitution $\xi_{\bf k}:=\big( \omega^{(0)}_{\bf k}\big)^{\frac{1}{2}} W_{\bf k}^{-\frac{1}{2}}$, where we choose $\omega_{\bf k}^{(0)}=k$ and $\omega_{\bf k}^{(0)}=1$, respectively to consider the case where the Mukhanov-Sasaki Hamiltonian is mapped to the harmonic oscillator with frequency $k$ and $1$, respectively. As shown above, rewritten in terms of $\xi_{\bf k}$ the Mukhanov-Sasaki equation merges into the Ermakov equation. Hence, if we express the Ermakov equation in terms of $W_{\bf k}$ we can rewrite the Mukhanov-Sasaki equation in terms of $W_{\bf k}$. For this we consider the second derivative $\xi^{\prime\prime}_{\bf k}$ expressed in terms of $W_{\bf k}$. We obtain:
\begin{equation*}
\xi_{\bf k}^{\prime\prime}=-\frac{\sqrt{\omega_{\bf k}^{(0)}}}{2}\left(\frac{W_{\bf k}^{\prime\prime}(\eta)}{W^{\frac{3}{2}}_{\bf k}(\eta)}-\frac{3}{2}\frac{
(W_{\bf k}^\prime(\eta))^2}{W^{\frac{5}{2}}_{\bf k}(\eta)}\right).
\end{equation*}
Reinserting this back into the Ermakov equation yields:
\begin{equation}
-\frac{\sqrt{\omega_{\bf k}^{(0)}}}{2}\left(\frac{W_{\bf k}^{\prime\prime}(\eta)}{W^{\frac{3}{2}}_{\bf k}(\eta)}-\frac{3}{2}\frac{
(W_{\bf k}^\prime(\eta))^2}{W^{\frac{5}{2}}_{\bf k}(\eta)}\right)  +\omega_{\bf k}^2(\eta)\frac{\sqrt{\omega_{\bf k}^{(0)}}}{W_{\bf k}^{\frac{1}{2}}(\eta)} - \frac{(\omega_{\bf k}^{(0)})^2}{(\omega_{\bf k}^{(0)})^\frac{3}{2}}W_{\bf k}^\frac{3}{2}(\eta)=0.  
\end{equation}
Multiplying the entire equation by $\big( \omega_{\bf k}^{(0)} \big)^{-\frac{1}{2}}W^{\frac{1}{2}}_{\bf k}$ we end up with:
\begin{equation}
W^2_{\bf k}=\omega^2_{\bf k}-\frac{1}{2}\left(\frac{W_{\bf k}^{\prime\prime}(\eta)}{W_{\bf k}(\eta)}-\frac{3}{2}\left(\frac{
W_{\bf k}^\prime(\eta)}{W_{\bf k}(\eta)}\right)^2\right),
\end{equation}
and this agrees precisely with the defining differential equation for $W_{\bf k}$ in (\ref{eq:DefWk}). The adiabatic condition required for the modes in this context carries over to a condition on the large ${\bf k}$ behavior of the function $\xi_{\bf k}$, being a solution of the Ermakov equation. As usual for adiabatic vacua, they do depend on the chosen extension to the infraed sector. In the formalism presented in this work this arbitrariness is encoded in the choice of how the unitary transformation is modified for the modes ${\bf k}$ with $||{\bf k}||\leq ||{\bf k}_\epsilon ||$. From this we can conclude that the ansatz for adiabatic vacua and the framework of the Lewis-Riesenfeld invariant leads to equivalent solutions for possible vacuum states if one reformulates the adiabatic condition in terms of the the solution $\xi_{\bf}$ of the Ermakov equation. Furthermore, we can understand our solution obtained for quasi-de Sitter and de Sitter in this context now. For the modes associated with the Mukhanov-Sasaki equation on a de Sitter background, the adiabatic condition needs to be satisfied for $k^2 \gg \eta^{-2}$, that is $k\eta \gg 1$. Using the explicit solution for $\xi_{\bf }$ in the case of de Sitter given by $\xi_{\bf k}=\big( 1+\frac{1}{(k\eta)^2} \big)^{\frac{1}{2}}$ we obtain $\lim\limits_{k\eta\to \infty}\xi_{\bf k}=1$. This corresponds to $\lim\limits_{k\eta\to \infty}W_{\bf k}=\omega^{(0)}_{\bf k}=k$, where we only considered the map with $\omega^{(0)}_{\bf k}=k$ here because the second one with $\omega^{(0)}_{\bf k}=1$ was not unitarily implementable on Fock space. In the case of de Sitter, the integral can be easily computed and the solution is given by
\begin{equation}
\label{eq:vkLR}
v_{\bf k}(\eta)=\frac{1}{\sqrt{2k}} \sqrt{1+\frac{1}{(k\eta)^2}}e^{-ik\eta}e^{i\arctan(k\eta)}.
\end{equation}
In case the solution for $W_{\bf k}$ cannot be determined in a simple manner, one uses a WKB approximation for the integral involved in the adiabatic ansatz in (\ref{eq:AdiabVacua}), yielding adiabatic vacua of a certain order at which the expansion is truncated, see for instance \cite{Winitzki:2005rw,Casadio:2004ru} for applications. However, since we have determined an analytical solution for the Ermakov equation for $\omega_{\bf k}^{(0)}=k$  we did not get an approximate solution for $W_{\bf k}$ up to some adiabatic order, but obtained the full solution for $W_{\bf k}$. This way of relating the two formalisms also provides the possibility to have a very clear interpretation of the Fourier mode associated with the Bunch-Davies vacuum in the Lewis-Riesenfeld invariant formalism. Now comparing the phase factors of Fourier modes associated with the Bunch-Davies vacuum with the ones obtained from the ansatz for the adiabatic vacua in (\ref{eq:AdiabVacua}),  we realize the following: The Fourier modes we obtain  from the Lewis-Riesenfeld invariant formalism, that agree with the conventional one, can be understood as an adiabatic vacuum of non-linear adiabatic order, that is without any truncation, using the relation between the Ermakov equation and the defining differential equation for adiabatic vacua.
Considering the solution in (\ref{eq:vkLR}) in the limit $k \eta \gg 1$, we realize that these modes merge into the standard Minkowski modes up to an irrelevant phase and thus satisfy the adiabatic condition.  Note that we have chosen the normalization of the Wronskian in such a way that the final mode functions $v_{\bf k}$ agree, regardless of whether we chose the map that relates the MS system with a harmonic oscillator to have frequency $\omega^{(0)}_k=k$ or $\omega^{(0)}_k=1$, respectively. However, our analysis shows that on Fock space, the map that intertwines between the harmonic oscillator with $\omega^{(0)}_k=1$ and the Mukhanov-Sasaki equation cannot be implemented unitarily due to ultraviolet divergences and thus the latter choice cannot be obtained in a natural way in the Lewis-Riesenfeld formalism. For the reason that the solution in (\ref{eq:vkLR}) was obtained from a unitary transformation that maps the Mukhanov-Sasaki Hamiltonian into the harmonic oscillator Hamiltonian for all modes ${\bf k}$ with $\norm{\bf k}>\norm{{\bf k}_\epsilon}$, we can interpret this adiabatic vacuum as the natural one associated to this unitary transformation.
\\

We summarize these results of the last two sections in the diagram \ref{diagrammRes} below. We have seen that we can obtain a solution of the Mukhanov-Sasaki equation at the level of the mode functions (and find the associated vacuum) by means of the solution of the Ermakov equation $\xi_\mathbf{k}(\eta)$ combined together with a time-dependent phase that corresponds to the time rescaling from the classical theory in equation (\ref{eq:timescaling}). In our formalism we have the freedom of choosing the target frequency $\omega^{(0)}_\mathbf{k}$ as we map our Hamiltonian, where we considered two different choices in this work here. One natural choice is to just remove the time dependence and keep the time-independent $\mathbf{k}^2$ term in the frequency, which gives a transformation that is implementable for all but the infrared modes , where one can choose to modify the map appropriately as has been discussed above. It is in this sense natural to do so, since in the limit at past conformal infinity, this transformation is the identity as one would expect. Contrary to that, mapping all frequencies to unity results in a residual squeezing at very early times and most importantly in an ultraviolet divergence in the integral of the Shale-Stinespring condition. Using our results it can be shown that the non-squeezed adiabatic vacua are unitarily inequivalent to the generalized Bunch-Davies vacuum because the time-independent squeezing map that relates a harmonic oscillator with frequency $\omega^{(0)}_{\bf k}=k$ to the one with frequency $\omega^{(0)}_{\bf k}=1$ cannot be implemented as a unitary operator on Fock space.

\begin{figure}[H]
\centering	
\begin{tikzpicture}
\node (N0) at (0.0,6.5) {};     
\node (N9) at (1.0,5.2) {};     
\node (N8) at (-1.0,5.2) {};    
\node (N1) at (0,5.5) {Mukhanov-Sasaki Hamiltonian};
\node (N2) at (-4.2,2.0) {harm. oscillator $\omega^{(0)}_\mathbf{k} 
= \Norm{\mathbf{k}}$};
\node (N3) at (+4.2,2.0) {harm. oscillator $\omega^{(0)}_\mathbf{k} 
= 1$};
\node (N4) [align=left] at (-4.2,-2.8) {Bogoliubov transformation \\[0.3em] unitarily implementable \\[0.3em] (for $\Norm{\bf k} \leq \epsilon$ identity map)};
\node (N5) [align=left] at (+4.2,-2.8) {Bogoliubov transformation \\[0.3em] not unitarily implementable \\[0.3em] (divergent in the ultraviolet)};
\node (N10) [align=left] at (-4.7, -0.3) {$v_{\rm k}=N_{\bf k} \xi_{\bf k}\exp \bigg\{ -i\omega^{(0)}_{\bf k}\int^\eta \frac{d\eta'}{\xi_{\rm k}^2(\eta')} \bigg\}$ \\[0.9em] $N_\mathbf{k} = \sqrt{\ddfrac{W(v_\mathbf{k}, \overline{v}_\mathbf{k})}{2i \Norm{\mathbf{k}}}}$};
\node (N11) [align=left ]at (5.2,-0.3) {$v_{\rm k}=N_{\bf k} \xi^{(sq)}_{\bf k}\exp \bigg\{ -i\omega_{\bf k}^{(0)}\int^\eta \frac{d\eta'}{(\xi^{(sq)}_{\rm k})^2(\eta')} \bigg\}$ \\[0.9em] $N_\mathbf{k} = \sqrt{\ddfrac{W(v_\mathbf{k}, \overline{v}_\mathbf{k})}{2i}}$};
\node (N6) at (0,-3.8) {};      
\node (N7) at (-7.8,6.0) {};    
\path[line width=0.4pt,->]
(N2) edge [bend left=77] node [left] {} (N4)
(N3) edge[bend right=75] node [left] {} (N5)
(N8) edge node[left,font=\scriptsize]{$\hat{\Gamma}^\mathbf{k}_\xi$ with $\xi_\mathbf{k}(\eta_0) = 1$ \qquad} (N2)
(N2) edge node[above,font=\scriptsize]{} (N8)
(N9) edge node[right,font=\scriptsize]{\quad $\hat{\Gamma}^\mathbf{k}_{\xi^{(sq)}}$ with $\xi^{(sq)}_\mathbf{k}(\eta_0) = \big( k \big)^{-\frac{1}{2}}$} (N3)
(N3) edge node[above,font=\scriptsize]{} (N9)
(N2) edge node[above,font=\scriptsize]{$t$-indep. squeezing} (N3)
(N3) edge node[below,font=\scriptsize]{$\omega^{(0)}_\mathbf{k} \to 1$ or $\omega^{(0)}_\mathbf{k} \to \Norm{\mathbf{k}}$} (N2)
(N5) edge node[above,font=\scriptsize]{non-unitary} (N4)
(N4) edge node[below,font=\scriptsize]{residual squeezing} (N5);
\end{tikzpicture} \vspace{-0.5cm}
\caption{Graphical summary of the two different maps analyzed on Fock space.}
\label{diagrammRes}
\end{figure}
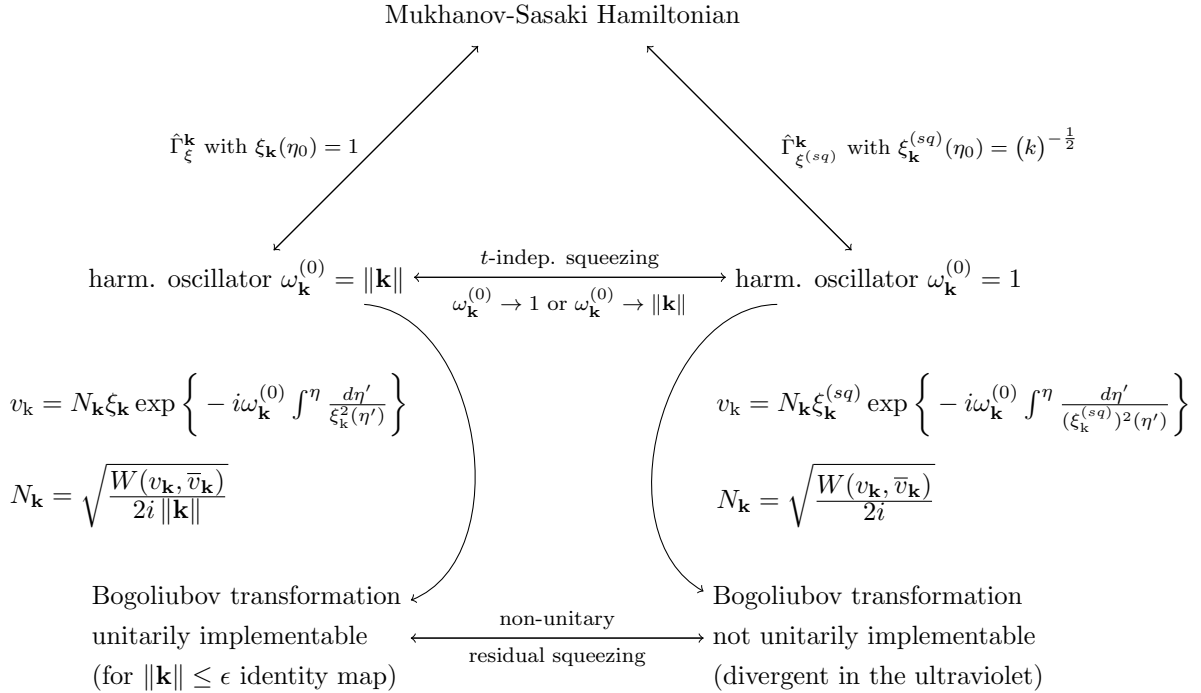

\section{Applications} 
\label{sec:applications}
\subsection{Solution of the Ermakov equation on quasi-de Sitter spacetime}
In the following we will derive and investigate a specific solution to the Ermakov equation on a quasi-de Sitter background. This leaves us with the opportunity to simplify this solution to the case of de Sitter, where the solution is known, and perform a quantitative comparison of the behavior of $\xi(\eta)$ for these two spacetimes. Our starting point is again the Mukhanov-Sasaki equation shown in (\ref{eq:Mukhanov_slow_roll}), which will be restated for the reader's convenience:

\begin{align*}
    v''_{\mathbf{k}}(\eta) +  \bigg( \Norm{\mathbf{k}}^2
- \frac{4\nu^2 - 1}{4\eta^2} \bigg) v_{\mathbf{k}}(\eta) = 0.
\end{align*}
Naturally, in the limit of vanishing slow-roll parameters $\varepsilon$, $\tau$ and $\kappa$ the equation merges into  the Mukhanov-Sasaki equation on de Sitter, i.e. we find that $\nu^2-\tfrac{1}{4} \to 2$ as expected. Next, we will bring the equation into a slightly different but also frequently used form. This is done by multiplying the entire equation by  $\eta^2$, which is possible since $\eta$ ranges from $-\infty$ to $0$, both obviously excluded. Further, we introduce new functions  $w(- k \eta)$ with $k = \Norm{\mathbf{k}}$ that are related to the original mode functions by $w(- k \eta)=\frac{v_{\mathbf{k}}}{\sqrt{-\eta}}$. This leads to the following differential equation for $w(- k \eta)$:
\begin{align}
    \chi^2 \frac{d^2 w(- \chi)}{d\chi^2} + \chi \frac{d w(- \chi)}{d\chi} + \big( \chi^2 - \nu^2 \big) w(-\chi) = 0, \quad \chi = k \eta,
    \label{eq:Bessel}
\end{align}
This is of advantage because (\ref{eq:Bessel}) precisely corresponds to the generalized Bessel differential equation with a well-studied framework of solution techniques. Primarily, we are interested in a set of two linearly independent solutions of this equation in order to construct a solution for the Ermakov equation following the path taken in \cite{Ermakov:Leach}. The most general solution to the Bessel equation is given in terms of Bessel functions $\BesselJ(-k\eta), \BesselY(-k\eta)$ of the first and second kind, respectively. These can be rewritten in terms of Hankel functions $\Hankelf(-k\eta), \Hankels(-k\eta)$ of the first and second kind, which are given by:
\begin{equation}
    \Hankelf(-k\eta) \defeq \BesselJ(-k\eta) + i \BesselY(-k\eta), \quad
    \Hankels(-k\eta) \defeq \BesselJ(-k\eta) - i \BesselY(-k\eta).
    \label{eq:Hankel_def}
\end{equation}
These functions form a linearly independent set of solutions to equation (\ref{eq:Bessel}). Introducing constants $\alpha, \beta \in \mathbb{C}$ we can give a general solution to the Mukhanov-Sasaki equation on quasi-de Sitter by means of resubstituting $v_{\mathbf{k}} = \sqrt{-\eta} \, w(- k \eta)$ and inserting the previously found basis of solutions for the Bessel equation:
\begin{align*}
    v_{\mathbf{k}}(\eta) = \sqrt{-\eta} \big( \alpha \Hankelf(-k\eta) + \beta \Hankels(-k\eta) \big)
\end{align*}
Now we follow \cite{Ermakov:Leach} and can start to construct a (unique) solution of the Ermakov equation with either the use of $\BesselJ(-k\eta), \BesselY(-k\eta)$ or $\Hankelf(-k\eta), \Hankels(-k\eta)$, respectively. This can be achieved by the following procedure, which can be straightforwardly verified by direct computation. We have
\begin{equation}
\xi_\mathbf{k}(\eta) = \sqrt{A_\mathbf{k} u_\mathbf{k}^2 + 2B_\mathbf{k} u_\mathbf{k} v_\mathbf{k} + C_\mathbf{k} v_\mathbf{k}^2}, \qquad
A_\mathbf{k} C_\mathbf{k} - B_\mathbf{k}^2 = \Norm{\mathbf{k}}^2 W(u_\mathbf{k}, v_\mathbf{k})^{-2},
\label{eq:Ermakov_prescript}
\end{equation}
where $u_\mathbf{k}, v_\mathbf{k}$ are two linearly independent solutions of the Ermakov equation, and $W(u_\mathbf{k}, v_\mathbf{k})$ denotes the Wronskian determinant. As an additional 'initial' condition next to the Wronskian, we impose the well-definedness of the solution in the limit of past conformal infinity where for each mode the Mukhanov-Sasaki equation reduces to an harmonic oscillator with constant frequency. The function $\xi_\mathbf{k}$ should also solve the Ermakov equation in this limiting case of constant frequency. Consequently, we can insert the linear independent solutions (\ref{eq:Hankel_def}) into formula in (\ref{eq:Ermakov_prescript}) and investigate the behavior for $\eta \to -\infty$. Analyzing the asymptotic behavior of the Bessel functions (and correspondingly Hankel functions) according to \cite{Watson:Bessel}, we get:
\begin{align*}
    \Hankelf(-k\eta) &\sim \sqrt{-\frac{2}{\pi k \eta}} 
    \exp \Big\{ -i \Big(k \eta + \frac{\pi}{4} (2 \nu + 1) \Big) \Big\} \qquad \textrm{for} \qquad |\eta| \gg 1, \\[0.5em]
    \Hankels(-k\eta) &\sim \sqrt{-\frac{2}{\pi k \eta}} 
    \exp \Big\{ +i \Big(k \eta + \frac{\pi}{4} (2 \nu + 1) \Big) \Big\} \qquad \textrm{for} \qquad |\eta| \gg 1.
\end{align*}
We realize that in this limit the summand under the square root in (\ref{eq:Ermakov_prescript}) is only well-defined for vanishing coefficients $A_\mathbf{k}, C_\mathbf{k}$ such that only the mixed term remains. In order to determine the coefficient $B_\mathbf{k}$ we need to find an expression for the Wronskian determinant of Hankel functions, which is non-trivial to obtain in a straightforward manner. However, we know that the Wronskian of solutions of the harmonic oscillator equation is constant in time and we have an relation for the asymptotic behavior of the Hankel functions. Given this we have 
\begin{align}
    W(\sqrt{-\eta}\Hankelf(-k\eta), \sqrt{-\eta}\Hankels(-k\eta))
    = -\eta \, W(\Hankelf(-k\eta), \Hankels(-k\eta)).
    \label{eq:Wronskian_qdS}
\end{align}
As the next step, let us rewrite the derivative with respect to conformal time of $W(\Hankelf(-k\eta), \Hankels(-k\eta))$ in terms of a differential equation by the use of its anti-symmetry and the Bessel differential equation (\ref{eq:Bessel}) obeyed by $\Hankelf, \Hankels$:
\begin{align}
    W'(\Hankelf, \Hankels) 
    = - \frac{1}{\eta} W(\Hankelf, \Hankels)  \quad
    \Longrightarrow \quad W\big( \Hankelf(-k \eta), \Hankels(-k \eta) \big) \propto 
    \frac{D_\mathbf{k}}{\eta}, \label{eq:Wronskian_Hankel}
\end{align}
where we allowed for that the constant $D_\mathbf{k}$ can vary for each mode. Note that the proportionality of the Wronskian of the Hankel functions in (\ref{eq:Wronskian_Hankel}) is in accordance with the fact that it is conserved on solutions of the Mukhanov-Sasaki equation, as seen in equation (\ref{eq:Wronskian_qdS}). Finally, after insertion of the asymptotic behavior of the Hankel functions, we find:
\begin{align*}
    W\big( \Hankelf(-k\eta), \Hankels(-k\eta) \big) \sim \frac{4i}{\pi \eta} \quad \textrm{for} \quad |\eta| \gg 1 \quad \Longrightarrow \quad D_\mathbf{k} = \frac{4i}{\pi}.
\end{align*}
Note that the Wronskian is purely imaginary, which is expected due to the negative sign of the $B_\mathbf{k}^2$ term in the condition presented in the second equation in (\ref{eq:Ermakov_prescript}) for the coefficients. If we had chosen a different route and had taken Bessel instead of Hankel functions, we would have to choose $B_\mathbf{k} = 0$ for consistency, but with a corresponding purely real Wronskian determinant. As a final result, we can determine $B_\mathbf{k}$:
\begin{equation}
    W(\sqrt{-\eta}\Hankelf(-k\eta), \sqrt{-\eta}\Hankels(-k\eta)) =
    -\eta \frac{4i}{\pi \eta} = -\frac{4i}{\pi} = \textrm{const} \quad \Longrightarrow \quad B_\mathbf{k} = -\frac{k \pi}{4}
\end{equation}
Due to the requirement that the transformation induced by $\Gamma_\xi$ should be unitary, we need $B_\mathbf{k}$ to be chosen such that the final solution $\xi_\mathbf{k}(\eta)$ is real, which is always possible in this case due to the involved squares:
\begin{equation}
    \xi_\mathbf{k}(\eta) = \sqrt{-\frac{k \pi \eta }{2} \Hankelf(-k\eta) \Hankels(-k\eta)} = \sqrt{-\frac{k \pi \eta }{2} \Big( \big( \BesselJ(-k \eta)\big)^2 + \big( \BesselY(-k \eta)\big)^2 \Big)}
    \label{eq:Ermakov_qdS}
\end{equation}
Another important aspect is the correct limit at past conformal infinity, which we can immediately deduce from the asymptotic forms of the Hankel functions above. This suggests that for each Fourier mode $\xi_\mathbf{k}(\eta)$ solves the Ermakov equation in the case where the Mukhanov-Sasaki frequency becomes a constant $\omega^{(0)}_{\mathbf{k}} \defeq \lim_{\eta \to -\infty} \omega_{\mathbf{k}} (\eta)= \Norm{\mathbf{k}}$, that is:
\begin{equation*}
    \lim_{\eta \to -\infty} \xi_\mathbf{k}(\eta) 
    = \lim_{\eta \to -\infty} \sqrt{-\frac{k \pi \eta }{2} \frac{2}{\pi k |\eta|}}
    = \lim_{\eta \to -\infty} \sqrt{- \textrm{sgn}(\eta)} = 1.
\end{equation*}
At this point we still need to investigate whether given the solution $\xi_k(\eta)$ on quasi-de Sitter we can rediscover the solution for de Sitter 
 in the case of vanishing slow-roll parameters. For this purpose, we consider the half-integer expressions for the Bessel functions:
\begin{align*}
    J_{n+\tfrac{1}{2}}(x) &= (-1)^n \sqrt{\frac{2}{\pi}} x^{n + \tfrac{1}{2}} \Big( \frac{\d}{x \d x} \Big)^n \, \frac{\sin(x)}{x} \qquad \forall \; n \in \mathbb{N}, \\[0.5em]
    Y_{n+\tfrac{1}{2}}(x) &= (-1)^{n+1} \sqrt{\frac{2}{\pi}} x^{n + \tfrac{1}{2}} \Big( \frac{\d}{x \d x} \Big)^n \, \frac{\cos(x)}{x}
    \quad \forall \; n \in \mathbb{N}.
\end{align*}
Form this we obtain an expression for $\xi_\mathbf{k}(\eta)$ on de Sitter where $\nu = 3/2$:
\begin{align*}
    \xi_\mathbf{k}^{(dS)} &= \sqrt{-\frac{k \pi \eta }{2} \bigg( \bigg( \sqrt{\frac{2}{\pi}} \frac{k \eta \cos(k \eta) - \sin(k \eta)}{(-k \eta)^{\tfrac{3}{2}}} \bigg)^2 + \bigg( \sqrt{\frac{2}{\pi}} \frac{k \eta \sin(k \eta) + \cos(k \eta)}{(-k \eta)^{\tfrac{3}{2}}} \bigg)^2 \bigg)} = \sqrt{1 + \frac{1}{(k \eta)^2}},
\end{align*}
which of course retains the same limit at past conformal infinity as the more complicated solution for non-vanishing slow-roll parameters. The solution $\xi_\mathbf{k}^{(dS)}(\eta)$ can be obtained in full analogy to the procedure outlined above using the well-known solution for the Mukhanov-Sasaki frequency on de Sitter. Depending on the particular choice of basis for the space of solutions, one needs to eliminate either of the coefficients in (\ref{eq:Ermakov_prescript}) due to the required well-definedness of the limiting case $|\eta| \to \infty$. The outcome precisely corresponds to $\xi_\mathbf{k}^{(dS)}(\eta)$ found in the limit above.
\subsection{Eigenstates of the Lewis-Riesenfeld invariant}
\begin{figure}[H]
\centering
\includegraphics[scale=0.41]{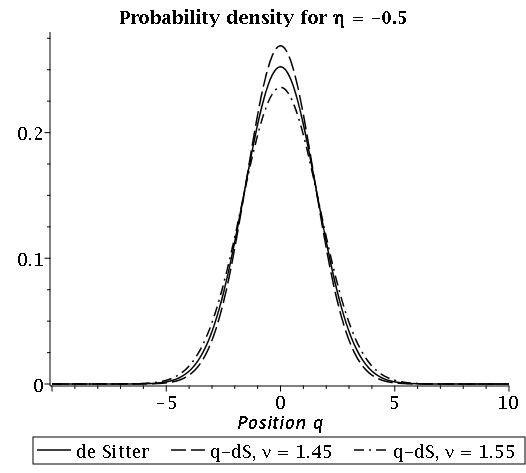}
\includegraphics[scale=0.41]{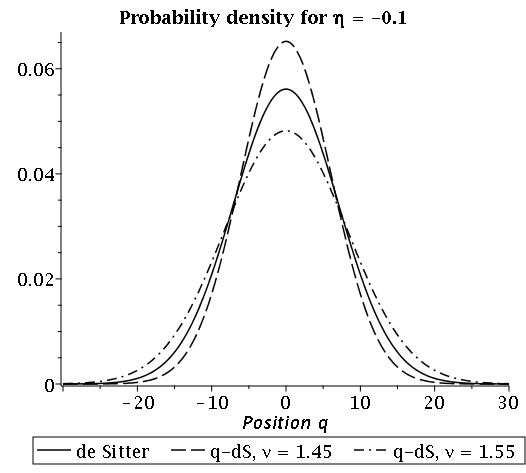}
\includegraphics[scale=0.41]{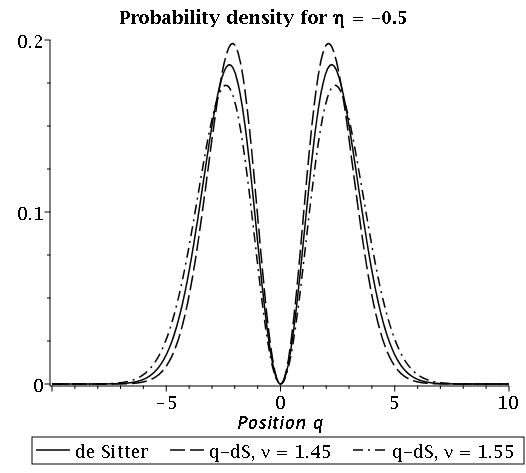}
\includegraphics[scale=0.41]{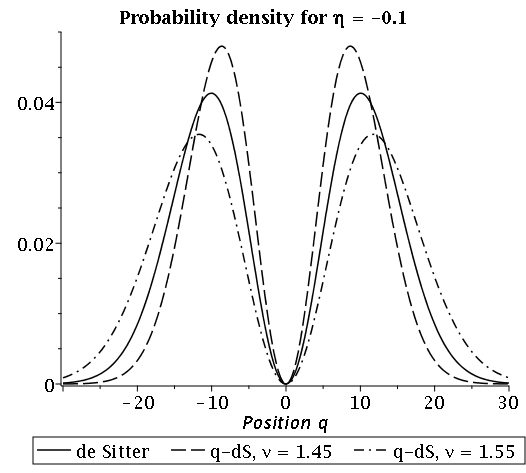}
\caption{Single-mode probability densities $|\Psi_0(q,\eta)|^2$ (upper line) and $|\Psi_1(q, \eta)|^2$ (lower line) according to the solutions in (\ref{eq:Psi_0_solution}) and (\ref{eq:Psi_1_solution}) on quasi-de Sitter for three different values of the effective slow-roll parameter $\nu$ from equation (\ref{eq:Mukhanov_slow_roll}), including de Sitter with $\nu = 3/2$ at two different conformal times and with $\omega_0 = k = 1$. The used slow-roll parameters are to be understood as an example, consider \cite{Planck_inflation:2018} for the allowed parameter space and constraints on them according to the Planck mission.}
\label{fig:Psi_solutions}
\end{figure}

As a test scenario for the formalism outlined in this work we construct and analyze the explicitly time-dependent eigenstates of the Lewis-Riesenfeld invariant. This will happen at the level of a quantum mechanical toy-model and serve the purpose of exhibiting the mathematical convenience of the formalism as well as the (squeezing) properties of the unitary transformation obtained in the context of the Lewis-Riesenfeld invariant. These eigenstates can be easily found by applying the previously obtained (inverse) Bogoliubov transformation $\hat{\Gamma}^\dagger_\xi$ to the defining property of the vacuum, that is $\hat{A} \ket{0} = 0$. We obtain:
\begin{equation*}
    \hat{\Gamma}^\dagger_\xi \hat{A} \hat{\Gamma}_\xi \hat{\Gamma}^\dagger_\xi \ket{0} 
= \textrm{Ad}_{\hat{\Gamma}^\dagger_\xi} (\hat{A}) \hat{\Gamma}^\dagger_\xi \ket{0} 
= e^{-\nu(\xi)} \big( \hat{A} - \delta_+(\xi) \hat{A}^\dagger \big) \hat{\Gamma}^\dagger_\xi \ket{0} 
\eqdef \hat{B} \hat{\Gamma}^\dagger_\xi \ket{0} = 0,
\end{equation*}
with the BCH coefficients $\nu(\xi)$ and $\delta_+(\xi)$ determined in section \ref{sec:BCH}. That is, the vacuum state of the Bogoliubov transformed annihilation operator $\hat{B}$ corresponds to the unitarily transformed initial vacuum state. Recall that $\hat{\Gamma}_\xi$ was capable of relating the time-independent Hamiltonian $\hat{H}_0$ and the Lewis-Riesenfeld invariant $\hat{I}_{LR}$ via the adjoint action, in other words, the Lewis-Riesenfeld invariant factorizes in terms of $\hat{B}, \hat{B}^\dagger$. Reexpressing the operators above in position representation we end up with a first-order differential equation for the transformed vacuum state. Understandably, this equation contains explicitly time-dependent coefficients due to the explicit time dependence of the Bogoliubov transformation. We obtain the following solution for the ground state $\Psi_0(q, \eta)$:
\begin{align}
    \Psi_0(q, \eta) &= \bigg(\frac{\omega_0}{\pi \, \xi^2(\eta)} \bigg)^\frac{1}{4} \exp \bigg\{ \bigg( \frac{i}{2} \frac{\xi'(\eta)}{\xi(\eta)} - \frac{\omega_0}{2\, \xi^2(\eta)} \bigg) q^2 \bigg\},
    \label{eq:Psi_0_solution}
\end{align}
where $\xi'(\eta)$ denotes the derivative with respect to conformal time, we again used that the mass $m=1$ here and conveniently have set $\hbar = 1$ as before. The first excited state can be obtained from $\textrm{Ad}_{\hat{\Gamma}^\dagger_\xi} (\hat{A}^\dagger) \hat{\Gamma}^\dagger_\xi \ket{0} = \hat{B}^\dagger \hat{\Gamma}^\dagger_\xi \ket{0}$ and is found to be:
\begin{align}
     \Psi_1(q, \eta) &= \bigg(\frac{\omega_0}{\pi \, \xi^2(\eta)} \bigg)^\frac{1}{4} \sqrt{\frac{2 \omega_0}{\xi^2(\eta)}} q \exp \bigg\{ \bigg( \frac{i}{2} \frac{\xi'(\eta)}{\xi(\eta)} - \frac{\omega_0}{2\, \xi^2(\eta)} \bigg) q^2 \bigg\}.
     \label{eq:Psi_1_solution}
\end{align}
Note that for a time-independent frequency $\omega(\eta) = \omega_0$ the solution merges into the standard quantum harmonic oscillator since $\hat{I}_{LR}$ and $\hat{H}_0$ coincide in this limit by construction due to $\xi(\eta) =\xi_0=1$. The details of the underlying spacetime, i.e. what determines the values  of the various slow-roll parameters enters through the solution $\xi(\eta)$ of the Ermakov equation, which is sensitive to the background via the Mukhanov-Sasaki frequency $\omega(\eta)$ and consequently through the real index $\nu$ of the Hankel and Bessel functions in the final solution in equation (\ref{eq:Ermakov_qdS}). The plots in figure (\ref{fig:Psi_solutions}) display the absolute squares of the solutions in equations  (\ref{eq:Psi_0_solution}) and (\ref{eq:Psi_1_solution}), respectively, at two different conformal times. 

Considering the explicit form of the generator of the Bogoliubov transformation $\hat{\Gamma}_\xi$, we realize that it represents a generalized squeezing operator with explicitly time-dependent coefficient functions. These coefficient functions on the other hand are sensitive to the background spacetime via the Ermakov equation and consequently the Mukhanov-Sasaki frequency $\omega_\mathbf{k}(\eta)$ involved in equation (\ref{eq:Mukhanov_slow_roll}). In this way it is expected that the eigenstates of the Lewis-Riesenfeld invariant, which are, up to a phase, eigenstates of the single-mode time-dependent Mukhanov-Sasaki Hamiltonian, show a time-dependent spread which approaches the time-independent case for very large absolute values of conformal time $|\eta| \gg 1$, that is, close to the Big Bang. 
\\

A comparison with the work in \cite{Bertoni:1997} bears a strong resemblance to the eigenstate of the Lewis-Riesenfeld invariant found there, however let us compare the results from \cite{Bertoni:1997} and ours in more detail. Firstly, the derivation in \cite{Bertoni:1997} is performed in cosmological time, whereas we have made the transition to conformal time beforehand, so the explicit occurrences of the scale factor are absent in our work. Secondly, when having a closer look at the Ermakov equation in \cite{Bertoni:1997} it becomes evident that in the context of the canonical transformation $\Gamma_\xi$, the time-independent frequency of the Hamiltonian $H_0$ is unity. As a consequence, this means for the case of field theory that every mode with $\omega_\mathbf{k}(\eta)$ would be mapped to exactly the \textit{same} frequency $\omega^{(0)}_{\mathbf{k}} = 1$, which modifies the solution of the Ermakov equation by an additional $k^{-\frac{1}{2}}$, leading to an ultraviolet divergence in the Shale-Stinespring condition (\ref{eq:Shale_condition}) and diminishing the ability to implement it by a separate treatment of the infrared modes with $\norm{\mathbf{k}} \leq \norm{{\bf k}_\epsilon}$. Thirdly, the authors in \cite{Bertoni:1997} claim that the creation- and annihilation operators that decompose the time \textit{dependent} Hamiltonian are related to the ones associated with the Lewis-Riesenfeld invariant via a Bogoliubov transformation. According to our analysis, while this is true, their relation is more subtle: As it is true that $H(t)$ can be mapped into $I_{LR}$ in the classical theory by means of an extended symplectic map (which in a sense corresponds to a Bogoliubov transformation quantum mechanically), this might not be straightforwardly implementable in the quantum theory even on the one-particle Hilbert space. It can be implemented if and only if the  time-rescaling function is chosen such that $\xi(t)^{-2}$ has an analytic anti-derivative, which is for example the case on a de Sitter background. The exponential sandwiched between $\hat{\Gamma}^\dagger_\xi$ and $\hat{\Gamma}_{\xi,0}$ in (\ref{eq:time_evolution_op}) can then be rewritten as the exponential of an analytic function in the time \textit{operator}, conjugate to the momentum operator $\hat{p}_t$. This is the reason why we perform a reduced phase space quantization of Dirac observables, where this problem is absent, rather than Dirac quantization. Consequently, the transformation $\hat{\Gamma}_\xi$ acts as a one-parameter (i.e. time-dependent) family of unitary transformations on the reduced (physical) phase space. This transformation is suitable for transforming the Hamiltonian within the Schr\"odinger equation into the  independent one $\hat{H}_0$, which again can be related to the invariant $\hat{I}_{LR}$ by means of a time-dependent Bogoliubov transformation.

\section{Conclusions}
\label{sec:Conclusions}
In this work we used the method of the Lewis-Riesenfeld invariant in order to analyze in which sense the dynamical properties of the Mukhanov-Sasaki equation select possible candidates for initial states in the context of inflation.
We started in the classical theory and rederived a time-dependent canonical transformation that relates the system of a time-dependent harmonic oscillator to the one of a time-independent harmonic oscillator, where in our case the explicit time-dependence is determined by the frequency involved in the Mukhanov-Sasaki equation. Using this map, the entire time dependence of the oscillator can be removed leading to a simplification as far as finding solutions for the time-dependent system is concerned. As a first step, this was done for systems with finitely many degrees of freedom using an extended phase space in which time and its corresponding momentum are part of the phase space, following the work in \cite{Struckmeier}. This has the advantage that the time rescaling involved in this transformation can be naturally embedded in the framework of the extended phase space, whereas in early work such as \cite{Hartley:1982ygx}, the corresponding phase factor needs to be introduced with a less clear physical motivation. The ansatz we used employed techniques introduced in \cite{Struckmeier} and was based on an idea of \cite{Chung}, which was generalized to arbitrary even and finite phase space dimensions. This transformation revealed the relationship between the Lewis-Riesenfeld invariant, the time-dependent and time-independent harmonic oscillator Hamiltonian already implicitly used in \cite{Hartley:1982ygx}. 
\\

Since at the level of the extended phase space the system of the time-dependent harmonic oscillator is a constrained system, we could choose between Dirac and reduced quantization in order to later extract the physical sector of the quantum theory. Because the time rescaling included in the canonical transformation involves an integral over a time interval, a Dirac quantization might be problematic if an analytic expression for the anti-derivative does not exist. In order to circumvent this problem, we chose reduced phase space quantization for which it was necessary to construct Dirac observables and consider the physical phase space according to the methods introduced in \cite{Dittrich:Observables} and \cite{Thiemann:ReducedPS}. Fortunately, the Dirac observables satisfy the standard canonical algebra so that representations thereof can be easily found. Their dynamics is generated by the Dirac observable associated with the time-dependent Hamiltonian that consequently becomes the physical Hamiltonian of the system. 
As a preparation for the quantum theory, we constructed the associated generator of the canonical transformation on the physical phase space, giving rise to a corresponding flow that represents the canonical transformation on the physical degrees of freedom. A crucial ingredient for the construction of the Lewis-Riesenfeld invariant as well as the corresponding canonical transformation that removes the time dependence from the Hamiltonian is a time-dependent auxiliary function $\xi(t)$ that has to satisfy the Ermakov differential equation.  This requirement ensures that the Lewis-Riesenfeld invariant is a quadratic polynomial in the elementary configuration and momentum variables, that can be interpreted as a Dirac observable in the extended phase space because it commutes with the first class constraint. Given a solution of the Ermakov equation, we can use it to construct the canonical transformation on the finite dimensional physical phase space, whose generator is shown in (\ref{eq:ClassGen}). 
\\

For the quantum theory we first restricted our analysis to the case of finitely many degrees of freedom and as a special case considered the one-particle Hilbert space. All results obtained in this context can be easily generalized to more but finite degrees of freedom. On the one-particle Hilbert space, the time-dependent canonical transformation can be implemented as a unitary operator $\hat{\Gamma}_\xi$ whose explicit form is given in (\ref{eq:GammaXi}). With the help of $\hat{\Gamma}_\xi$ we can remove the explicit time dependence of the Schr\"odinger equation, analogous to the treatment in \cite{Schroedinger_1D} and map it to a Schr\"odinger equation involving a time-independent harmonic oscillator Hamiltonian. In this way we further obtain a time evolution operator $\hat{U}(t_0,t)$ which can be shown to correspond to the Dyson series of the time-dependent theory, consisting of three individual unitary operators. When having a closer look at the occurring exponentials, there is an obvious explanation of the required additional phase factor in the solutions of the Schr\"odinger equation in \cite{Hartley:1982ygx}, amounting to some function multiplying either the exponentiated time-independent Hamiltonian or the Lewis-Riesenfeld invariant, depending on the relative ordering we chose in $\hat{U}(t_0,t)$. The immediate effect of the time-rescaling $\xi(t)$ becomes evident in this exponential, as the relative phase is sensitive to the background spacetime. Each of the operators in $\hat{U}(t_0,t)$ corresponds to an exponentiated unitary representation of the $\mathfrak{sl}(2, \mathbb{R})$ Lie algebra, as can be shown by explicitly evaluating the Lie brackets of the generators. For practical computations and later applications, we used a generalized Baker-Campbell-Hausdorff decomposition of unitary representations of non-compact groups shown in \cite{BCH_Decomp} to decompose $\hat{\Gamma}_\xi$ into normal- or anti-normal ordered contributions, respectively. This gave us the possibility to rewrite the unitary transformation $\hat{\Gamma}_\xi$ on the one-particle Hilbert space as an explicitly time-dependent Bogoliubov transformation, where the time dependence enters through the solution $\xi(t)$ of the Ermakov equation. 
\\

This was an important preparation for the generalization to the field theory context we were mainly interested in in this work. The crucial criterion for the existence of a unitary implementation of a Bogoliubov transformation on the Fock space is the Shale-Stinespring condition \cite{Solovej_Diagonalization}, which essentially denotes that  the product of off-diagonal entries of the Bogoliubov map needs to be Hilbert-Schmidt. As our results show a straightforward generalization to Fock space where the time-dependent oscillator is described by the Mukhanov-Sasaki equation and the target system is for each mode a harmonic oscillator with constant frequency does not work because either infrared or infrared and ultraviolet divergences occur, leading to a violation of the Shale-Stinespring condition. Here we considered two common choices used in the existing literature, where the constant frequency is either $\omega^{(0)}_{\bf k}=k$ or $\omega^{(0)}_{\bf k}=1$ respectively, which corresponds to two slightly different Ermakov equations in our case. Both choices yield an infrared divergence caused by the infrared modes, whereas for the second choice in addition an ultraviolet divergence occurs. If we compare our results obtained with the existing results in the literature, the work in \cite{invariant_vacuum} takes as the starting point a charged massive scalar field in a de Sitter space time and hence the Ermakov equation in this case includes an additional friction term and cannot directly be compared to our result. The author of \cite{invariant_vacuum} also uses the map to a harmonic oscillator with frequency $\omega^{(0)}_{\bf k}=k$ and also obtains no ultraviolet divergences for his slightly different map. However, as far as we can see, a careful analysis of the Shale-Stinespring condition is not presented in \cite{invariant_vacuum} and thus we expect that, similar to our case, infrared singularities are present. In \cite{Villasenor} the map with constant frequency $\omega^{(0)}_{\bf k}=1$ was considered and in agreement with our results, they also obtain an ultraviolet divergence for the operator $\hat{\Gamma}_\xi$. In their work the theory is defined on a torus allowing them to isolate the zero mode and exclude it from their analysis, hence no infrared divergences occur. Thus, we can conclude that the second choice, where the target frequency is chosen to be $\omega^{(0)}_{\bf k}=1$, cannot be implemented unitarily on Fock space, whereas for the first choice there might be a chance to find a unitary implementation for the first case with target frequency $\omega^{(0)}_{\bf k}=k$ if we are able to consistently modify the map for the infrared modes such that the infrared divergence are no longer present. One possibility can be to also formulate a model where the spatial slices have the topology of a torus chosen in  such a way that experimentally one cannot distinguish between a model whose spatial slices have the topology of a torus and one with non compact spatial slices. In this case we could also exlucde the zero mode and modify the map for this specific mode in a way that the Shale-Stinespring condition is satisfied. The corresponding Bolgoliubov transformation can then be defined for all but the zero mode. Note that this case further allows to identify the background with the zero mode, as it is for example usually done in hybrid loop quantum cosmology, see for instance \cite{ElizagaNavascues:2016vqw}. However, if we stick to non-compact spatial slices we have to consider a slightly different strategy. 
\\

As possible solutions in this direction we discussed the quantum Arnold transformation introduced in \cite{Arnold_trafo:Guerrero}. The goal was to apply it to the modes below a certain infrared cutoff, that is for the sphere with $\norm{{\bf k}} \leq \norm{{\bf k}_\epsilon}$ in Fourier space. We found that the Arnold transformation cannot be understood as a Bogoliubov map, since it renders the creation and annihilation operators in the transformed picture equal up to a global sign. A closer look at the involved terms shows that the reason for this pathological behavior is the absence of a the $q^2$ contribution in the Lewis-Riesenfeld invariant. It is precisely this aspect that disrupts the commutator algebra and hence does not qualify as an infrared extension of $\hat{\Gamma}_\xi$. Our proposal for an infrared extension is to use the identity map within the cutoff region, the reasons are twofold. Firstly, the identity map can be trivially regarded as a Bogoliubov transformation that exists on this sector of the Fock space. Secondly, by not altering the form of the Hamiltonian in the infrared regime, the off-diagonal terms remain as in the standard case, which means that they are subdominant for very early times where $k\eta \ll 1$. Hence, by adopting this strategy we are able to define a unitarily implementable Bogoliubov transformation on the entire Fock space as depicted in equation (\ref{eq:full_bogoliubov}), that performs a Hamiltonian diagonalization on all modes with norm greater than the infrared cutoff $\norm{{\bf k}_\epsilon}$.
\\

In section \ref{sec:RelAdVacua} we showed how the solution of the Ermakov equation can be used to construct a solution of the Mukhanov-Sasaki equation and as expected, the time-rescaling plays a pivotal role here. If we rewrite the solution to the Mukhanov-Sasaki equation in a polar representation as shown in (\ref{eq:polar_modes}), the Mukhanov-Sasaki equation requires that the real part in the polar representation needs to be a solution of the Ermakov equation, opening a clear connection between the two formalisms. Following this route further, we can also recover the defining differential equation for adiabatic vacua from the Ermakov equation, meaning that if we have a solution of the Ermakov equation given, from this we can easily construct a non-linear solution of the adiabatic vacua differential equation, where non-linear refers to the fact that it is a full solution without truncating the solution at any adiabatic order. The adiabatic condition usually required in this context carries over to a condition on the solution of the Ermakov equation for each mode. This in turn can be directly related to specific properties of the unitary map corresponding to the time-dependent canonical transformation between the time-dependent and time-indepdendent harmonic oscillator. Hence, there is an interesting interplay between the characteristic properties of the unitary map and the choice of adiabatic vacua. Considering this and the fact that we set our initial condition at the limit of conformal past infinity, the Lewis-Riesenfeld method leads to mode functions that can be interpreted as a non-squeezed adiabatic vacuum of non-linear order, that is without performing any truncation. The property of being non-squeezed reflects our freedom of choice of mapping to a target frequency $\omega_{\bf k}^{(0)}=k$ that causes no residual squeezing if we apply the unitary operator to an already time-independent harmonic oscillator. Furthermore, the time rescaling involved in the mode function becomes $e^{-ik\eta}$ in the limit of large (negative) conformal times, showing that the mode function obtained here are compatible with the condition used in the Bunch-Davies case.
\\

 Finally, in section \ref{sec:applications} we have illustrated how the formalism we used throughout this work can be used in terms of computing eigenstates of the Lewis-Riesenfeld invariant associated to a particular system. In our case, this was a time-dependent harmonic oscillator corresponding to the Mukhanov-Sasaki equation on a quasi-de Sitter spacetime for a single mode. Together with the construction of these eigenstates via $\hat{\Gamma}_\xi$, we outlined how to find a solution $\xi(\eta)$ of the Ermakov equation with this particular time-dependent frequency $\omega(\eta)$ corresponding to the background geometry of quasi-de Sitter. This was done by using the asymptotic behavior of the Hankel functions (which solve the Mukhanov-Sasaki equation on quasi-de Sitter analytically) and the fact that the Wronskian of two linearly independent solutions of this equation is a constant. Finally, we provided a visualization of the time-dependent squeezing operation $\hat{\Gamma}_\xi$ in terms of the probability densities of two time-dependent eigenstates of $\hat{I}_{LR}$ for different values of the slow-roll parameters, which precisely correspond to the probability densities of solutions of the Schr\"odinger equation of the associated time-dependent Hamiltonian $H(t)$. As far as the computation of the power spectrum is concerned, we do not expect new insights from our obtained results because what enters into the computation is the final Fourier mode that we constructed in both cases in such a way that the results agree with the standard result for the Mukhanov-Sasaki mode. Our results however, give new insights on whether there exists a time-independent harmonic oscillator Hamiltonian associated with the Mukhanov-Sasaki Hamiltonian that is unitarily equivalent in the field theory context. In future work we want to analyze applications of this formalism to other than quasi de Sitter spacetimes. This requires in particular to find solutions of the Ermakov equation in this more general case and analyze whether the corresponding transformation can be implemented unitarily. Furthermore, we plan to investigate in future research how the transformation in the classical theory on the extended phase space can be lifted to the field theory context. This might be realizable in the framework of the Gaussian dust model presented in \cite{Giesel:2012rb} where the dust fields can be used as reference fields for physical temporal and spatial coordinates. 

\section*{Acknowledgements}
M.K. and K.G. would like to thank Hanno Sahlmann for illuminating and productive discussions during the project as well as Beatriz Elizaga Navascués and Thomas Thiemann for fruitful discussions towards the end of this work. M.K. thanks the Heinrich-B\"oll Foundation for financial support.

\bibliography{PaperMSHam.bib}

\begin{thebibliography}{10}

\bibitem{Mukhanov:1990me}
Viatcheslav~F. Mukhanov, H.~A. Feldman, and Robert~H. Brandenberger.
\newblock {Theory of cosmological perturbations. Part 1. Classical
  perturbations. Part 2. Quantum theory of perturbations. Part 3. Extensions}.
\newblock {\em Phys. Rept.}, 215:203--333, 1992.

\bibitem{Mukhanov}
Viatcheslav Mukhanov.
\newblock {\em Physical Foundations of Cosmology}.
\newblock Cambridge University Press, 2005.

\bibitem{Langlois:1994ec}
D.~Langlois.
\newblock {Hamiltonian formalism and gauge invariance for linear perturbations
  in inflation}.
\newblock {\em Class. Quant. Grav.}, 11:389--407, 1994.

\bibitem{Giesel:2017roz}
Kristina Giesel and Adrian Herzog.
\newblock {Gauge invariant canonical cosmological perturbation theory with
  geometrical clocks in extended phase-space — A review and applications}.
\newblock {\em Int. J. Mod. Phys.}, D27(08):1830005, 2018.

\bibitem{Giesel:2018opa}
Kristina Giesel, Adrian Herzog, and Parampreet Singh.
\newblock {Gauge invariant variables for cosmological perturbation theory using
  geometrical clocks}.
\newblock {\em Class. Quant. Grav.}, 35(15):155012, 2018.

\bibitem{Giesel:2018tcw}
Kristina Giesel, Parampreet Singh, and David Winnekens.
\newblock {Dynamics of Dirac observables in canonical cosmological perturbation
  theory}.
\newblock 2018.

\bibitem{Danielsson:2002mb}
Ulf~H. Danielsson.
\newblock {On the consistency of de Sitter vacua}.
\newblock {\em JHEP}, 12:025, 2002.

\bibitem{ArmendarizPicon:2003gd}
C.~Armendariz-Picon and Eugene~A. Lim.
\newblock {Vacuum choices and the predictions of inflation}.
\newblock {\em JCAP}, 0312:006, 2003.

\bibitem{Handley:2016ods}
W. J. Handley, A. N. Lasenby, and M. P. Hobson.
\newblock {Novel quantum initial conditions for inflation}.
\newblock {\em Phys. Rev.}, D94(2):024041, 2016.

\bibitem{Struckmeier}
Jürgen Struckmeier.
\newblock Hamiltonian dynamics on the symplectic extended phase space for
  autonomous and non-autonomous systems.
\newblock {\em Journal of Physics A: Mathematical and General}, 38(6):1257,
  2005.

\bibitem{Chung}
Angel Garcia-Chung, Daniel~Guti{\'e}rrez Ruiz, and J~David Vergara.
\newblock Dirac's method for time-dependent hamiltonian systems in the extended
  phase space.
\newblock {\em arXiv preprint arXiv:1701.07120}, 2017.

\bibitem{Hartley:1982ygx}
John~G. Hartley and John~R. Ray.
\newblock {Coherent states for the time-dependent harmonic oscillator}.
\newblock {\em Phys. Rev.}, D25(2):382, 1982.

\bibitem{Villasenor}
Daniel Gómez-Vergel and Eduardo Villaseñor.
\newblock The time-dependent quantum harmonic oscillator revisited:
  Applications to quantum field theory.
\newblock 324:1360--1385, 03 2009.

\bibitem{invariant_vacuum}
Salvador Robles-Perez.
\newblock {Invariant vacuum}.
\newblock {\em Phys. Lett.}, B774:608--615, 2017.

\bibitem{Arnold_trafo:Guerrero}
Julio Guerrero and Francisco~F López-Ruiz.
\newblock The quantum arnold transformation and the ermakov–pinney equation.
\newblock {\em Physica Scripta}, 87(3):038105, 2013.

\bibitem{Schroedinger_1D}
M~Fernández Guasti and H~Moya-Cessa.
\newblock Solution of the schrödinger equation for time-dependent 1d harmonic
  oscillators using the orthogonal functions invariant.
\newblock {\em Journal of Physics A: Mathematical and General}, 36(8):2069,
  2003.

\bibitem{Dittrich:Observables}
B.~Dittrich.
\newblock {Partial and complete observables for Hamiltonian constrained
  systems}.
\newblock {\em Gen. Rel. Grav.}, 39:1891--1927, 2007.

\bibitem{Thiemann:ReducedPS}
Thomas Thiemann.
\newblock {Reduced phase space quantization and Dirac observables}.
\newblock {\em Class. Quant. Grav.}, 23:1163--1180, 2006.

\bibitem{RovelliPartial}
C.~Rovelli.
\newblock Partial observables.
\newblock {\em Phys. Rev. D}, 65:124013, 2002.
\newblock \href{https://arxiv.org/abs/gr-qc/0110035}{gr-qc/0110035}.

\bibitem{RovelliObservable}
C.~Rovelli.
\newblock What is observable in classical and quantum gravity?
\newblock {\em Class. Quant. Grav.}, 8(2):297, 1991.

\bibitem{Giesel:2007wi}
K.~Giesel, S.~Hofmann, T.~Thiemann, and O.~Winkler.
\newblock {Manifestly Gauge-Invariant General Relativistic Perturbation Theory.
  I. Foundations}.
\newblock {\em Class. Quant. Grav.}, 27:055005, 2010.

\bibitem{Giesel:2007wk}
K.~Giesel, S.~Hofmann, T.~Thiemann, and O.~Winkler.
\newblock {Manifestly Gauge-invariant general relativistic perturbation theory.
  II. FRW background and first order}.
\newblock {\em Class. Quant. Grav.}, 27:055006, 2010.

\bibitem{Giesel:2007wn}
K.~Giesel and T.~Thiemann.
\newblock {Algebraic quantum gravity (AQG). IV. Reduced phase space
  quantisation of loop quantum gravity}.
\newblock {\em Class. Quant. Grav.}, 27:175009, 2010.

\bibitem{Domagala:2010bm}
Marcin Domagala, Kristina Giesel, Wojciech Kaminski, and Jerzy Lewandowski.
\newblock {Gravity quantized: Loop Quantum Gravity with a Scalar Field}.
\newblock {\em Phys. Rev.}, D82:104038, 2010.

\bibitem{Giesel:2012rb}
Kristina Giesel and Thomas Thiemann.
\newblock {Scalar Material Reference Systems and Loop Quantum Gravity}.
\newblock {\em Class. Quant. Grav.}, 32:135015, 2015.

\bibitem{Husain:2011tk}
Viqar Husain and Tomasz Pawlowski.
\newblock {Time and a physical Hamiltonian for quantum gravity}.
\newblock {\em Phys. Rev. Lett.}, 108:141301, 2012.

\bibitem{Han:2015jsa}
Yu~Han, Kristina Giesel, and Yongge Ma.
\newblock {Manifestly gauge invariant perturbations of scalar–tensor theories
  of gravity}.
\newblock {\em Class. Quant. Grav.}, 32:135006, 2015.

\bibitem{Giesel:2016gxq}
Kristina Giesel and Almut Oelmann.
\newblock {Reduced Loop Quantization with four Klein-Gordon Scalar Fields as
  Reference Matter}.
\newblock 2016.

\bibitem{BCH_Decomp}
D~Rodney~Truax.
\newblock Baker-campbell-hausdorff relations and unitarity of su(2) and su(1,1)
  squeeze operators.
\newblock 31:1988--1991, 05 1985.

\bibitem{Solovej_Diagonalization}
Phan~Th{\`a}nh Nam, Marcin Napi{\'o}rkowski, and Jan~Philip Solovej.
\newblock Diagonalization of bosonic quadratic hamiltonians by bogoliubov
  transformations.
\newblock {\em Journal of Functional Analysis}, 270(11):4340--4368, 2016.

\bibitem{Arnold:1978}
V.~I. Arnold.
\newblock {Supplementary chapters to the theory of ordinary differential
  equations}.
\newblock {\em Nauka, Moscow, 1978, Springer. English transl., Geometrical
  methods in the theory of ordinary differential equations -Verlag, New
  York-Berlin, 1983}.

\bibitem{Planck_inflation:2018}
Y.~Akrami et~al.
\newblock {Planck 2018 results. X. Constraints on inflation}.
\newblock 2018.

\bibitem{Winitzki:2005rw}
Sergei Winitzki.
\newblock {Cosmological particle production and the precision of the WKB
  approximation}.
\newblock {\em Phys. Rev.}, D72:104011, 2005.

\bibitem{Casadio:2004ru}
Roberto Casadio, Fabio Finelli, Mattia Luzzi, and Giovanni Venturi.
\newblock {Improved WKB analysis of cosmological perturbations}.
\newblock {\em Phys. Rev.}, D71:043517, 2005.

\bibitem{Ermakov:Leach}
PGL Leach and K~Andriopoulos.
\newblock The ermakov equation: a commentary.
\newblock {\em Applicable Analysis and Discrete Mathematics}, pages 146--157,
  2008.

\bibitem{Watson:Bessel}
George~Neville Watson.
\newblock {\em A treatise on the theory of Bessel functions}.
\newblock Cambridge university press, 1995.

\bibitem{Bertoni:1997}
C.~Bertoni, F.~Finelli, and Giovanni Venturi.
\newblock {Adiabatic invariants and scalar fields in a de Sitter space-time}.
\newblock {\em Phys. Lett.}, A237:331--336, 1998.

\bibitem{ElizagaNavascues:2016vqw}
Beatriz Elizaga~Navascués, Mercedes Martín-Benito, and Guillermo~A.
  Mena~Marugán.
\newblock {Hybrid models in loop quantum cosmology}.
\newblock {\em Int. J. Mod. Phys.}, D25(08):1642007, 2016.

\end{thebibliography}
\bibliographystyle{unsrt}

\end{document}